\documentclass[12pt]{article}

\usepackage{amsmath,amssymb}
\usepackage{authblk}
\usepackage{enumitem}  
\usepackage[linktocpage]{hyperref}  
\usepackage[usenames, dvipsnames]{color} 
\usepackage{graphicx}  
\usepackage[titletoc,title]{appendix}

 \usepackage{xcolor}
 
\usepackage{latexsym}
\usepackage{amssymb}
\usepackage{amsmath}
\usepackage{graphicx}
\usepackage{enumitem}
\usepackage{slashed}
\usepackage{hyperref} 
\usepackage{endnotes}

\newcommand\mathC{\mkern1mu\raise2.2pt\hbox{$\scriptscriptstyle|$}
        {\mkern-7mu\rm C}}              

\def\be{\begin{equation}}
\def\ee{\end{equation}}
\def\bear{\begin{eqnarray}}
\def\eear{\end{eqnarray}}

\def\a{\alpha}
\def\b{\beta}

\def\d{\delta}

\def\Tr{{\rm Tr}}

\def\dd{\mbox{d}}

\def\a{\alpha}
\def\b{\beta}
\def\d{\delta}

\def\f{\phi}

\def\m{\mu}
\def\n{\nu}

\newcommand{\ti}[1]{\tilde{#1}}

\newcommand{\sm}[1]{\mbox{\scriptsize #1}}
\newcommand{\tn}[1]{\mbox{\tiny #1}}
\renewcommand{\@}[1]{\sqrt{#1}}
\renewcommand{\le}[1]{\label{#1}\end{eqnarray}}
\newcommand{\bea}{\begin{eqnarray}}
\newcommand{\eea}{\end{eqnarray}}

\newcommand{\eq}[1]{(\ref{#1})}

\def\ffract#1#2{\raise .35 em\hbox{$\scriptstyle#1$}\kern-.25em/
\kern-.2em\lower .22 em \hbox{$\scriptstyle#2$}}

\begin{document}

\pagestyle{empty}
\numberwithin{equation}{section}

\centerline{{\Large\bf  Conceptual Analysis of Black Hole Entropy}} 
\vskip.25truecm
\centerline{{\Large\bf in String Theory}}
\vskip.95truecm

\begin{center}
{\large Sebastian De Haro,$^{1,2,3,5}$ Jeroen van Dongen,$^{4,5}$ }
\vskip.20cm
{\large Manus Visser,$^4$ and Jeremy Butterfield$^1$}\\
\vskip .75truecm
$^1${\it Trinity College, Cambridge}\\
$^2${\it Department of History and Philosophy of Science, University of Cambridge}\\
$^3${\it Black Hole Initiative, Harvard University}\\
$^4${\it Institute for Theoretical Physics, University of Amsterdam}\\
$^5${\it Vossius Center for the History of Humanities and Sciences, University of Amsterdam}

\vskip 1cm
\today
\\
\end{center}
\vskip 1cm

\begin{center}
\textbf{ \bf Abstract}
\end{center}
The microscopic state counting of the extremal Reissner-Nordstr\"om black hole performed by Andrew Strominger and Cumrun Vafa in 1996 has proven to be a central result in string theory. Here,  with a philosophical readership in mind, the argument is presented in its contemporary context and its rather complex conceptual structure is analysed. In particular, we will identify the various inter-theoretic relations, such as duality and linkage relations, on which it depends. We further aim to make clear why the argument was immediately recognised as a successful accounting for the entropy of this black hole and how it engendered subsequent work that intended to strengthen the string theoretic analysis of black holes. Its relation to the formulation of the AdS/CFT conjecture will be briefly discussed, and the familiar  reinterpretation of the entropy calculation in the context of the AdS/CFT correspondence is given. Finally, we discuss the heuristic role that Strominger and Vafa's microscopic account of black hole entropy played for the black hole information paradox. A companion paper analyses the ontology of the Strominger-Vafa black hole states, the question of emergence of the black hole from a collection of D-branes, and the role of the correspondence principle in the context of string theory black holes.\\
\newpage
\pagestyle{plain}
 \tableofcontents

\newpage

\section{Introduction}\label{sec:intro}

On 9  January   1996, Andrew Strominger and Cumrun Vafa posted a short article in the high energy theory, or `hep-th', section of the physics preprint webserver (the `arXiv', then xxx.lanl.gov). As the title of their article---`Microscopic Origin of the Bekenstein-Hawking Entropy'---indicates, it contained a microscopic calculation of the entropy of a black hole. This, in the eyes of many string theorists, provided a first microphysical account of black hole entropy, and was taken to confirm  the Bekenstein-Hawking entropy formula\footnote{Named after Jacob Bekenstein (1972, 1973) and Stephen Hawking (1975).}, which states that the entropy of a black hole is equal to a quarter of its horizon area in Planckian units.  
The calculation followed closely a series of key developments in string theory, which were spawned by Edward Witten's duality conjectures and Joseph Polchinski's (re)introduction of D-branes, both in 1995. With these in hand, Strominger and Vafa could finally count a set of quantum states that, through intricate identifications and approximations, they identified with the microstates of a black hole. 
This microstate counting of black hole entropy is perceived by string theorists as one  of the major successes of their theory, and its publication is among the most highly  cited articles in high energy theoretical physics.\footnote{For example, Becker et al.~(2007:~p.~14) write: ``[T]hese studies have led to a much deeper understanding of the thermodynamic properties of black holes in terms of string-theory microphysics, a fact that is one of the most striking successes of string theory so far.'' According to a Google Scholar citation count, Strominger and Vafa (1996) has been cited over 3000 times. The 2017 edition of the `Top Cited Articles of All Time' for the hep-th preprint archive of the INSPIRE high energy website places the article at no. 16; see  http://inspirehep.net/info/hep/stats/topcites/2017/eprints/to\_hep-th\_alltime.html  (both data retrieved on 2 February 2019).}

Recently, historians and philosophers of science have begun to focus on quantum gravity and string theory. As the above may illustrate, the microstate counting of black hole entropy is a key result in these subjects, in particular because it is intimately tied to the black hole information paradox, which has been of interest to philosophers for some time.\footnote{See e.g.~Belot, Earman and Ruetsche (1999); Dongen and De Haro (2004);  Maudlin (2017); Wallace (2018, 2019). See also J.~van Dongen and S.~De Haro  (Forthcoming), `History and Philosophy of the Black Hole Information Paradox' for a historical account of the debate.} 
Indeed, 
many physicists saw the microphysical entropy calculation provided by Strominger and Vafa, through its identification  of quantum states for the black hole, 
as a first strong indication that non-unitary scenarios for black hole evolution would be incorrect.

The black hole state counting depends on string theory duality, which 
is currently also receiving considerable attention in the philosophy of physics literature.\footnote{On duality, see for example Rickles (2011); Matsubara (2013); Dieks, Dongen and De Haro (2015); De Haro (2017a, 2017b); De Haro and Butterfield (2018); Castellani (2017); Huggett (2017); Read and M\o ller-Nielsen (2018); for a general history of string theory, see Cappelli et al.~(2012) and Rickles (2014); for a discussion of its methodology, see Dawid (2013).}  
The counting result was an essential element in the developments that led to the AdS/CFT duality; and should thus be of great interest to those who study the epistemology of duality. Strominger and Vafa's calculation raises important questions of interpretation---e.g.~what states, that is, states of \emph{what system}---are being counted? Can they justifiably be identified as `black hole quantum states'? Clearly, issues of ontology, epistemology and the interpretative practices of modern physics stand at the centre here, and merit the attention of historians and philosophers of modern physics.

This paper offers an introduction to Strominger and Vafa's 1996 result by outlining its conceptual structure and its place in the string theory literature of the mid-1990s. This will aid the philosophical interpretation of the argument, and illuminate its role in understanding black hole entropy. The present article offers primarily a conceptual analysis and contextualization of the entropy calculation; a first treatment of larger philosophical and interpretative questions is given in a companion paper, entitled `Emergence and Correspondence for   String Theory Black Holes'.  \\
\\
Our analysis here focusses on two questions in particular:  
\\
\\
(i)~~Why did Strominger and Vafa's entropy calculation have such  persuasive power in the mid-1990s and has continued to do so, so that most string theorists believe that it has provided a microscopic underpinning of the Bekenstein-Hawking black hole entropy formula,
and thus promises a resolution of the black hole information paradox?\\
\\
(ii)~~What was its significance for further developments in string theory in the (roughly) two years that followed its formulation, which eventually saw the formulation of the AdS/CFT correspondence?\\
\\
To address issue (i), we will identify the inter-theoretic relations (dualities and linkage relations) that 
the Strominger-Vafa entropy counting argument
depends on. `Factorizing' the argument in this way will aid its conceptual assessment: in particular because the reader of Strominger and Vafa's paper is only cursorily alerted to some of their key assumptions (if at all)---which of course reflects the fact that their article was written primarily for an audience of specialists who had been voraciously consuming and producing preprints on dualities and black holes throughout the previous year. We will sketch these contemporary developments---both for the sake of historiography, and to further our notion of `understanding' in modern quantum gravity.\footnote{See also De Haro and de Regt (2018).} 

We will also discuss how the result was generalised.  This will show how some of its weaknesses were addressed, and illustrates its larger influence within the string theory literature. In particular, we will discuss, albeit briefly, how the entropy calculation and its generalisations played a role in the formulation of the AdS/CFT correspondence at the end of 1997 (issue (ii)).

After the first formulation of AdS/CFT, the field refocussed on elaborating this correspondence, since this promised a successful non-perturbative formulation of quantum gravity. Existing and novel accounts of black holes were (re)developed in its context, which became the paradigmatic perspective from which string theory was practiced in the following two decades. Our discussion here ends with a presentation of how the Strominger-Vafa entropy calculation was redone in the context of AdS/CFT, which reflects how the field today understands and presents the result.

Our conceptual analysis in Section \ref{argument} will distinguish four main theoretical contexts that we introduce in Section \ref{sec:history}: \\
\\
(1)~~Supergravity, i.e.~Einstein's theory of general relativity with specific matter fields. The black hole is a solution of this theory;\\
(2)~~A theory of strings that interact with higher-dimensional objects called `D-branes';\\
(3)~~A theory describing the dynamics of the D-branes themselves, without strings; and:\\
(4)~~The conjectured fundamental `M-theory'. \\
\\
The relations between these four theories will be a conjectured duality between (1) and (2), and linkage relations among the others, as we vary two quantities: namely, a distance scale at which the system of interest is probed, and the string coupling constant. 

With these notions in place, the essence of Strominger and Vafa's article can be stated straightforwardly. First, it presents a calculation of the number of microstates of a system of D-branes as in (3), which are weakly interacting and which are probed at short distance scales. The calculation is then extrapolated and compared to the Bekenstein-Hawking formula for the corresponding black hole in supergravity (1), which is probed at large distances, and where the string coupling is strong (both calculations take place at low energies). The cogency of this argument relies on two assumptions: first, the {\it conjectured duality between open and closed string theory}. This suggests the identification of (1) and (3), via (2), as being `the same theories, for different ranges of the relevant quantities'.  

Second, the {\it supersymmetry} of the various systems further justifies the comparison of their entropy at different values of the relevant quantities (namely, distance scale and coupling), because the entropy is guaranteed to be invariant under suitable changes of distance scale and coupling, due to supersymmetry. In other words, the black hole is {\it extremal}. We will also address, in Section \ref{IG}, three potential weaknesses of the Strominger and Vafa argument: the inferential limitations coming (i) from the assumption of supersymmetry (in particular, extremality), (ii) from the use of higher dimensions (the Strominger-Vafa black hole is five-dimensional), and (iii) from the approximations used.

The plan of the paper is as follows. In Section \ref{sec:history}, we introduce the reader to the conceptual developments relevant to the Strominger-Vafa calculation. In Section \ref{argument}, we first present the argument as contained in the original Strominger-Vafa paper, which we then analyse and `factorize' into its different components. 
Then, in Section \ref{IG}, we discuss the potential weaknesses of the argument, as well as some of the main generalisations that appeared in the literature in the next two-year period. In Section \ref{rads}, we briefly address the relationship between the Strominger-Vafa entropy calculation and AdS/CFT: how the former prepared the way for the latter, and the latter prompted a reinterpretation of the former. Section \ref{dconc} offers a discussion of our findings. Finally, to aid our readers in further navigating the relevant literature, the Appendix contains additional mathematical details of the Strominger-Vafa black hole and compares it with the three-charge extremal black hole familiar from recent textbooks.

\section{D-branes and their Rediscovery}\label{sec:history}
  
First, we briefly discuss the relevant conceptual and historical background that will allow us, in later Sections, to understand the microscopic black hole entropy counting argument by Strominger and Vafa, and in particular why it has been seen as offering an explanatory, microphysical account of black hole entropy. Subsection \ref{thsinvolved} is a brief outline 
of string theory and its key concepts relevant to our subject; 
and subsection \ref{ocsd} introduces two developments from 1995 that were instrumental to Strominger and Vafa's 1996 calculation: Edward Witten's (1995) `web of dualities' and M-theory conjecture, and Joseph Polchinski's (1995) formulation of D-branes.

\subsection{String theory}\label{thsinvolved}

Superstring theory, as presented in modern texts, is a theory of quantum gravity whose primary objects are strings. Today it is often presented as a `framework' rather than a `theory', very much in the sense in which quantum field theory is seen as a framework---and as opposed to specific theories like the Standard Model. It is considered to offer more than just one model of reality; but along with various such models, it is primarily considered a theoretical `recipe book' containing methods and concepts fit to address a host of physical systems and problems. Still, its main aspiration has been to offer a consistent quantum theory of gravity: much like, indeed, the Standard Model has offered a consistent perspective for the particle world. The approach and a number of its claims (lately particularly in the realm of cosmology) have remained controversial, in large measure due the absence of direct connections to the empirical.\footnote{See  e.g.~Ellis and Silk (2014) for a critical appraisal of string theory and its cosmologies.} The latter also applies to the subject we discuss here, the string theory black hole: string theory accounts of Hawking radiation and black hole entropy have yet to find any empirical confirmation.
Yet, here we will not be concerned with an appraisal of string theory as a scientific programme, but only with the conceptual analysis of the state counting performed by Strominger and Vafa---and how this was perceived as a successful account of black hole entropy, as assessed from \emph{within the string paradigm}. 

String theory may contain open or closed quantum strings. {\it Open strings} have endpoints carrying non-abelian charges (namely, analogues of the quark charges in the Standard Model). In current  understanding, open string endpoints are often thought of as being attached to higher-dimensional hyperplanes called \emph{D-branes} (see Section \ref{ocsd}). D-branes are dynamical objects and they can be seen as the source of the string's non-abelian charges, i.e.~the charges of the string endpoints are the D-brane's charges. {\it Closed strings,} on the other hand, are the carriers of the gravitational force. Their excitations contain a state with the properties of the hypothetical graviton (a massless, spin-2 quantum state); and at low energies this massless, spin-2 string excitation reproduces the Einstein equations. Thus the distinction between open and closed strings is crucial because it corresponds, roughly, to the distinction between nuclear-type forces and gravitational forces.

Two fundamental constants play an important role in string theory: the string length scale $l_{\tn s}$ and the string coupling constant, $g_{\tn s}$. The string length determines the tension of   strings $T \sim 1/l_{\tn s}^2$,   and is often replaced by the parameter $\alpha'$, defined by: $\alpha':= l_{\tn s}^2$. A perturbative expansion in terms of   $\alpha'$ measures `string-like' effects: for $\alpha' \rightarrow 0$, the string behaves like a particle. The string coupling constant $g_{\tn s}$ is the  expansion parameter for string scattering amplitudes, which takes into account  quantum loop effects. 
This is not just a number: it is determined by one of the   excitation modes of the string, called the dilaton field $\phi$. The string coupling is given by the vacuum expectation value of $e^{\phi}$, i.e.~$g_{\tn s} = \left \langle e^{\phi} \right \rangle$.  In closed string perturbation theory, every loop (that is: hole in the string worldsheet) introduces an extra factor of $g_{\tn s}^2$. Open string perturbation theory, on the other hand, uses an expansion in terms of $g_{\tn s}  N$---now often referred to as the `'t Hooft coupling'---in which $N$ is the number of D-branes. This is because every D-brane carries one unit of elementary charge, and in the Feynman diagram expansion one sums over all charges---analogously to how, in QCD, one sums over all colour charges.\footnote{See 't Hooft (1974) for the QCD case.} These two different perturbative expansions (in $\alpha'$ or $g_{\tn s}$) will play a central role in our discussion.

String theories come in five types, of which only one (`type I' superstring theory) was originally thought to contain both open strings and closed strings. The other four superstring theories were understood to contain only closed strings.  However, the discovery of D-branes allowed for the possibility of open strings in two of these other four theories, namely in the two `type II' string theories, identified as `type IIA' and `type IIB' string theory. In this article, we will be concerned with these type II string theories. In these, the endpoints of the string are not `freely moving', but are restricted to move on the surface of a D-brane. Thus, the type II string theories that we will deal with contain both open strings (attached to D-branes) and closed strings. 

Today, string theory jargon, when it discusses `open string theory', actually refers to the {\it open string sector} of type II (A or B) theory; we will follow this usage of the term.\footnote{Occasionally, the open string theory still includes some weakly-interacting closed strings. This plays a role in Maldacena's (1998c) argument, and we will return to it in Section \ref{rads} and in our companion paper.} 
Likewise, when discussing  `closed string theory', we will mean the {\it closed string sector} of type II theory, i.e.,~only the closed strings. This jargon is motivated by a duality that is conjectured to exist between open and closed strings and that will play a significant role in what follows (see Section \ref{ocsd}). Roughly put, it implies that the open strings attached to the D-branes can be described in a dual representation containing only closed strings. Under this duality, the D-branes, on which the open strings end, are replaced by a curved background geometry---often a higher-dimensional black hole. The validity of this conjectured `open-closed' duality in certain regimes makes the Strominger-Vafa argument possible, as will become clear in Section \ref{argument}.

At low energies, i.e.~at energies much lower than the string mass (which is proportional to the inverse of the string length), closed superstring theories are well approximated by semi-classical {\it supergravity theories.} Supergravity theories are versions of general relativity coupled to additional fermionic fields that make the theory supersymmetric, i.e.~the theory contains the same number of bosons and fermions, and these can be transformed into each other in  a way that leaves the equations of motion invariant. Their action thus contains the usual Einstein-Hilbert term, plus in addition 
matter field terms (both bosonic and fermionic) that contribute to the energy-momentum tensor; their details depend on the superstring theory in question. In this way, a semi-classical supergravity theory effectively sets the Einstein tensor, $G_{\m\n}$, equal to the 
stress-energy tensor, $T_{\m\n}$, of the supersymmetric matter fields.\footnote{Einstein's equations appear in closed string theory as the 1-loop contribution to the beta function. The validity of Einstein's equations is the condition for the conformal invariance of the string theory to one loop. Higher-loop contributions appear as higher-derivative terms in Einstein's equations. See Callan et al.~(1985). Sigma model loops (i.e.~$\alpha'$ corrections) are considered in Callan et al.~(1986), and string loops (i.e.~$g_s$ corrections) are considered in e.g.~Green (1999). For the role of the spacetime metric as a coherent state of gravitons, see p.~165 of Green et al.~(1987).}

As is well-known, string theory requires ten spacetime dimensions for its mathematical consistency.\footnote{In fact, the condition that the dimension $D=10$, and the condition that Einstein's equations coupled to matter fields are satisfied, are both part of a single set of consistency conditions: the preservation at the quantum level of the classical conformal symmetry of the string. See~Green et al.~(1987:~pp.~164-181).} As a consequence, the supergravity actions that one derives in a low energy limit are ten-dimensional. In this paper, we will consider five- and four-dimensional black holes. These are obtained by a procedure of {\it compactification} of the supergravity action: one assumes that $n$ out of the 10 dimensions are taken up by a compact space, so that one can derive an effective $(10-n)$-dimensional action via a Kaluza-Klein procedure. In this paper, we will work directly in the 5- or 4-dimensional perspective. 

\subsection{Open-closed string duality and D-branes}\label{ocsd}

The counting of black hole microstates by Strominger and Vafa was made possible by a number of key innovations in string theory: Witten's (1995) formulation of {\it string theory dualities} together with Polchinski's (1995) description of the dynamics of {\it D-branes}. Equally important were general debates on the black hole information paradox to which string theorists had increasingly contributed. These debates---originating in ideas of Stephen Hawking from the mid-1970s and flaring up in the 1990s---centred on whether black hole evaporation takes place according to the unitary laws of quantum mechanics, or whether these are violated due to the presence of the black hole horizon.\footnote{See Hawking (1975, 1976).}
Some critics of Hawking's original proposal of non-unitary evolution, such as Gerard 't Hooft and Leonard Susskind, were convinced that an exact counting of black hole microstates would strongly support  the unitary evolution scenario: this would suggest that black hole entropy is anchored in the usual quantum mechanical microphysics, with its unitary evolution laws, and is not attributable to some global spacetime properties.\footnote{See `t Hooft (1985, 1993) and Susskind (1995).} Indeed, upon learning of the results of Strominger and Vafa, Leonard Susskind (2008:~p.~394) believed that Hawking's ``jig was up". Strominger and Vafa, too, stated in the conclusion of their article that they found it ``hard to imagine how any calculation [of Hawking emission] based on our D-brane description of the extremal black hole could yield a non-unitary answer" (1996:~p.~103).

For more than two decades, four laws of black hole mechanics had been identified that were fully analogous to, but difficult to interpret as laws of thermodynamics, in particular since a microphysical picture---despite Hawking's evaporation result---was lacking.\footnote{See Bekenstein (1972,  1973), Bardeen, Carter,  Hawking (1973) and Hawking (1975). For reviews on black hole thermodynamics, the Hawking effect and   discussions  of the microscopic interpretation of black hole entropy,    see e.g.~Jacobson (1996, 2005), Wald (2001), Wallace (2018, 2019).} Strominger and Vafa saw their result as a key contribution to achieving such a microphysical understanding of black holes:
\begin{quote}\small
A missing link in this circle of ideas is a precise statistical mechanical interpretation of black hole entropy. One would like to derive the Bekenstein-Hawking entropy---including the numerical factor---by counting black hole microstates. The laws of black hole dynamics could then be identified with---and not just be analogous to---the laws of thermodynamics.\footnote{Strominger and Vafa (1996:~p.~99).} \end{quote}
The dynamics of the debate over the information paradox was indeed strongly affected by the result of Strominger and Vafa, and scholars increasingly moved away from Hawking's original position.\footnote{See J. van Dongen and S. De Haro (Forthcoming), `History and Philosophy of the Black Hole Information Paradox'.} 

The debate itself had been essential in drawing the attention of string theorists to the problem of black hole entropy and evaporation, and thus in framing the question that Strominger and Vafa---{\it what is the microphysical entropy of a black hole?}---had sought to answer. 
That they homed in on this question in late 1995 and raced to its answer was due, however, to recent technical and conceptual innovations internal to string theory: the duality relations suggested by Witten, and the D-branes that Polchinski (re)introduced subsequently. These developments were central to what is known as the `Second Superstring Revolution', of which the result by Strominger and Vafa is also considered to have been a part.\footnote{See the discussion in  Rickles (2014), Chapter ~10.}

In 1995, Edward Witten had presented evidence for a series of dualities that connected different ten-dimensional string theories with each other, and with eleven-dimensional supergravity. In a talk at the {\it Strings '95} conference at the University of Southern California, he further conjectured that underlying these dualities would be a still unknown eleven-dimensional theory, which he dubbed `M-theory', of which the known string theories and eleven-dimensional supergravity theories would be specific limits or approximations.\footnote{Witten (1998b:~p.~1129) famously justified the choice of letter $M$ by explaining that:  ``M stands for magic, mystery, or matrix, according to taste.'' Earlier, the list had included ``membrane'' (see Witten,  1996b:~p.~383), yet following work by Thomas Banks et al.~(1997), membranes were replaced as matrices. The interpretation of ten dimensional string theory in terms of eleven dimensional supergravity had been put forward earlier by Paul Townsend (1995), who had argued that
strings in ten dimensions descend from eleven dimensional membranes compactified on a circle.}
Witten (1995) had arrived at this conjecture by thinking about strongly interacting strings (in jargon: strings `at strong coupling'), and by analysing some already known dualities, such as T-duality in string theory and S-duality in quantum field theory, which he now extended to string theory. Thus he found that string theories form an interconnected web, connected by dualities, the common origin of which should be sought in the more fundamental M-theory.

A key ingredient of Witten's argument for M-theory concerned classical states in the low-energy and semi-classical limit of {\it closed string theory}, i.e., states of supergravity with interesting properties. The existence of such states followed from the supersymmetry of the supergravity fields, and they were associated with solutions of the equations of motion of supergravity that had horizons; thus, they were black holes. Witten (1995:~pp.~89-90)  now suggested  that these classical solutions should 
correspond to supersymmetric quantum states;  and that these corresponding quantum mechanical counterparts would have the same electric charge, because the charge is independent of the string coupling, which depends on $\hbar$. Witten did not clearly distinguish microscopics from macroscopics in his conjecture: he only referred to these states as the quantized versions of the classical supergravity solutions.

The mass of the classical black hole solutions is proportional to their electric charge, and inversely proportional to the string coupling constant: $M=c\,{\frac{|Q|}{g_{\tn s}}}$. Here, $M$ and $Q$ are the mass and electric charge of the black hole, $g_{\tn s}$ is the string coupling constant, and $c$  is a numerical constant. Because of this inverse proportionality to the coupling constant, the objects become {\it infinitely massive} at small coupling, i.e.,~as $g_{\tn s}\rightarrow0$. This implied that they cannot be made of or identified with ordinary, weakly-coupled, string states. They were rather seen as solitonic solutions in string theory---soon, as we will see, these quantum objects acquired the interpretation of `D-branes'.

This kind of dependence on the coupling constant was new to string theory---which suggested that a new class of supersymmetric states had been found in the theory. They were identified as `BPS states', after  Evgeny Bogomol'nyi (1976), and Manoj Kumar Prasad and Charles Sommerfield (1975; `BPS') who had derived similar bounds for solitonic states in field theory.\footnote{More specifically, the former for magnetic monopoles, and the latter for monopoles and dyons.}  The charges associated with these string states are called {\it Ramond-Ramond charges}.\footnote{This name refers to the {\it periodic} boundary conditions that all the fermions on the strings satisfy, for such D-branes. This contrasts with Neveu-Schwarz boundary conditions, which are anti-periodic.\label{Ramond}} The corresponding black holes that Witten discussed were known as compactifications of `$p$-brane' solutions of ten-dimensional supergravity theory. They were `extremal' black holes: they have the minimum mass possible for a given charge in supergravity theories.\footnote{Hull and Townsend (1995); Witten (1995), in particular~p.~89.}  The fact that their mass is minimal implies that such black holes cannot lose energy. So, they do not Hawking radiate and they have zero temperature. Yet, their entropy need not be zero: their horizon area is non-zero. Their quantum equivalents might then be degenerate ground states. The $p$-branes, finally, are extended over $p$ spatial dimensions, so that a 0-brane is a particle (that is, a black hole), a 1-brane is a black string, a 2-brane is a black membrane, et cetera. 
\\
\\
Joseph Polchinski (1995) defined `D-branes' in the  {\it open string theory} (i.e.~the open string sector of a type II theory; cf.~Section \ref{thsinvolved}) as hyperplanes on which open strings can end. Invoking a transformation using the conjectured open-closed string duality (see below), he suggested identifying these D-branes with Witten's quantum BPS states that would correspond to the black $p$-brane solutions of the supergravity limit of closed string theory.\footnote{Polchinski (2017:~p.~90) wrote in his memoirs: ``As Andy Strominger has reminded me, he, Gary Horowitz and I had lunch together nearly every day for three years, without realizing that their black $p$-branes [originally introduced by Horowitz and Strominger (1991)] and my D$p$-branes were the same.''} Thus D-branes were BPS states.

D-branes (`D' stands for `Dirichlet', a type of boundary condition) had already been formulated by Polchinski in earlier work with Jin Dai and Robert Leigh (1989)---yet, it was not until Witten's lecture at \emph{Strings '95} that Polchinski understood their full potential significance, and studied their properties in detail, especially their dynamics.\footnote{See Rickles (2014:~pp.~208-221, esp.~p.~215) and Polchinski (2017:~Sec.~8.5).} Polchinski (2017:~pp.~93-95) recalled in his memoirs: 
\begin{quote}\small
    At the end of Witten's talk, Mike Green and I looked at each other and said `that looks like D-branes'. [...] It was a shock wave, for me and the rest of the field. I had been living with D-branes for eight years, but never taking it too seriously [...]. But for almost everyone else, it was a new thing: string theory was no longer just string theory, it had D-branes as well.
\end{quote}
In 1995, Polchinski calculated the mass of D-branes and found agreement with Witten's result. Furthermore, their calculations agreed about the number of supersymmetries that D-branes preserved. Polchinski also showed that D-branes couple to the appropriate gauge fields on their world-volume, that the electric and magnetic charges associated with these gauge fields satisfy Dirac's quantization condition, and that D-branes carry one unit of elementary charge---thus correctly accounting for Ramond-Ramond charges in string theory (cf.~footnote \ref{Ramond}). In other words, D-branes were the appropriate {\it sources} for Ramond-Ramond charges. Since the number of dimensions of D-branes and $p$-branes was equal, Polchinski originally called them D$p$-branes.\footnote{Recall, from Section \ref{thsinvolved}, that there are two different kinds of type II string theories: type IIA features D$p$-branes with even $p$, and type IIB features branes with odd $p$. This distinction plays an important   role in the technical details of the entropy calculation by Strominger and Vafa; see Section \ref{heartofdarkness}.\label{typesII}}  In this paper, however, we use the term `D-branes' in order to distinguish them clearly from black $p$-branes.

\begin{figure}
\begin{center}
\includegraphics[height=6cm]{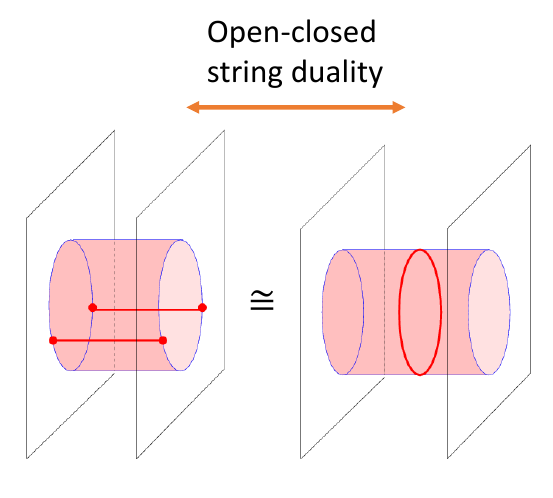}
\caption{\small The conjectured open-closed string duality, for specific stringy diagrams. A one-loop diagram for
a pair of open strings that is created, propagates vertically, and annihilates (left), is argued to be equivalent to the tree-level closed string diagram, for propagation of a single closed string moving horizontally (right).}
\label{OpenClosed}
\end{center}
\end{figure}

Because string theory is a theory of gravity, D-branes had to be dynamical. With Dai and Leigh, Polchinski had   shown in 1989 that the fluctuations of D-branes in directions normal to the brane could be described by massless excitations of open strings that propagate on it; this already gave a qualitative picture of the dynamics of D-branes. In 1995,  Polchinski provided further details: he calculated the one-loop amplitude for the interaction between two parallel D-branes via the open strings stretched between them. The same amplitude could equivalently be interpreted (by a reinterpretation of the string diagram) as a tree-level closed string amplitude: namely, an exchange of a single closed string between the D-branes, hence as a gravitational interaction: see the corresponding stringy diagrams in Figure \ref{OpenClosed}. 

This symmetry of the D-brane amplitude is a manifestation of the conjectured {\it open-closed string duality}. It was known since the mid-1980s that perturbative closed string amplitudes could be written in terms of appropriate perturbative open string amplitudes (cf.~Kawai, Lewellen, and Tye, 1986), and Polchinski's work now suggested such a relation between open and closed strings in the presence of D-branes. The correspondence between open string calculations that use D-branes and closed string calculations that produce supergravity is a key ingredient of the Strominger-Vafa argument, as we will see in Section~\ref{argument}.
When interpreting his novel D-branes, Polchinski (1995:~p.~4726) argued that ``presumably one should think of them as an alternate representation of the black $p$-branes."
Later, in particular in the context of the Strominger-Vafa entropy calculation, it will be helpful to think of the D-branes as  `microscopic' candidates (defined in open string theory) for `macroscopic', BPS black hole states in the supergravity limit of closed string theory---yet, such an interpretation is not explicitly given in these articles.

D-branes are non-perturbative objects---like $p$-branes, they become highly massive at weak string coupling. So, they are not `seen' in ordinary string perturbation theory. D-branes interact at very high energies and small length scales; they probe distances smaller than the string length, $l_{\tn s}=\sqrt{\a'}$, in other words: they are `fine grained'.\\
 
\noindent Thirteen days after Polchinski published his preprint on the arXiv, Witten (1996a) circulated another article that gave a concrete open string description of D-branes. Polchinski had studied the case of {\it two} parallel D-branes: Witten now gave a very general description of an arbitrary number $N$ of D-branes, possibly intersecting along various directions. His method showed how to derive the conformal field theory living on the world-volume of a stack of intersecting D-branes (what is called the `D-brane world-volume theory').\footnote{A conformal field theory is a quantum field theory with conformal symmetry, i.e.~it is invariant under the special class of diffeomorphisms that are local scale transformations (and which, in flat space, give an extension of the Poincar\'e group).} Specifically, Witten analysed {\it bound states} of D-branes,\footnote{Witten (1996a:~p.~341) defined a bound state as follows: ``A bound state is a state that is normalizable except for its center of mass motion''.\label{boundSt}}  whose world-volume theories turned out to be non-abelian. First, he identified the excitations of the D-branes in type IIB string theory, in the directions {\it parallel} to their world-volume, in the 10-dimensional ambient space in which the D-branes move, as massless gauge bosons with a U(1) gauge symmetry on the D-brane world-volume. That is, these were ordinary electromagnetic fields on the D-brane. Second, he identified the excitations {\it transverse} to the D-branes with scalars on the world-volume. He then argued that the low-energy description of the D-branes, valid when the branes are close to each other, is a theory with an unbroken $\mbox{U}(N)$ gauge symmetry group:
\begin{quote}
{\it The low-energy D-brane effective action is obtained from the action of ten-dimensional $\mbox{U}(N)$ supersymmetric Yang-Mills theory, dimensionally reduced to the $(p+1)$-dimensional world-volume of the D-branes.} 
\end{quote}
This is a central principle concerning D-branes, which also underpins the Strominger-Vafa calculation, as we will see. The forces between the D-branes are modelled by strings stretched between them. Thus strings whose endpoints are on the same D-brane are massless, and have an abelian gauge symmetry group, U(1). Strings that stretch between different D-branes are massive (they acquire their mass from the tension produced by the different D-branes pulling at their ends), and their gauge symmetry group is likewise an abelian U(1). But if the branes are brought close together, the strings are light, and the symmetry is non-abelian, viz.~$\mbox{U}(N)$.\footnote{The potential energy function between the D-branes is obtained from the dimensional reduction of the ten-dimensional supersymmetric Yang-Mills theory. It is a quartic function of the distances between the D-branes, which are scalars in the adjoint representation of the $\mbox{U}(N)$ gauge group, and so they are non-commuting matrices. The potential is minimized when the D-branes are parallel to each other (whether they coincide or not). Witten also showed that the mass of a string stretching between two nearby D-branes is indeed linear in the separation between the D-branes.} 
Thus Witten had brought the world-volume theories of D-branes in to the realm of gauge theories with gauge group $\mbox{U}(N)$, familiar from the nuclear interactions.\\

\noindent The connection between the open and closed string descriptions, which had not been the focus of Witten's (1996a) paper, was developed in further work: of which we will discuss two papers that brought into focus the way that distances and energies map across open-closed string duality. These developments were important for understanding the physics of D-branes, and how it relates to gravity.

A key element of the putative open-closed string duality is the exchange of high and low energies, or of ultra-violet and infrared limits, across the duality.  In a widely read set of lecture notes on D-branes, Polchinski, together with Shyamoli Chaudhuri and Clifford Johnson (1996:~p.~16), offered the following heuristic explanation for this inversion, in connection with the diagram in Figure \ref{OpenClosed}: ``Consider the limit $t\rightarrow0$ of the [open string] loop amplitude [on the left, where $t$ is the radius of the circle, i.e.~one is considering a short-lived interaction between open strings]. This is the ultra-violet limit for the open string channel [because the open string is long, i.e.~very massive, and it exerts a large force on the D-branes, hence large exchange of momentum.] [...] [B]ecause of duality, this limit is correctly interpreted as an \emph{infrared} limit'' of the closed string channel on the right of Figure \ref{OpenClosed}. This is because $t$ is now interpreted as the radius, i.e.~a measure for the {\it length}, of the closed string, which therefore appears very small: and so, it is a large-distance limit. Today, this connection between high and low energies goes under the name of `the UV/IR connection'.\footnote{See Susskind and Witten (1998).} 
The connection thus follows from the fact that exchange of long, open strings  is highly energetic and thus corresponds to short distances: which under the duality corresponds to the exchange of short, i.e.~pointlike and thus light, closed strings; and hence to long distances, when looked at from the closed string theory side. 

Michael Douglas, Daniel Kabat, Philippe Pouliot and Stephen Shenker (1997) exhibited, in August of 1996 (that is, after the work of Strominger and Vafa, and building on other contributions that had been made in the meantime) the consequences of open-closed duality for the relation between the {\it world-volume theory} of the D-branes and {\it supergravity}. Namely, they argued that short-distance phenomena in open string theory are described by the {\it infrared behaviour of the world-volume theory}. As Witten (1996a) had shown, the world-volume theory was an ordinary quantum field theory. The novelty lay in the evidence that this low-energy effective action could describe short-distance phenomena in string theory and M theory: such as scattering between D-branes at Planckian energies. Yet, because D-branes are fine-grained non-perturbative objects, the infrared dynamics of the associated world-volume theory turned out to capture well the short-distance behaviour of string theory. This came about as follows. 

Douglas et al.~(1997) used the conjectured open-closed string duality to compare the D-brane's world-volume dynamics, associated with the open string theory, with the spacetime dynamics of the closed string theory. Namely, they showed that for {\it large separations, $r$, between the D-branes}, compared to the fundamental string length (i.e.~for $r\gg l_{\tn s}$), the interaction between the D-branes is most easily described in closed string theory: as an exchange of {\it closed strings}. For the $r\gg l_{\tn s}$ limit is the low-energy limit of the closed string theory, and the effects are then easily captured by supergravity. This echoed Polchinski's interpretations of his one-loop amplitude calculation (cf.~the above quote from Polchinski, Chaudhuri, and Johnson). Since open string effects do not manifest themselves in this approximation, they can be ignored in it.
However, for {\it small separations between the D-branes,} i.e.~for $r\ll l_{\tn s}$, the interaction is best described by quantum loops of {\it open string} states. Closed string effects are very small at small $r$. And since the mass of the open strings stretched between the D-branes is proportional to $r$, and the dynamics of the lightest open string states is encoded in the world-volume theory, short distances between D-branes translate into low energies in the D-brane's world-volume theory. Thus, Douglas et al.~(1997:~p.~89) arrived at the following principle:  
\begin{quote}
{\it The leading behaviour, as $r\rightarrow 0$, of D-brane interactions is determined by the low-energy behaviour of the world-volume theory on the D-brane.}
\end{quote}
\noindent This implies that D-branes, being non-perturbative objects, give an efficient description of short distances in string theory.

To sum up: long-distance interactions between D-branes through the exchange of open strings were shown to be described by supergravity (itself the low-energy limit of closed string theory), while the short-distance interactions were shown to be described by the D-brane world-volume theory of the lightest open strings. However, both the D-brane and the supergravity theories are low-energy effective descriptions of string theory, because of the exchange of high energies and low energies under open-closed duality.\footnote{The discussion in the last four paragraphs has ignored the role of the string coupling. We will bring this into the discussion in Section \ref{SVanalyse}.} Even though the result of Douglas et al.~was not yet fully stated at the time of the article by Strominger and Vafa, its logic was part of the latter's concrete calculation.

\section{The Strominger-Vafa Argument}\label{argument}

In this Section, we will review the Strominger-Vafa  (1996) counting of black hole microstates. In Section \ref{SVcalculation}  we review the original calculation, and we will do so in considerable technical detail which some readers may choose to skip; in Section \ref{SVanalyse}  we  give an analysis of the overall structure of their argument, and thus the main message of this article. But who are Andrew Strominger and Cumrun Vafa, and how did they arrive at their subject? We address this first in Section \ref{gearup}.

\subsection{Gearing up to calculate black hole entropy}\label{gearup}

Andrew Strominger (born 1955) graduated from MIT, submitting a thesis in 1981 on `The large symmetry approximation in quantum field theory', supervised by particle theorist Roman Jackiw. Jackiw had recommended Strominger to steer clear of quantum gravity---advice which the latter already did not follow in his years as a graduate student, when he developed a distinct interest in black holes, and particularly Hawking's work on their evaporation and evolution.\footnote{Andrew Strominger, in interview with J.~van Dongen and S.~De Haro, Harvard University, 20 November 2018.} As a fellow at the Institute for Advanced Study and faculty member in Santa Barbara, Strominger fully moved into string theory, and contributed to influential work on Calabi-Yau compactifications in 1985.\footnote{Candelas et al.~(1985).} He continued developing his interests in black holes and was one of the main voices in the contentious debates on the information paradox that flared up in the early 1990s\footnote{In 2018, Strominger prided himself as ``not being stuck to one point of view'' regarding the information paradox  (interview with J.~van Dongen and S.~De Haro, Harvard University, 20 November 2018). This is reflected in the positions that his papers at the time endorsed: an article by Callan, Giddings, Harvey, and Strominger (1992) was instrumental in introducing the black hole information paradox to string theorists, while suggesting that black hole evolution is unitary; in Banks, O'Loughlin, and Strominger (1993), it  was argued, on the other hand, that information is stored in long-lived black hole remnants, while Polchinski and Strominger (1994) advocated baby universes as remnant candidates. Finally, Fiola, Preskill, Strominger, and Trivedi (1994) presented a model consistent with information loss. For a review of the various approaches to information loss, see Strominger (1995:~Sect.~4).}---clearly, counting black hole entropy was a result he had a keen and long-standing interest in. Strominger brought particular expertise to the subject:
black hole and solitonic solutions in semi-classical low energy approximations to string theory had already been one of his main subjects for at least half a decade. Indeed, he had shown in 1991 with Gary Horowitz that low energy approximations to various string theories contained a whole class of black hole and black string solutions, among which was an extremal five-brane with an event horizon, i.e.~an extremal black $p$-brane in five dimensions.\footnote{Horowitz and Strominger (1991); see also e.g.~Strominger (1990).}         

Strominger had begun to undertake attempts to count black hole entropy in the early 1990s, realising that ``it would be good to count [the entropy] for extremal black holes, because they were BPS-protected."
BPS objects were attractive since many of their properties remain unchanged as one goes from weak to strong values of the coupling, i.e.~as the strength of the gravitational force is turned up or down.\footnote{See Garfinkle, Horowitz, and Strominger (1991:~p.~3142), Giddings and Strominger (1992:~p.~627).} In spite of various attempts, however, the work did not bear fruit: ``I was missing some crucial ingredients", Strominger said in 2018.\footnote{He had been ``trying to compute indices on multi-black hole moduli spaces and get the entropy there''; Andrew Strominger, interview with J.~van Dongen and S.~De Haro, Harvard University, 20 November 2018.} In particular, he lacked understanding of D-branes.

Joseph Polchinski and Strominger were both working in Santa Barbara in 1995. 
Strominger already started calculating black hole entropy using D-branes even before Polchinski's article on them had appeared. First, he calculated the entropy of the near-extremal three-brane. ``It came out as the area and everything; in particular, the $N^2$ [i.e.~the gauge theory result for the area, in string units] was there, which seemed like the hard part." Yet, the result was off by a power of 4/3 when compared with the field theory value: here, the entropy was not a BPS-protected quantity.\footnote{This mismatch was also found by Gubser, Klebanov and Peet (1996:~Eq.~(32)). Their reference [20] and the reference [23] in Maldacena (1998c) are references to this unpublished work by Strominger.}
Strominger was however convinced that a positive result could soon be attained. ``At that point I knew that our understanding of non-perturbative string theory was enough that it was clearly going to be possible to calculate these things, and I did nothing else at that time." He also ``knew there was going to be some hard maths involved.''\footnote{Andrew Strominger, in interview with J.~van Dongen and S.~De Haro, Harvard University, 20 November 2018.} Strominger tried to recruit Polchinski for the project, who however demurred,\footnote{In his Memoirs, Polchinksi (2017:~p.~100) recalled that ``Strominger came to me excited that he would be able to calculate the microscopic density of states of black holes. Having learned GR from [Stephen] Weinberg, I had not given this question much thought, but Strominger, a more gravitational physicist, told me that this was just as important as the information problem. His calculation was just off by a constant, and he was looking for help. This was all too new to me, and I had nothing to contribute.''} and, towards the end of 1995, he left Santa Barbara for Cambridge, MA, on a trip to visit his family (his father, Jack L.~Strominger, is a noted biochemist at Harvard). Despite his confidence, Strominger had not yet found a matching result for black hole entropy.  \\
\\
Cumrun Vafa (born 1960) grew up in Iran and moved to the US to attend university.  He started at MIT, where he did a double major in physics and mathematics, and continued his studies at Princeton University, obtaining a PhD in 1985 under the supervision of Edward Witten. Vafa had written a thesis on quantum field theory, entitled `Symmetries, inequalities and index theorems', while also picking up an interest in strings. This quickly led to highly successful collaborations on string compactifications on `orbifolds' and a fellowship and faculty career at Harvard University.\footnote{Vafa (1985); Dixon et al.~(1985, 1986).}

Vafa, too, had zeroed in on BPS objects, already before concretely thinking about counting black hole microstates. ``I was really fascinated by their precise properties, that they can be so robust.''\footnote{C.~Vafa, in interview with S.~De Haro, Harvard University, 30 November 2018.} In 1992, he had shown with Sergio Cecotti that for a certain class of two-dimensional quantum field  (`Landau-Ginzburg') theories, the energy of solitons was bounded from below by the  ``Bogomol'nyi'' (i.e.~BPS) bound, which would not be renormalised due to its topological properties.\footnote{Cecotti and Vafa (1993:~p.~572).}   Bogomol'nyi-saturated states were supersymmetric, could be explicitly counted, and played an important role in the classification of vacua in these theories, as they interpolated between different vacua.\footnote{Cecotti and Vafa (1993). This article was instrumental for Seiberg and Witten's work (1994) on electric-magnetic duality of ${\cal N}=2$ supersymmetric Yang-Mills theory in four dimensions, which uses the BPS condition. }
The potential of such states for possible entropy calculations for black holes did not go unnoticed: the next year, Leonard Susskind, in a first attempt to calculate black hole entropy in string theory, pointed out that extremal black holes may be specifically suited for the task as their gravitational and electromagnetic energies would cancel; he particularly credited Vafa for the insight.\footnote{See Susskind (1998 [1993]:~pp.~128, 130.)}   

In the summer of 1994, together with Witten, Vafa performed a strong coupling test of S-duality in ${\cal N}=4$ supersymmetric Yang-Mills theory\footnote{{$\cal N$}-supersymmetry relates each bosonic state to ${\cal N}$ different fermionic states. S-duality relates a strongly coupled theory, with coupling constant $g \gg 1$, to an equivalent  weakly coupled theory, with coupling constant $1/g \ll 1$.} on a four-dimensional space called K3, using the invariances of BPS states.
Vafa and Witten found that the partition function of 
supersymmetric Yang-Mills theory on K3 gives the partition function of a bosonic string. Yet, it was unclear why there was such a connection between super Yang-Mills theory and the bosonic string---it looked like a coincidence. 
The connection nevertheless became ``completely obvious'' to Vafa
after Polchinki's rediscovery of D-branes in 1995.\footnote{C.~Vafa, in interview with S.~De Haro, Harvard University, 30 November 2018.} 
In particular, he quickly reinterpreted the bosonic string oscillations that he and Witten had found in supersymmetric Yang-Mills theory in terms of bound states of D0-branes. Subsequently, he reinterpreted the supersymmetric Yang-Mills theory on K3 as the world-volume theory of $N$ D4-branes on K3, and checked the consistency of the result with one of the new string dualities (`heterotic-type IIA' duality).\footnote{Vafa (1996a; 1996b).} So, by year's end in 1995, Cumrun Vafa was leading the pack in expressing the physics of BPS D-branes in terms of field theories living on their world-volume.   \\
\\
Strominger and Vafa soon met. 
``Strominger was visiting Harvard, and we started talking about black holes'', Vafa recalled in 2018. 
Their plan was to apply his ideas about BPS states and K3 techniques to an appropriately chosen black hole system. ``Around that time people were writing feverishly about black holes and dualities because we knew dualities could teach you about aspects of quantum gravity, and black holes were one of the prominent things that had not been solved in that context.''\footnote{C.~Vafa, in interview with S.~De Haro, Harvard University, 30 November 2018.} Andrew Strominger found that 
``once I got going with Cumrun, it was very quick. [...] 
 I knew Vafa knew the kind of mathematics that was needed to solve this problem, so I translated it [...] and it turned out he did have just the right piece of mathematical information about dimensions of moduli spaces of 3-branes in K3.''\footnote{A.~Strominger, in interview with J.~van Dongen and S.~De Haro, Harvard University, 20 November 2018.} 
A division of labour was agreed upon. Vafa: 
\begin{quote}\small
I was thinking about the BPS objects. [...] This was a particular computation that I understood in the context of ${\cal N}=4$ Yang-Mills. I was trying to construct from these objects a gas of D-branes: objects that looked like black holes.
[...] Strominger said 5d [BPS] black holes should have entropy, he went and did that calculation. `All right, why don't you concentrate on that calculation, I will try to compute what I can from the string side, [and see] what entropy I get.'\footnote{C.~Vafa, in interview with S.~De Haro, Harvard University, 30 November 2018.}
\end{quote}
Thus, Strominger calculated the entropy of  a black hole using   supergravity techniques and Vafa computed the entropy of a configuration of D-branes, and they compared their results---and these matched. But what exactly had they done?

\subsection{Strominger and Vafa's entropy calculation}\label{SVcalculation}

The article by Strominger and Vafa derived the entropy of a certain class of black holes by counting microstates in string theory. In particular, it related the area of the horizon of these black holes to the number of microscopic states of a D-brane configuration in string theory. The result is based on the connection between black $p$-brane solutions of supergravity theory and D-branes in string theory that had been proposed by Polchinski: Strominger and Vafa showed that the expression for the classical horizon area equals the entropy of a certain system of D-brane states. Consequently, confidence grew in the hypothesis that D-branes and string theory could offer microscopic and quantum descriptions of classical black hole spacetimes. 

The derivation hinged on establishing a connection between (A) a theory of gravity considered in the strongly coupled, black hole regime (i.e.,~in the range of parameters for which a supergravity description is valid, $g_{\tn s}N\gg1$; more on this later) and (B) an open string theory, which is valid in a weakly coupled regime (i.e.~at~$g_{\tn s}N\ll 1$; $N$ is here the number of D-branes).
There was (and is) insufficient understanding of string theory at strong coupling to count directly the number of D-brane configurations in this regime. Strominger and Vafa believed that when going from weak to strong coupling, \emph{the D-brane configuration turns into a classical black hole spacetime}.\footnote{For a discussion of this point, see our companion paper `Emergence and Correspondence for String Theory Black Holes'.}
Therefore, they counted the number of states at weak coupling in string theory and then extrapolated that result to the supergravity regime. This procedure is justified, they argued, since the D-branes are BPS states: this implies that their total number does not change under suitable changes of the coupling (see Section \ref{SVanalyse} for an explanation of this fact). 

We will focus on (A) above (strongly coupled supergravity) in Section \ref{sec:sugraperspective}; on (B) above (weakly coupled open string theory) in Section \ref{heartofdarkness}; and on Strominger and Vafa's extrapolation of their calculation from within (B) to (A), in Section \ref{SVanalyse}.

\subsubsection{(A)~The black hole from the supergravity perspective}
\label{sec:sugraperspective}

The black hole considered in the article by Strominger and Vafa is five-dimensional. During their project, they did consider working with four-dimensional black holes, but they could not compute the entropy for the dual D-brane system. Vafa wanted to wrap a D6-brane on the six-dimensional internal manifold $\mbox{K3} \times T^2$, where $T^2$ is the two-torus. This implied having to count membrane states in six dimensions, rather than string states. 
``I told Strominger, `I'm trying to do the calculation, but it's too difficult. I can't use the same techniques [i.e.~those of Vafa and Witten (1994) and Vafa (1996a)] to get the four-dimensional [cases]'."\footnote{C.~Vafa, in interview with S.~De Haro, Harvard University, 30 November 2018.}

He noticed that the D-brane calculation would be considerably easier for five-dimensional black holes. This was because Vafa's earlier method started in six dimensions: so that compactifying a string down to five dimensions would have given him a D0-brane, i.e.~a configuration that looks like a black hole in five dimensions; and Vafa's intuition was that for strings one knows how to count states. Hence, Vafa asked Strominger if they could do the calculation in five dimensions, and whether there are usable black holes known in five dimensions. ``He thought about it, and said `yes, of course, there are these BPS black holes also in five dimensions'."\footnote{C.~Vafa, in interview with S.~De Haro, Harvard University, 30 November 2018.}  
However, in 1994 Stephen Hawking, Gary Horowitz and Simon Ross had claimed that extremal Reissner-Nordstr\"{o}m black holes (which are BPS black holes from a supergravity perspective) have zero entropy.\footnote{Hawking, Horowitz and Ross (1995).} Strominger did not trust that result and constructed for himself the BPS black hole solution in five dimensions: indeed showing that the horizon area is finite, and thus, by virtue of the Bekenstein-Hawking formula, he inferred they have a nonzero entropy.  
\\
\\
In what follows, we will first give an outline of this black hole, since it is also contained in the publication by Strominger and Vafa. 
It is a five-dimensional black hole solution of type IIA supergravity theory, which is~the low energy limit of type IIA closed string theory, compactified on a five-dimensional internal manifold $\mbox{K3} \times S^1$. Here,  $S^1$ is a circle and K3 is a specific compact, Ricci-flat, hyperk\"ahler four-dimensional manifold, also called a `Calabi-Yau' 4-manifold. Since the ten-dimensional supergravity theory is compactified on a five-dimensional internal manifold, the ten-dimensional supergravity action simplifies to a five-dimensional action, of which Strominger and Vafa consider the following terms:\footnote{Cf.~Eq.~(2.1) in Strominger and Vafa (1996).}
\begin{equation}\label{SVaction}
 S = \frac{1}{16 \pi } \int \dd^5 x ~\sqrt{- \tilde g} \left [ e^{-2 \phi} \left ( R + 4\,  ( \ti\nabla \phi )^2 - \frac{1}{4} \tilde{H}^2 \right) - \frac{1}{4} F^2 \right] \; .
 \end{equation} 
Here, $\phi$ is the dilaton field, $F$ is a Ramond-Ramond 2-form field strength (related to the D-brane charges), and $\tilde{H}$ is a 2-form field strength (related to the electric field of a fundamental string); we can think of these fields, $F$ and $\tilde{H}$, as two different electric fields. Strominger and Vafa present their black hole solution, Eq.~\eq{SVBH} below, with a metric tensor, $g_{\m\n}$, that is rescaled by an exponential of the dilaton, relative to the tilded metric used above, $\ti g_{\m\n}$ (the two metrics are related by $ g_{\m\n}=e^{-4\f/3} \ti g_{\m\n}$; see also the Appendix). Finally, note that here conventions are used in which Newton's constant, speed of light, and the string length are equal to one:\footnote{According to Eq.~\eq{gnewton} in the Appendix, the five-dimensional Newton's constant and the string length can be independently set equal to one, if the string coupling is not fixed.} $G_{\tn N} = l_{\tn s} =c=1$. 

Solving the equations of motion of the above action, and assuming spherically symmetric configurations, gives the following black hole solution for the metric:\footnote{Cf.~Eq.~(2.8) in Strominger and Vafa (1996).}
\begin{equation}
  \begin{aligned}\label{SVBH}
  \dd s^2 &= - f(R)\, \dd t^2 + \frac{\dd R^2}{f(R)}  + R^2\, \dd \Omega_3^2~,  \\
f(R)&=\left(1- \frac{R^2_{\tn S}}{R^2}\right)^2~.
  \end{aligned}
  \end{equation}
The horizon has a Schwarzschild radius $R_{\tn S}$ whose size is determined by the electric charges:\footnote{Cf.~Eq.~(2.9) in Strominger and Vafa (1996).}
    \begin{equation}\label{SchwarzschildR}
    {R_{\tn S}}= \left (  \frac{8 Q_{\tn H}  Q_{\tn F}^2}{\pi^2}  \right)^{1/6} .
   \end{equation}
Here, $Q_{\tn H}$ is the (electric) charge associated with the field $\tilde{H}$, and $Q_{\tn F}$ is the electric charge associated with the field $F$. 

The above solution is an extremal Reissner-Nordstr\"{o}m black hole in five dimensions, with its overall charge and horizon radius determined by the two electric charges.\footnote{See the Appendix   for more on the mathematical details of this black hole solution, and in particular for a discussion of how the various charges are related to the single charge of the traditional Reissner-Nordstr\"{o}m geometry.} The adjective `extremal' implies that, in the units chosen, the mass of the black hole is equal to its charge; this equality of mass and charge is the same condition as Witten's (1995) BPS condition for the states of the supersymmetric theory (see Section \ref{ocsd}). A charged black hole has an inner and an outer horizon, and in the extremal case the two coincide.\footnote{For a discussion, see Hawking and Ellis (1973:~pp.~156-161) and Townsend (1997:~Chapter 3; extremal case:~pp.~72-74). The ordinary, five-dimensional, Reissner-Nordstr\"om solution has $f(R)=1- 2m/ R^2 + q^2 / R^4$. The extremal black hole is obtained by setting $m=q$, in which case the solution simplifies to Eq.~\eq{SVBH}, with Schwarzschild radius $R_{\tn S}=\sqrt{q}$. Comparing this value to Eq.~\eq{SchwarzschildR}, we see that the overall power of the electric charges, i.e.~$1/2=(1+2)/6$, matches, just as they should on dimensional grounds. The difference between the supergravity solution, Eq.~\eq{SVBH}, and the ordinary extremal Reissner-Nordstr\"om solution, is that the supergravity solution has two electric fields rather than one. \label{RNnote}}
The black hole has a finite horizon area, given by:\footnote{Cf.~Eq.~(2.10) in Strominger and Vafa (1996).}
\begin{equation}\label{AreaFirst}
\mbox{Area} = 8 \pi\, \sqrt{\frac{Q_{\tn H}\, Q_{\tn F}^2}{2}}.
\end{equation}
This black hole description is only valid in a certain regime of parameters: it is derived in an effective theory, namely supergravity. For the supergravity theory to provide an accurate description of the system, one needs to suppress both string loops (i.e.~assume that the interactions are weak) and $\alpha'$-corrections (i.e.~assume that the strings are probed only at large distance scales, so that the string can be seen as point-like; cf.~the discussion at the start of Section \ref{thsinvolved}). String loop diagrams can be suppressed by taking the string coupling to be very small, i.e.~$g_{\tn s} \ll 1$. The string coupling is determined by the asymptotic value of the dilaton $\phi$, i.e.~$g_{\tn s} = \langle e^{\phi_\infty} \rangle$, and the dilaton is a constant everywhere for the black hole solution    \eqref{SVBH},   given by  $e^{\phi } \sim Q_{\tn F}/Q_{\tn H}$. Thus the string coupling is proportional to
\begin{equation}\label{stringcplingratiocharge}
g_{\tn s}\sim \frac{Q_{\tn F}}{Q_{\tn H}} \, ,
\end{equation}
and small coupling implies  $Q_{\tn F}\ll Q_{\tn H}$.
The $\alpha'$-corrections can be neglected when the black hole horizon is larger than the string scale,\footnote{The Schwarzschild radius of interest, for the discussion of  $\a'$-corrections, is obtained from the string-frame metric, $\ti g_{\m\n}$, rather than directly from the Einstein frame metric, Eq.~\eq{SVBH}. Thus Eq.~\eq{SchwarzschildR} is rescaled by a factor of $g_{\tn s}^{2/3}$, and we get the following value of interest for the Schwarzschild radius: $\ti R_{\tn S}\sim\sqrt{Q_{\tn F}^2/Q_{\tn H}}$. } i.e.~$\ti R_{\tn S}\gg\sqrt{\a'}$. This means that $Q_{\tn F}^2\gg Q_{\tn H}$, since we set $\alpha'=1$. Putting the two conditions together, we get the following requirement: 
  \begin{equation}\label{sugraC}
  \text{Supergravity regime:}~~~~~~~~Q_{\tn F}^2 \gg Q_{\tn H}\gg Q_{\tn F} \gg 1, 
  \end{equation}
which also implies that: $g_{\tn s} N  \gg 1$, where $N:=Q_{\tn F}$. In the D-brane calculation in Section \ref{heartofdarkness}, $N$ will be reinterpreted as the number of D-branes, because every D-brane carries one unit of elementary electric charge (and the elementary unit of electric charge will be set to one).

\subsubsection{(B)~The heart of darkness? The D-brane calculation}
\label{heartofdarkness}

The black holes of the previous subsection are now taken, following Polchinski's 1995 D-brane proposal,  to be approximations to D-brane configurations in superstring theory (see Section \ref{ocsd}). However, this  D-brane picture only applies---and can only be used to do calculations---in a regime that has different values for the parameters above, namely:
  \begin{equation}\label{psr}
\text{D-brane regime:} \qquad Q_{\tn F}\ll Q_{\tn H} ~~\mbox{and}~~  Q_{\tn F}^2\ll Q_{\tn H}~.
  \end{equation}
The first condition implies that the string coupling, Eq.~\eqref{stringcplingratiocharge}, is small, and the second condition  states that $\alpha'$-corrections cannot be neglected, which 
can be rewritten as $g_{\tn s} N  \ll 1$, with again $N=Q_{\tn F}$.
As we explained in Section \ref{thsinvolved}, the effective coupling in open string perturbation theory is actually given by $g_{\tn s} N$, i.e.~the 't Hooft coupling, because each brane contributes a factor of $g_{\tn s}$ to the total open string amplitude. So, Eq.~\eqref{psr} represents the weakly coupled regime of open string theory, while the supergravity calculation is valid only for large values of the 't Hooft coupling.

The D-brane configuration that Strominger and Vafa considered is a solution to type IIB superstring theory---which only contains D$p$-branes with odd $p$ (cf.~footnote \ref{typesII})---compactified on K3\,$\times \,S^1$. The configuration is any combination of D1-, D3-, and D5-branes with momentum, suitably supersymmetric, intersecting along the $S^1$, as long as the total Ramond-Ramond charge is $Q_{\tn F}$. The D-branes intersect along the $S^1$, and their remaining spatial directions are compactified on supersymmetric cycles (i.e.~closed submanifolds, for example, the two circles that together make up a two-torus) of the four-dimensional internal manifold K3 (for more on the significance of the $S^1$, see point (ii) below). To be specific, D3- and D5-branes are   wrapped on supersymmetric 2- and 4-cycles inside K3, with their remaining direction along the $S^1$. Thus, the world-volumes of all the branes take the form $S^1\times C$, where $C\subset\mbox{K3}$ is a 0-, 2- or 4-cycle. 

The fact that different D-brane configurations are allowed (i.e.~D1, D3, and D5-branes) is significant: indeed the Strominger-Vafa entropy calculation is the same for different D-brane configurations, and the authors do not make any statement singling out one of them as preferred. In the subsequent literature, however (following Callan and Maldacena, 1996), most studies have been about configurations of intersecting D1- and D5-branes, for which calculations are simpler (see Sections \ref{SVanalyse} and \ref{NEBHs}). 

The identification between the D-brane system and the black hole system of Section  \ref{sec:sugraperspective} was made, principally, on the basis of the Ramond-Ramond charges of both systems (cf.~Section \ref{ocsd}), the level of supersymmetry preserved by both systems, and on a string duality that relates the type IIA and type IIB string theories.
This duality, called `T-duality', offers an exact map between the states of IIB and IIA string theory on the circle, in which  momentum modes   along the circle in the former theory are replaced with winding modes of the string in the latter theory, and a large circle is mapped to a small circle, according to $R\mapsto R'=l_{\tn s}^2/R$. Moreover, under T-duality, a D-brane wrapping the circle is mapped to a D-brane of one less dimension, i.e.~a D-brane of $p$ spatial dimensions is mapped to a D-brane of $p-1$ spatial dimensions. 
In the Strominger-Vafa analysis, T-duality on the circle $S^1$ maps the D1-D3-D5-P system in type IIB theory, where $P$ is the number of momentum modes on the $S^1$, to a D0-D2-D4-F1 system in type IIA theory, i.e.~an intersection of D0, D2, and D4-branes along a fundamental string.\footnote{In particular, the D-brane configuration considered by Callan and Maldacena (1996), i.e.~the D1-D5-P system in type IIB theory, maps under T-duality on the $S^1$ to a  D0-D4-F1 system in type IIA theory.}  The D4-branes are wrapped on the compactified four-dimensional $\mbox{K}3$ manifold, the D2-branes are compactified on a 2-cycle, and the fundamental strings F1 are wound along the circle. 
Interestingly, Strominger and Vafa did not actually perform the microstate counting for the D0-D2-D4-F1 system---which would have been more natural from the type IIA supergravity theory perspective---but they instead performed an equivalent calculation for the D1-D3-D5-P system, and then related the type IIB D-brane calculation to the type IIA supergravity calculation through T-duality (together with open-closed string duality).

To sum up: in the regime of weak 't Hooft coupling $g_{\tn s} N$, the D-brane configuration in type IIB theory which is dual to  the Strominger-Vafa black hole  is the   `D1-D3-D5-P system', i.e.~a bound state of D1-, D3- and D5-branes with momentum modes $P$ along the circle. The D1-, D3- and D5-branes carry a total of $Q_{\tn F}$ elementary units of electric charge, while the number of momentum modes, $P$, is equal to the charge $Q_{\tn H}$.\footnote{In the type IIB configuration, the energy is given by the momentum, which after T-duality to type IIA gets mapped to the winding of the string on the $S^1$: but since in type IIA the string is static, all of its energy is in its charge (it is BPS-saturated), and so the winding number is also equal to the string's electric charge, hence $P=Q_{\tn H}$.\label{momentumcharge}} Strominger and Vafa (1996:~p.~102) concluded:
\begin{quote}\small
Thus the BPS states of the D-brane world-volume theory we are considering carry precisely the charges $Q_{\tn F}$ and $Q_{\tn H}$ for which the corresponding extremal black hole solutions were found in the previous section.
\end{quote} 
\noindent Next, the degeneracy of the black hole system as a function of $Q_{\tn F}$ and $Q_{\tn H}$ is computed by counting the number of BPS bound states. 
This calculation of the degeneracy of  microscopic D-brane states is the heart of Strominger and Vafa's article. It proceeds in four steps, as follows:\\
\\
(i)~~{\it Supersymmetric ground states.}
First, Strominger and Vafa used Witten's (1996a) result to identify the nature of the states in the D-brane world-volume theory that they needed to count, and the amount of supersymmetry that these states should preserve.\footnote{Witten (1996a) had related BPS states of D-branes and strings in open string theory (dubbed states that preserve ``spacetime supersymmetry'', because they preserve the supersymmetry of the ambient spacetime, which is larger than the world-volume supersymmetry) 
to {\it supersymmetric ground states} of the world-volume theory of the corresponding D-branes  (see Section \ref{ocsd}). BPS states that preserve half of the spacetime supersymmetries (for example, a stack of D1-branes) correspond to fully supersymmetric ground states of the world-volume theory of the D-branes, since the ambient spacetime has twice as many supersymmetries as the world-volume of the D-branes.\label{qsusy}} They considered BPS states for intersections of different D-branes that together break 1/4 of the spacetime supersymmetries, which hence correspond to supersymmetric ground states of the world-volume theory that preserve {\it half} of the supersymmetries (cf.~footnote \ref{qsusy}).\\
\\
(ii)~~{\it The world-volume.} After having identified the states of interest in the world-volume theory, Strominger and Vafa specified the relevant {\it world-volume} itself on which this theory lives. To do so, they took the four-dimensional K3 to be much smaller than the $S^1$: so that, effectively, the world-volume of the D-branes is $S^1\times\mathbb{R}$, where $\mathbb{R}$ is time, i.e.~the world-volume theory is effectively two-dimensional.\footnote{Strominger and Vafa did not explain why they took the $S^1$ radius to be large compared to the K3. However, since T duality maps a circle of large radius to a circle of small radius, dualizing from type IIB with a {\it large} circle back to the type IIA black hole, will give back the small circle (recall that the five-dimensional black hole was compactified on $\mbox{K3}\times S^1$). 
Thus, in type IIA, the black hole is {\it compactified} on $\mbox{K3}\times S^1$ while, in type IIB, the D-branes are {\it wrapped} on $\mbox{K3}\times S^1$, but compactified only on (cycles of the) K3.}\\
\\
(iii)~~{\it The theory on the world-volume.} Third, Strominger and Vafa further specified the {\it theory} on the $S^1\times\mathbb{R}$ world-volume. To do this, they used earlier work by Vafa (1996a), which made clear that the relevant theory is a supersymmetric `sigma model': roughly speaking, a quantum field theory whose fields are maps from some base manifold to a target space ${\cal M}$ that is endowed with a metric. This target space is usually referred to as a `moduli space', because it is the space swept out by the possible values that moduli (i.e.~parameters) of certain D-brane solutions can take: so we will adopt this usage in what follows.

Vafa's (1996a) argument to derive the sigma model is astonishingly simple, even if the mathematical details are highly technical. The theory on the D-brane intersection $S^1\times\mathbb{R}$, following Witten (1996a; see our Section \ref{ocsd}), is simply the dimensional reduction of the ten-dimensional type IIB open string configuration. Consider first a single D1-brane intersecting a single D5-brane. Each brane has a world-volume theory which is the abelian (i.e.~charged under U(1)), supersymmetric Yang-Mills theory in ten dimensions, dimensionally reduced to the corresponding dimension of the D-brane's world-volume: viz.~to two dimensions for the D1-branes, and to six dimensions for the D5-branes. The interactions between the D-branes are given by the open strings stretching between them, which give additional matter multiplets on the common two dimensional space, charged under the two U(1)'s. The argument can be generalised to multiple D-branes: and to do so, Vafa dualized from type IIB to type IIA, where he considered $k$ D0-branes and one D4-brane. The $k$ D0-branes then have a {\it non-abelian} $\mbox{U}(k)$ symmetry, where the eigenvalues of the $\mbox{U}(k)$ field can be thought of as the relative positions of the D0-branes in the internal space. The sigma model arises once again as the dimensional reduction of the (now non-abelian) 10-dimensional open string theory, as indicated before. The moduli space arises from the degrees of freedom of the D0-branes in the bound state: namely, from their positions. Since each D0-brane is a point particle and can move inside K3, the total moduli space is the Cartesian product of $k$ K3's, up to permutations of the D0-branes, since they are indistinguishable---thus one quotients by the symmetric group $S_k$. So, in a slogan: 
\begin{quote}
{\it The number of microstates of the Strominger-Vafa black hole is the number of independent ways in which the D0-branes can move inside K3}.\footnote{For more details of the above argument, see Vafa (1996a:~pp.~416-418).}
\end{quote}
The upshot was that, as just argued, the moduli space ${\cal M}$ is the Cartesian product of $k$ copies of K3, quotiented by the symmetric group of $k$ elements, $S_k$, that permutes the $k$ copies of K3:\footnote{Cf.~Eq.~(3.1) in Strominger and Vafa (1996). This moduli space had already appeared in Vafa and Witten (1994:~pp.~44-45), in the context of electric-magnetic duality of supersymmetric Yang-Mills theory on K3. For more on this connection, see footnote \ref{bosonicS}.} 
\begin{equation}
    \begin{aligned}
\label{moduliS}
{\cal M} \quad &\,= \quad \underbrace{(\mbox{K3}\times\cdots\times \mbox{K3})}_{\text{$k$ times}}/S_k\\
k \quad &:= \quad \left[\frac{1}{2} Q_{\tn F}^2+1\right]~,
\end{aligned}
\end{equation}
where the square brackets in the definition of $k$ restrict to the integer part. 

T-dualised back to type IIB string theory, this was Vafa's conjecture for the moduli space of intersecting D1, D3, and D5-branes on $\mbox{K3}\times S^1$, with total electric charge $Q_{\tn F}$. Notice that the above expression reinterprets the square of the charge, $Q_{\tn F}^2$, {\it geometrically} as the self-intersection number of the cycles inside the K3, i.e.~as the total number of times that the D-branes self-intersect.\footnote{The above formula for the moduli space depends only on the total D-brane charge $Q_{\tn F}$ and not on the details about the D-branes, which can be supersymmetric intersections of D1-, D3- or D5-branes on $S^1$. This reflects what we said in Section \ref{heartofdarkness}: Strominger and Vafa cited a number of earlier works that verified Vafa's conjecture for the moduli space, i.e.~for particular D1-D3-D5-brane intersections, for self-intersecting D3-branes, and for D1-D5-brane intersections. The latter was due to Vafa (1996b). For the connection of this calculation to the phenomenon of electric-magnetic duality of Vafa and Witten (1994): see the discussion in Section \ref{svads}.} This point, and the next one, are two points where Vafa's expertise was crucial.\\
\\
(iv)~~{\it Extracting the central charge from the dimension of the moduli space and Cardy's formula.} The last step consisted of counting the BPS states of the supersymmetric sigma-model, which on the moduli space ${\cal M}$ given in Eq.~\eq{moduliS} is a (1+1)-dimensional conformal field theory (CFT) 
on the circle, $S^1$. 

The degeneracy of the states with fixed energy in a generic, two-dimensional CFT, was known to be given by the Cardy  formula, which takes the form:\footnote{Cf.~Eq.~(3.2) in Strominger and Vafa (1996). This counts only the left-moving sector of the CFT, whose states have eigenvalues $L_0=n$, because for BPS states the right-moving sector is in the vacuum, i.e.~their energy eigenvalue $\bar L_0$ vanishes, and thus does not contribute to the entropy. Cardy's formula for the degeneracy of states is the ordinary statistical mechanical degeneracy of states, defined from the partition function: $Z=\Tr\, e^{-\b H}=\sum_nd(n,c)\,e^{-\b E_n}$, where $H$ is the Hamiltonian with eigenvalues $E_n$, and $\b$ is the inverse temperature. See Cardy (1986:~p.~194).}
\begin{equation}\label{cardyF}
d(n,c) \simeq \exp \left ( 2\pi \sqrt{\frac{n \, c}{6}} \right)    \, ,
\end{equation}
where $n$ is the energy 
of the 
states, and $c$ is the {\it central charge}, which measures the number of degrees of freedom, and is a fixed number. This formula applies in the so-called `Cardy regime': namely, when $n \gg c$.

For the D1-D3-D5-P world-volume sigma model discussed above, the central charge, $c$, is determined by the dimension of the moduli space ${\cal M}$: so that to get the entropy all one really needs to know about this space is its asymptotic dimension, i.e.~its dimension in the regime Eq.~\eq{psr}.\footnote{The moduli space itself plays an important role in entropy calculations beyond the leading classical result. See Ooguri et al.~(2004), and Section \ref{subleadingS}.} Strominger elaborated in an interview:\footnote{A.~Strominger, in interview with J.~van Dongen and S.~De Haro, Harvard University, 20 November 2018.}
\begin{quote}\small
That was the key thing: the calculation was set up in a way that you only needed to know the dimension of a moduli space, rather than the dimensions of the cohomologies of the moduli space [i.e.~roughly, the closed and non-exact forms on the space], which is a much harder problem.    
\end{quote}
The dimension of the moduli space of Eq.~\eq{moduliS} is $4k$, where $k=[\frac{1}{2} Q_{\tn F}^2+1]$. One may replace the square brackets by round brackets, since by convention the electric charge is normalised such that the value $Q_{\tn F}^2/2$ is an integer (see Strominger and Vafa (1996:~p.~100)).
The central charge is $c=6k$. The energy is equal to the momentum, which equals the electric charge (cf.~footnote \ref{momentumcharge}):\footnote{Cf.~Eq.~(3.3) in Strominger and Vafa (1996).}
\begin{equation}
n=P=Q_{\tn H}~,\qquad
c = 6 \left ( \frac{1}{2} Q_{\tn F}^2 +1 \right)    \, . 
\end{equation}
For the calculation of the black hole entropy, Strominger and Vafa were interested in the degeneracy of the ground state {\it of the D-brane world-volume theory} (i.e.~the number of BPS states). However, there are $Q_{\tn H}$ units of momentum, which is why the above does not correspond to the ground state {\it of the CFT,} but rather to a state of high energy, viz.~$n=Q_{\tn H}$. In other words, what the uncompactified D-brane world-volume theory of step (i) above sees as its ground state is not the same (and need not be the same) as the ground state of the CFT, which is the world-volume theory compactified on K3.

Plugging the above into the Cardy formula \eqref{cardyF} yields the following expression for the Boltzmann  entropy:\footnote{Cf.~Eq.~(3.4) in Strominger and Vafa (1996).}
   \begin{equation}\label{SVarelikeLudwig}
 S_{\tn{stat}} = \ln\, d \left( Q_{\tn F}, Q_{\tn H} \right) \simeq 2 \pi\, \sqrt{Q_{\tn H} \left (\frac{1}{2}\, Q_{\tn F}^2 + 1 \right) } \, .
 \end{equation} 
Upon translating the Cardy regime condition $n\gg c$ into a requirement on the charges, we see that this equation is derived in the regime of weak 't Hooft-coupling: $g_{\tn s} N  \ll 1$ or  $Q_{\tn F}^2\ll Q_{\tn H}$, with $Q_{\tn F}$ a fixed,  but not necessarily  large number. 

Next, Strominger and Vafa assumed that the Cardy formula also holds for $Q_{\tn F}^2 \gg 1$ or $c\gg 1$, i.e.~for a large number of degrees of freedom, while keeping the ratio $n/c$ fixed.\footnote{Recently, Hartman, Keller and Stoica (2014) have shown that the range of validity of the Cardy formula can be extended to the supergravity (or black hole) regime for two-dimensional CFTs with a large central charge and a  restricted number of low-energy states (a so-called `sparse light spectrum'). This confirms the assumption by Strominger-Vafa to apply the Cardy formula for a large number of degrees of freedom, along with the argument that the asymptotic degeneracy of BPS states is independent of the coupling, so is the same at strong 't Hooft coupling (see footnote \ref{topinv}).} Since in the large central charge limit the ratio $n/c$ is proportional to $Q_{\tn H}/Q_{\tn F}^2$, which is the same as the inverse of the 't Hooft coupling, $g_{\tn s}N$, with $N=Q_{\tn F}$, this means that the Cardy formula is taken to be valid for all couplings (in other words, it is independent of the 't Hooft coupling). In particular, this means that the Cardy formula can also be applied in the strong 't Hooft coupling $g_{\tn s}\gg1$ and weak string coupling $g_{\tn s}\ll1$ regime, and those requirements together imply $Q_{\tn F}^2\gg Q_{\tn H}\gg Q_{\tn F}\gg1$, which is the supergravity regime Eq.~\eq{sugraC}. Thus, working with the assumption above, Strominger and Vafa observed that in the regime $Q_{\tn F}^2\gg1$ (which includes the supergravity regime) the Cardy formula matches exactly the Bekenstein-Hawking area-entropy relation of the black hole:\footnote{Cf.~Eq.~(2.11) in Strominger and Vafa (1996).}
\begin{equation} \label{SVarelikeLudwig2}
 S_{\tn{BH}} = \frac{\mbox{Area}}{4 G_{\tn N}} = 2 \pi\, \sqrt{ \frac{Q_{\tn H}\, Q_{\tn F}^2}{2}} \, , 
 \end{equation}
 where the horizon area, Eq.~(\ref{AreaFirst}), has been inserted, and $G_{\tn N}=1$ in the last equality (further, units are chosen in which $c=k_{\tn B}=\hbar=1$).

This remarkable result was and still is viewed as strong evidence for the identification by Bekenstein and Hawking of the black hole horizon area with the object's thermodynamic entropy. Furthermore, it strengthened claims that string theory 
offers successful accounts of quantum gravity physics.
Even string critic Gerard 't Hooft was impressed: he stated soon after its formulation that the Strominger and Vafa result ``is clearly a point that [string theorists] scored. This I had not expected of string theory. It clearly shows that one can get quite far in describing black holes in string theory."\footnote{Gerard 't Hooft, interviewed in De Haro and Dongen (1998, p.~284, translated from the Dutch original). Conversely, a ``failure of the string counting of states to match the Bekenstein-Hawking entropy would have been a serious blow to the notion that string theory is a complete quantum theory of gravity", in the judgment of Horowitz, Maldacena and Strominger (1996:~p.~151).  }

\subsection{Analysis of the Strominger-Vafa argument}\label{SVanalyse}

The most important aspect of the above argument was that the entropy was derived by a microphysical analysis and could thus be seen as a statistical mechanical Boltzmann-type entropy. That analysis, by a web of approximations and conjectured dualities, was embedded in an intricate manner in a quantum theory of gravity: string theory. 
These approximations and dualities, in turn, allowed that the  state counting could be done in a field theory in which  gravity had in effect been turned off, while it still offered a reliable result: at weak 't Hooft coupling, supergravity did not have to be contended with, while microstates could be identified and counted in the D-brane world-volume theory. It is because of the numerical identity between entropies and the inter-theoretic relations that the calculation was taken to \emph{explain} black hole entropy---i.e.~it was taken to provide a microscopic, Boltzmannian account of the thermodynamic entropy---and to deliver a success for the string theory approach to quantum gravity. 

Clearly, such a result raises epistemological questions of various kinds. How reliable are the inferences that are being drawn? What is the epistemic status of the various regimes and their relation to one another? How are they related to the `real' world of four dimensional black holes? And finally, how does this argument compare with familiar methodological strategies in physics, in particular in cases where the fundamental theory itself---let alone its relation to the empirical world---is similarly uncertain and still under construction? All these concerns affect why and how the Strominger-Vafa calculation has been taken to be a convincing account of black hole entropy. 

In what follows, we will try to abstract the lines of inference of the above complex argument away from their details so as to address more general concerns: what is the inferential strategy and what is the general build-up of the argument? How are the various strands of the argument related to one another? We recommend our companion paper, `Emergence and Correspondence for String Theory Black Holes', for a study of the more interpretative concerns expressed above. \\

\noindent The derivation of black hole entropy from Section \ref{SVcalculation} crucially depends on connecting a strongly coupled supergravity description to a weakly coupled string regime, described by a D-brane world-volume theory. Strominger and Vafa gave two arguments for why both quantities, the horizon area and the degeneracy of D-brane states, should stay the same when changing the coupling constant. First, the degeneracy of states in the D-brane picture can be extrapolated from weak 't Hooft coupling to strong coupling without affecting the result, due to supersymmetry: this number is `protected' under a change of the coupling, because the degeneracy of the BPS states is a topological invariant in the theory considered.\footnote{That is the theory of the D-brane `moduli' space. On this point, Strominger and Vafa (1996:~p.~103) indicated that ``the asymptotic degeneracy of BPS states [...] is a topological quantity related to the elliptic genus''. The `elliptic genus' is a notion from algebraic topology (see Landweber, 1986:~pp.~1-54, 161-181) introduced in string theory by Witten (1987). Its use in physics is as a partition function of a quantum field theory in which fermionic states contribute with a minus sign (see Boer, 1999:~Eq.~3.1), so that bosons and fermions cancel each other out for all the states except for the BPS states. Thus, the elliptic genus is a partition function for the BPS states. The fact that the asymptotic degeneracy of BPS states is related to  the elliptic genus implies that the degeneracy---just like the elliptic genus---is a topological invariant in the theory, i.e.~it does not depend on continuous variations of parameters such as the coupling. The elliptic genus is important in calculations that go beyond the leading semi-classical result, see Section \ref{subleadingS}.  \label{topinv}}
So, one knows that the entropy values in different regimes will agree on the D-brane side, even if one can only do calculations in the weak coupling regime.

Similarly, Strominger and Vafa argued that the horizon area in the   $p$-brane black hole  remains the same under adiabatic changes of the coupling, i.e.~for modest variations of the dilaton field. Initially, in establishing the solution Eq.~(\ref{SVBH}), the dilaton had been taken to be a constant everywhere in spacetime, while its asymptotic value fixed the string coupling constant; see  the discussion leading to Eq.~(\ref{stringcplingratiocharge}). However, even if the dilaton is not the same at infinity and at the horizon, i.e.~$\phi_\infty \neq \phi_{\sm{hor}}$, it is a property of the  family of solutions of the supergravity theory to which Eq.~(\ref{SVBH}) belongs, that the near-horizon geometry does not change when the asymptotic value of the dilaton is varied. Hence, the near-horizon geometry is not affected if the coupling is changed, i.e.~the horizon area is the same at weak and at strong coupling.\footnote{Strominger and Vafa (1996:~p.~101) wrote: ``as the asymptotic value of the  [dilatonic] fields are adiabatically changed, the near-horizon geometry is unaltered". This is called the `attractor' mechanism in modern string theory literature.}

To sum up: the numerical match between Eq.~(\ref{SVarelikeLudwig}) 
and Eq.~(\ref{SVarelikeLudwig2}) is conceptually underpinned by the ``$p$-brane to D-brane relation'' due to Polchinski and others. Calculations done in these theories are valid in different regimes of the coupling; yet, the considerations above ensure that the results obtained remain valid when one crosses from one regime to another, making the numerical match more secure across all scales of the coupling.\\
\\
The Strominger-Vafa argument was formulated in the mid-1990s, and, as we saw, considered by many at the time \emph{to explain} black hole entropy.
To gain insight in its contemporary assessment we should of course present our analysis relative to the standards of rigour prevalent in theoretical high-energy physics of the period. So, we offer the following conceptual analysis of the Strominger-Vafa argument as it would have been understood and extended by the community of string theorists within which the original paper was received. 

\begin{figure}
\begin{center}
\includegraphics[height=8.5cm]{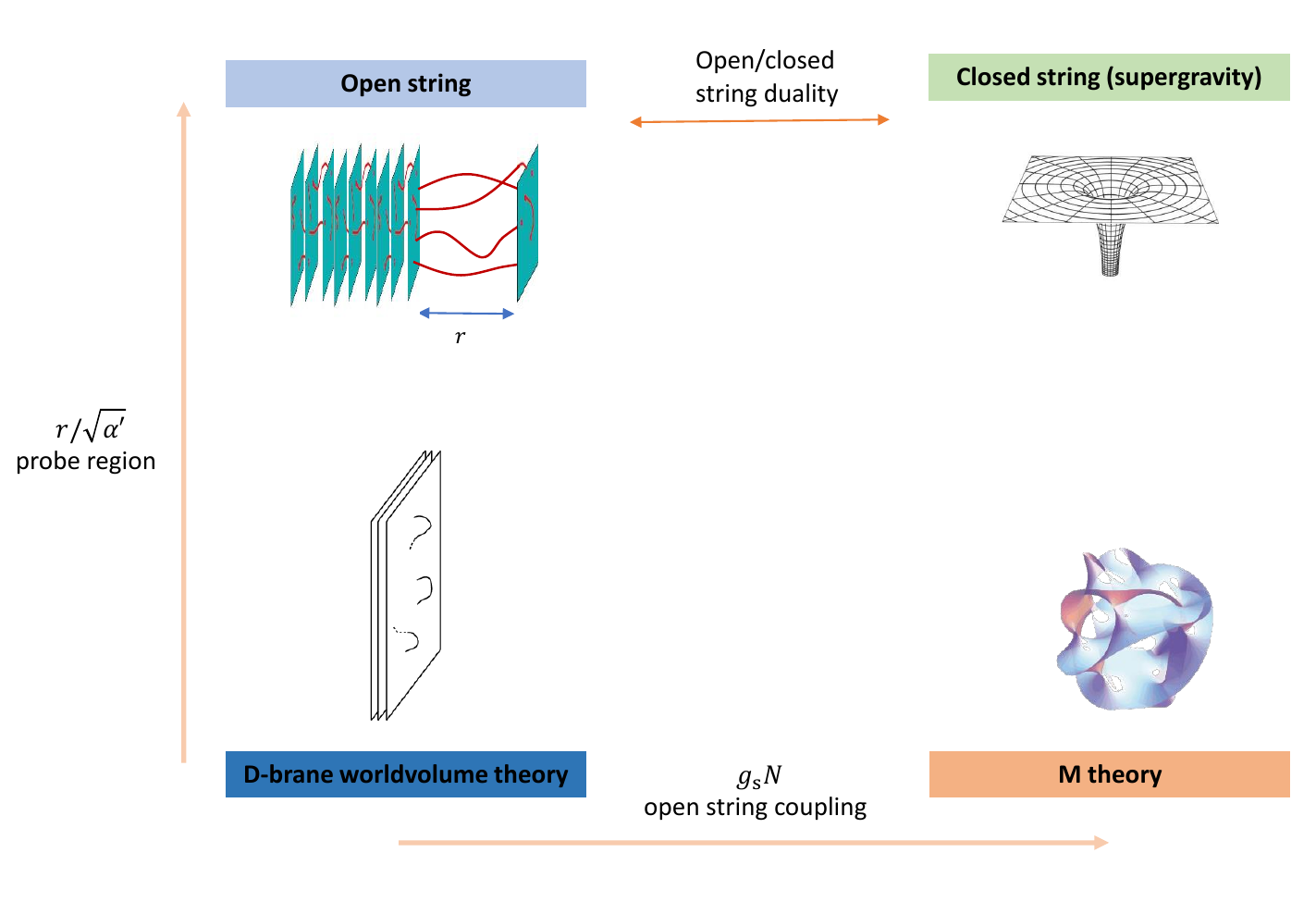}
\caption{\small The Strominger-Vafa argument: (i)~The 't Hooft coupling increases to the right. (ii)~The probe distance increases going up. Particular theories apply in the four corners:
(TopRight) Supergravity; (TopLeft)~Open string theory; (BottomLeft) D-brane field theory; (BottomRight)~M-theory. These corners can be imagined to be occupied by the following systems: (TopRight): 
a semiclassical black hole, observed from large distances. (TopLeft): a system with heavy, effectively fixed D-branes, probed by another D-brane at a large distance; between them, strings are attached that quantum mechanically interact. (BottomLeft): quantum mechanically interacting D-branes, stacked closely together and described using quantum field theory. (BottomRight): a still unknown non-perturbative quantum gravitational system. \emph{Source images}: Wikimedia.}
\label{SVdiagr}
\end{center}
\end{figure}

The article by Strominger and Vafa was very short: it is a brief statement of a result that hardly explains or reviews the concepts and resources that it depends on.\footnote{The arXiv (one-column) version of the Strominger-Vafa (1996) paper contains, including references, 11 pages,  and the published (two-column) version in {\it Physics Letters B} was 6 pages.} It was written during a frantic period of activity in string theory, in which many new key results followed each other in quick succession. After the work by Witten and Polchinski on dualities and D-branes, novel important papers appeared on a daily basis, and competition between authors was fierce.\footnote{Rickles (2014:~pp.~215-217) notes about the impact of Witten's and Polchinski's papers from 1995: ``Witten's paper itself caused a flurry of activity  (...) The impact of Polchinski's paper closely matches Witten's, and would have belonged to a pattern of co-citation." He substantiates this point with publication graphs, showing that in the following years these papers where cited between 100 and 200 times per year. Further, in an interview with Rickles (2014,~p.~216), Polchinski recalled: ``Within weeks of my paper, Vafa and Douglas and Sen had all pointed out important implications. I don't know of any episode like it in my experience where there had been such a change in a field.'' In an interview with the authors, Strominger recalled that interest in M-theory was actually larger at the time than interest in performing an entropy count (the latter ``was not a problem that was on the table"); he added that the result was however fairly quickly picked up (interview with J.~van Dongen and S.~De Haro, Harvard University, 20 November 2018).} The Strominger-Vafa entropy calculation itself prompted many additional attempts at black hole entropy counting using D-branes. This development, in turn, was one of the key components that led to the proposal of the celebrated AdS/CFT duality---more on that later. Given all the excitement and high pace of developments, it is hardly surprising that the entropy calculation by Strominger and Vafa was presented in a way that was accessible to a close knit group of experts, but hard to follow beyond that group.
Furthermore, some later reviews usually placed the result in the context of AdS/CFT or other later results, which may obscure rather than lay bare (a) the distinctive lines of inference of the original analysis, and (b) how this was at the time taken to be convincing and relevant---our topics, as announced in the Introduction.  \\
\\
{\it The Strominger-Vafa Diagram} \\
\\
To make the Strominger-Vafa argument more transparent, we present, in Figure \ref{SVdiagr}, a diagram that summarises the main regimes; its components will become clear in what follows. At its most basic level, the  argument compares two systems: a classical black hole and a configuration of D-branes. A priori they are physically distinct, but the argument establishes that they share quantitative properties of interest: namely, they have the same value for the Bekenstein-Hawking entropy and the number of states. These quantities are `protected' by invariances and symmetry principles that ensure that they have the same value in the two situations that are being compared. So, these are a priori \emph{physically different} systems that Strominger and Vafa showed to match up as regards a particular number because of invariances and symmetries in the underlying theories. 

These two facts---that these numbers match, and that the systems and theories in play are related via duality maps and approximation schemes---are taken to indicate that the microphysical account of the one system (the D-branes) ensures that in principle the same (or a sufficiently similar) microphysical picture can safely be assumed to exist for the other, gravitational black hole, system---even if that microphysical structure is at present, or even fundamentally, inaccessible to us. In the eyes of many string theorists, this indication is strong enough to be taken as \emph{evidence} of such a microphysical picture. Hence, they were confident in inferring that, e.g., non-unitary evolution of black holes is an unlikely scenario, as Strominger and Vafa did in 1996.

The two physical systems in play can only be properly described when appropriately choosing the values of two types of quantities in specific open and closed string theories: the symmetries and dualities that relate these theories then ensure that these two systems indicate properties about one another. The two types of quantities are: \\
\\
(i)~~~The {\it Parameters} of the solutions: the coupling constant $g_{\tn s}$, electric charges, and geometric quantities such as the Schwarzschild radius. Particular values fix a particular solution.\\
\\
(ii)~~The {\it Dynamical quantities}: the distances between the relevant objects, or the energies of the objects involved. One can also consider the scale at which a given object is probed. \\
\\
Choosing appropriate values (i.e.~large or small) for the quantities (i)-(ii) determines that the systems are (approximate) solutions of a particular string theory. Indeed, string theorists work with solutions of equations of motion which can be derived from string theory only in specific regimes of validity (for example, for low energies or small values of the couplings: cf.~Section \ref{thsinvolved}). The choices of values for parameters and quantities (i)-(ii) define the regimes of validity of these equations. In string theory jargon: such a solution can only be `trusted' in a certain regime of parameters and quantities  (see Section \ref{SVcalculation}).\footnote{For a detailed  discussion of the regimes of validity of the different string theory descriptions for black holes, see Callan and Maldacena (1996:~p.~604); Douglas, Kabat, Pouliot and Shenker (1997:~p.~89); Strominger (1998:~p.~7) and Skenderis (2000:~pp.~341-344, 355-358).}\\
\\
We will now discuss conditions (i) and (ii) for the two relevant cases, (A) and (B), of the Strominger-Vafa argument: \\
\\
(A)~~The black hole solution to the supergravity limit of type IIA string theory.\\ 
(B)~~The D-brane system as described by the world-volume theory derived from type IIB string theory.\\
\\
Regarding (i)-(A): The black hole calculation is valid for certain values of the string coupling constant, $g_{\tn s}$, and the Schwarzschild radius, $R_{\tn S}$.\footnote{See Strominger and Vafa (1996:~pp.~101-103).} These values are given by condition Eq.~\eq{sugraC}, which can be restated as: $g_{\tn s}N\gg1$, i.e.~the 't Hooft coupling is large (while we still have $g_{\tn s} \simeq Q_{\tn F}/Q_{\tn H} \ll 1$).\\
\\
Regarding (i)-(B): In the D-brane calculation, one is dealing with the `opposite' regime, as we saw in Eq.~\eq{psr}. This condition implies that the 't Hooft coupling is weak, i.e.~$g_{\tn s}N\ll 1$. This also means that, for given values of $N$, $g_{\tn s}$ will have to be {\it much} smaller for the D-brane system than it is for the black hole solution, (i)-(A).\\
\\
These two regimes, (i)-(A) with strong 't Hooft coupling, $g_{\tn s}N\gg1$, and (i)-(B) with weak 't Hooft coupling, $g_{\tn s}N\ll1$, are included in Figure \ref{SVdiagr}. Here, $g_{\tn s}N$ runs on the horizontal axis, increasing to the right. Therefore, the supergravity black hole, (A), appears on the right-hand side, and the D-brane system, (B), appears on the left-hand side.\\

\noindent We now discuss the vertical axis of Figure \ref{SVdiagr} or, in other words, the validity conditions for the quantities (ii) above. The quantity of the kind (ii) that we have singled out for this axis is the distance, $r$, compared to the string length $l_{\tn s}=\sqrt{\a'}$, at which the system is probed; or, in other words, the typical size of the system. More precisely, $r$ is the radial position of a probe object in the solution, and hence it is the spatial extent at which one is considering the solution. In particular, one is interested in the limits of this spatial extent, i.e.~one needs to establish the values of $r$ for which a given description can be trusted. In   Figure \ref{SVdiagr}, $r / \sqrt{\a'}$ increases going up.\\
\\
\noindent Regarding (ii)-(A): In the black hole description, the size of the system is fixed by the typical curvature scale of the black hole solution:  the Schwarzschild radius $R_{\tn S}$. So the range of  the radial coordinate $r$  is determined by the values for the length scale at which the supergravity solution can   be trusted to accurately describe the motion of a probe. Clearly, the black hole solution has a singularity at $r=0$, and at this point the solution is not valid. Yet, the classical supergravity solution can already not be trusted at the string scale, since `stringy' effects become important at this point (i.e., $\alpha'$-corrections to the geometry). Thus $r$ must be large compared to the string length, i.e.~$r/\sqrt{\alpha'}\gg 1$, in order for the solution to apply. This is why the black hole appears in the {\em upper}-right corner of Figure~\ref{SVdiagr}.\\
\\
Regarding (ii)-(B): In the D-brane description, the relevant distance scale is the separation between the D-branes and a probe brane. This determines the size of the system in the directions transverse to the D-branes. If probed at short distances (i.e.~if a probing D-brane is  up  close to other D-branes), only the massless states of the open strings that connect the D-branes contribute to the system's degrees of freedom. These states are described by a (1+1)-dimensional conformal field theory living on the D-branes, in which their number can be counted. 

As was explained later by Douglas et al.~(1997)---see our discussion at the end of Section \ref{ocsd}---the infrared regime of the world-volume theory on the D-branes gives the physics of short distance scales in the associated string theory. Strominger and Vafa   calculated the entropy at low energies in the D-brane world-volume theory. These energies correspond to short distances in string theory. Since   $r$  is thus taken to be small compared to the string length, i.e.~$r/\sqrt{\alpha'}\ll 1$, the D-brane world-volume theory   is found in the {\em bottom}-left corner of our diagram. This theory describes the quantum dynamics of the D-branes.\\
\\
We have so far described two corners of Figure~\ref{SVdiagr}: the (TopRight), i.e.~classical supergravity black hole at large distances, and (BottomLeft), i.e.~low-energy quantum D-brane theory at short distances. Yet, there are still two more corners in the diagram: (TopLeft) and (BottomRight). What is to be found there? And what role do they play in the argument? \\
\\
(TopLeft):~~The top-left corner contains D-branes as hyperplanes for Dirichlet boundary conditions for open strings, as they were found in perturbative open string theory in the original publications of Dai et al.~(1989) and Polchinski (1995)---cf.~Section \ref{ocsd}. The probe distance here is not chosen small and the string states are not restricted to massless states; it depicts the general situation for D-brane systems, that can be found at low values of the 't Hooft coupling.

The existence of this corner is a key ingredient of the Strominger-Vafa argument: it ensures that configurations of D-branes with specific charges can be identified with a black hole with given quantum numbers. This is where Polchinski's 1995 result enters the argument (see also Section \ref{ocsd}): he identified, via open-closed string duality, the D-branes in the top-left corner with the black $p$-branes in the top-right corner. The latter is here concretely given by the geometry of the Strominger-Vafa black hole. 

However, the microscopic entropy calculation of Strominger and Vafa is \emph{not} carried out in the top-left corner: it is carried out using the D-brane world-volume theory of the bottom-left corner. So, the argument not only involves the open-closed string duality that pairs the left and the right of the diagram. It also depends on the realisation that the D-brane world-volume theory is an effective description of open strings when the separation between the D-brane probe and the D-branes is small   compared to the string length; and it depends on the expectation that changing the D-brane probe distance would not affect the state counting. 

The top-left and bottom-left descriptions are equivalent, in the sense that they are similar to using lenses with different focal lengths, or antennas of different scales, when studying an object; the object's characteristics are not thereby changed, but may become properly visible with one choice of lens instead of another. For small values of the probe distance in the D-brane system, one can use the low-energy D-brane world-volume theory and thereby count the states of the stack of D-branes.  
Thus in our diagram, the top-left corner is also paired with the bottom-left corner, and their numbers of states are considered to be equal. That such identifications between open string theory and the D-brane world-volume theory were possible had been researched in the months prior to the Strominger-Vafa paper, in particular by Witten (1996a); see our Section \ref{ocsd}. 

So, Strominger and Vafa used the world-volume theory of the D-branes for their BPS state counting, as we spelled out in Section \ref{heartofdarkness}. 
This theory is obtained in the limit in which the distance between the probing D-brane and the D-brane system is much smaller than the string length.\footnote{See also Douglas, Polchinski and  Strominger~(1997).} 
Crucially, the number of states for such a BPS system does not change if that distance scale, measured in string length, is increased as one moves up on the left-hand side of our diagram.

As we move from the (BottomLeft) to the (TopLeft) corner, we increase the separation between the D-branes. The strings between them then become massive. This means that we go to high energies in the D-brane world-volume theory (recall the discussion at the end of Section \ref{ocsd}): and this is, in general, difficult, because an ultraviolet description of the D-brane world-volume theory, including the massive modes, is lacking. It is thus easier in the top-left corner to use the original string theory description of the D-branes, involving open strings with Dirichlet boundary conditions.  

In a sense, to make the Strominger-Vafa argument `work' one needed, in January 1996, to pick the right theories to fill the slots of Figure \ref{SVdiagr}. For the outline of the argument---the inter-theoretic relations represented by the structure of the diagram---and the likelihood of obtaining a successful result were already clear to many scholars contributing to the subject, just as we saw Strominger express in his 2018 interview.\footnote{In 1995 there were a number of attempts to calculate black hole entropy using D-branes: see for example Sen (1995), Larsen and Wilczek (1996), Cvetic and Tseytlin (1996). It proved difficult to obtain the correct factor of 1/4 for the Bekenstein-Hawking entropy formula from a well-defined microscopic configuration.} 

Only one month after the Strominger-Vafa paper, Callan and Maldacena (1996) already offered an alternative derivation of the entropy of the same system, illustrating again the quick pace of developments. This derivation also suggests that people were actively scouting out the possibilities of the various corners of the parameter space captured in 
 Figure \ref{SVdiagr}. For Callan and Maldacena
calculated the entropy directly in open string theory with Dirichlet boundary conditions, i.e.~in the top-left corner of our diagram. However, their argument was conceptually  different from Strominger and Vafa's. They faced a problem of combinatorics: counting the possible ways in which open strings can attach to the D1 and the D5-branes with given charges.\footnote{Clear presentations of the calculation mentioned here are found in Maldacena (1996:~Section 3), Zwiebach (2009:~Chapter 22.7) and Johnson (2003:~pp.~426-427).} Such a counting argument has more of a statistical mechanical flavour than the Strominger-Vafa calculation in which the number of BPS states followed rather abstractly from using the Cardy formula in  a complicated quantum field theory. See Section \ref{generalSV} for a further discussion of the Callan-Maldacena paper. \\
\\
{\it M-Theory Black Hole}\\
\\
(BottomRight):~~The bottom-right corner of Figure \ref{SVdiagr} represents whatever the black hole may be in the short-distance, high-energy description of the closed string theory. In this regime, quantum gravity effects should become directly visible. 

The coupling here has a large value, impeding direct calculations: of, for instance, the Hawking effect, so as to establish whether black hole evaporation is unitary or not. Strominger and Vafa, as indicated earlier, mentioned that this was a particular issue they wished to see resolved. They noted in their article that, indeed, a direct M-theory calculation of the effect was not possible, but that their entropy result made a non-unitary scenario considerably less likely. Clearly, then, their D-brane account was taken to have immediate epistemic import for the full quantum gravity theory: they inferred that the M-theory counterpart of the semiclassical black hole, i.e.~the object in our bottom-right corner, is most probably an object that obeys the traditional unitary quantum mechanical evolution laws.\footnote{Strominger and Vafa (1996:~p.~103).}  Clearly, then, they expected that translations of problems of non-pertubative  quantum gravity to D-brane descriptions at low values of the coupling may reveal more about the physics of full quantum gravity. As we will see in Section \ref{subleadingS}, the Strominger-Vafa calculation itself boosted efforts in subsequent months in this direction.

Yet, M-theory itself does not actually explicitly enter into the Strominger-Vafa argument: its presumed existence plays only a guiding role in the calculation. The formulation of M-theory and what this may bring, specifically about the  unitarity of black hole evaporation, provides the motivation for the Strominger-Vafa analysis. But what the diagram in Figure \ref{SVdiagr} illustrates is that the calculation, and the arguments that Strominger and Vafa use, do not necessarily require the formal \emph{existence} of M-theory---since when turning up the 't Hooft coupling one {\it at the same time} goes to large distances. Only the D-branes and classical gravity solution are explicitly involved in the calculation.

It is important to note the different limiting behaviour as a function of the probing distance, $r$, on the right and on the left of Figure \ref{SVdiagr}. 
In the closed string sector on the right-hand side, $r\gg\sqrt{\a'}$ means that the probing particle is at long distances and scans relatively low energies (supergravity is the infrared approximation of the string theory). But in the open string sector on the left-hand side, it is {\it for small D-brane separations} $r$ that we get the perturbative low energy limit of D-brane physics, viz.~at the bottom-left of the diagram. This, in turn, informs us about the states that are found at high energies  and large separation scales in the open string theory in the top-left corner of the diagram. 
So, `high' and `low' energies are inverted between the two theories on the left- and right-hand sides of Figure \ref{SVdiagr}: an effect that we already encountered in Section \ref{ocsd} (the `UV/IR connection'). This is a manifestation of the open-closed string duality that links the left- and right-hand side of the diagram: the duality maps states with high energy into states with low energy and vice versa.

Thus, in terms of the typical energy scales of the systems considered,  Figure \ref{SVdiagr} is asymmetric around its central vertical axis. 
Furthermore, the D-brane world-volume theory on the bottom-left is a `part' (a `special subsector') of the open string theory that is at the top, in the sense that the D-brane world-volume theory is an effective theory describing the low energy regime---the massless modes---of the full open string theory. That relation between top and bottom is reversed on the right-hand side: 
the semi-classical supergravity theory at the top is a special case or effective description of the full closed string theory or M-theory in the bottom-right corner.

\begin{figure}
\begin{center}
\includegraphics[height=7.6cm]{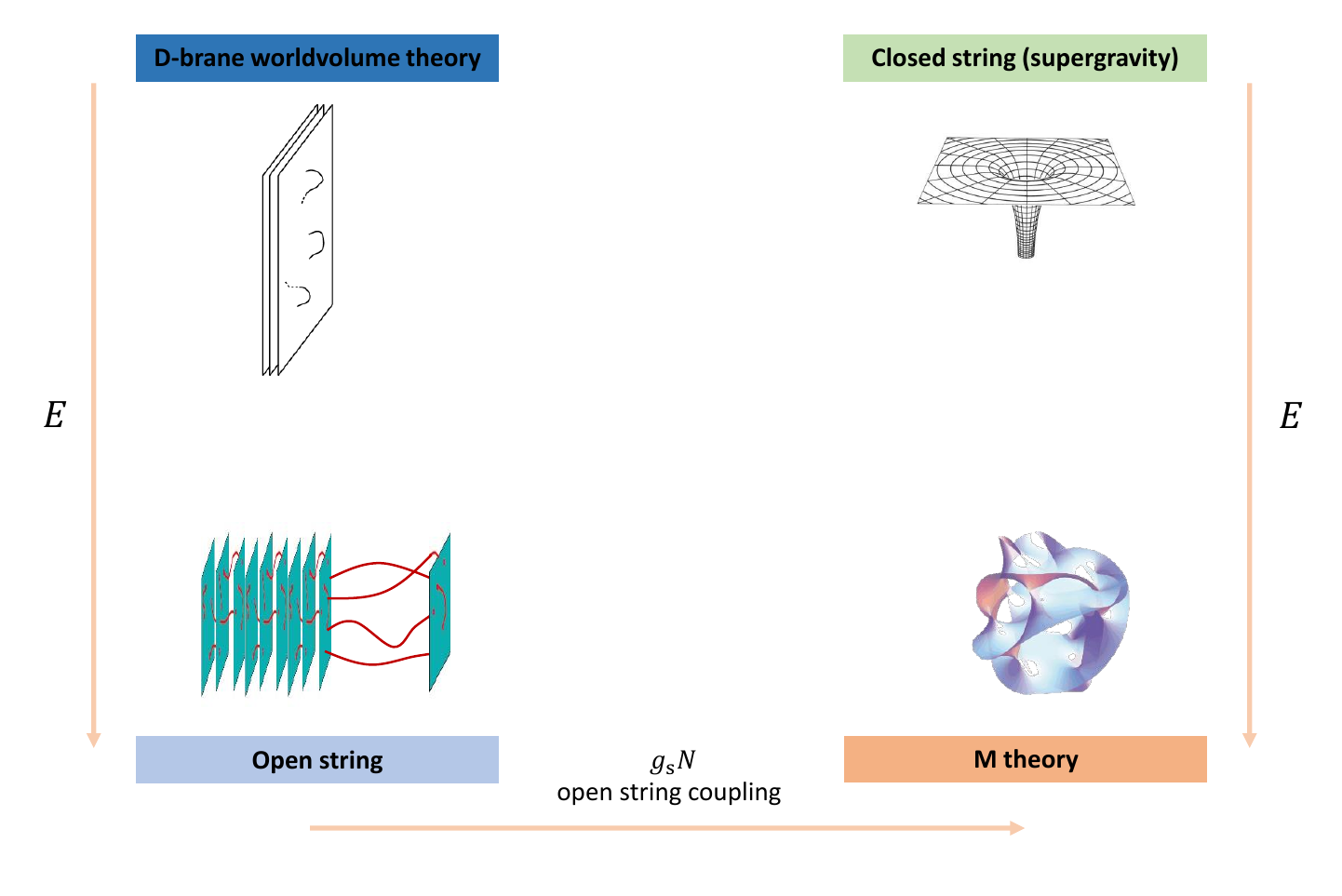}
\caption{\small Relations of energy and hierarchy in the various theories involved in the Strominger-Vafa argument. Compared with Figure \ref{SVdiagr}, the D-brane systems have changed places. In this diagram, energy increases going down, on both sides. The energy scales on the left-hand side are different compared with energy scales on the right-hand side. \emph{Source images}: Wikimedia.}
\label{SVwithEnergy}
\end{center}
\end{figure}

Accordingly, we present another Figure that captures a different set of interrelations between the various theories: Figure \ref{SVwithEnergy}. This particularly reflects the energy relations between the theories. Energy increases downwards on {\em both} sides (note that the energy on the left-hand side is at a different scale compared to the energy on the right-hand side). Consequently, both the more `fundamental' theories are now at the bottom of the Figure, and both effective theories are at the top. This Figure reflects an epistemic hierarchy between the theories: the theories at the top are special limiting cases derived from the more generally valid theories at the bottom, that are however less amenable to calculation.

\section{Idealisations and Generalisations}\label{IG}

Here we first address (Section \ref{idealised}) some of the idealisations that are inherent in the Strominger and Vafa argument and how these may limit its authority in relation to astrophysical black holes. Then we discuss (Section \ref{generalSV}) relevant subsequent work that applied similar state counting or related arguments beyond the idealised setting of the original Strominger and Vafa calculation, thereby increasing the plausibility and physical significance of the result and type of argument.\footnote{Although the idealizations we focus on are certainly among the important ones to examine, we agree that there are others we leave unexamined. In particular, we do not question the assumption of stationarity (Fredenhagen and Haag, 1990). We thank an anonymous referee for pressing this point.} We will be less detailed here than we were in the previous Section, since our primary aim is not to provide a technical analysis of the responses to various concerns that were expressed about the limited viability of the Strominger-Vafa result. Rather, our discussion is intended to invite further work, both foundational and philosophical, by outlining how various generalisations function within the overall enterprise of  black hole microstate-counting. 
In particular, we feel that the physical significance of the Strominger and Vafa (1996) argument cannot be properly addressed without engaging with these sources and the issues they address.

\subsection{Three idealisations made by the Strominger-Vafa set-up}\label{idealised}

We will first address three main idealisations that are required by Strominger and Vafa's  set-up.  Section \ref{generalSV} offers accounts of how people attempted to go beyond these idealisations in the period shortly after the original entropy calculation was presented. 

\subsubsection{Supersymmetry}
\label{supers}
One idealising assumption of the Strominger-Vafa argument is {\it supersymmetry}. This appears at two levels: (i) supersymmetry of the theory, and (ii) supersymmetry of the black hole solution, i.e.~the fact that the black hole is extremal. We treat these in order.\\
\\
(i)  Regarding the supersymmetry of the theory, i.e.~string theory:   The way supersymmetry functions in this argument does not differ essentially from how it functions elsewhere in string theory. String theory is a supersymmetric theory, and---in the absence of the experimental verification of string theory, or of any other theory of quantum gravity---the assumption that the underlying microscopic theory is supersymmetric does not seem to be a major obstacle for the physical salience of the argument, as long as the black holes that are described are themselves physically salient. That is, whether black holes are supersymmetric or not is what matters. And so, what needs to be secured, for the physical salience of this particular calculation, is the salience of the black hole solution itself (cf.~(ii) below). For the physical salience of the underlying microscopic description is far beyond what is now testable. 

This viewpoint is not to be construed as an expression of instrumentalism. Rather, it is a reflection of the empirical status of quantum gravity. Since the microscopic physics is inaccessible to current observations, the current role and relevance of the underlying microscopic string theory is primarily  conceptual. Therefore, the interest of microscopic explanations of black holes should depend first on the physical salience of the black hole solutions themselves. 
\\
\\
(ii)  Regarding the supersymmetric nature of the black hole solution itself  (i.e.~its extremal nature, see Sections \ref{ocsd} and \ref{SVcalculation}):  This {\it is} a threat to the physical---or, at least, the empirical---salience of the argument. For all the observed black holes in the universe are non-extremal.\footnote{But it has been suggested that some astrophysical black holes are near-extremal: cf.~McClintock et al.~(2006) and Guica et al.~(2009). These black holes are electrically neutral. We thank an anonymous referee for this point.} Extremality constitutes the boundary of the region of parameters that is considered to be physical: beyond extremality, i.e.~for $m<|q|$, the horizon disappears, and one is left with a naked singularity. Also, although an extremal black hole has an entropy (as we have seen), it has zero Hawking temperature, $T_{\tn H}=0$, and it is thus arguably an idealized thermodynamic object.\footnote{The Hawking temperature of a Reissner-Nordstr\"om black hole is $T_{\tn H}={(r_+-r_-)}/4\pi r_+^2$ in four dimensions and $T_{\tn H}={(r_+^2-r_-^2)/2\pi r_+^3}$ in five dimensions, where $r_{\pm}=m\pm\sqrt{m^2-q^2}$ and $r_{\pm}^2=m\pm\sqrt{m^2-q^2}$ are the radii of the outer and inner horizon in four and five dimensions respectively. In the extremal case $m=q$ so   $r_+=r_-=m$, so the temperature is zero in both four and five dimensions.\label{ftemperature}} To increase the physical relevance of black hole entropy calculations, one has sought to go away from extremality (see Section \ref{NEBHs}). Furthermore,  Reissner-Nordstr\"{o}m black holes are less relevant for astrophysics, since astrophysical black holes do not carry an intrinsic electric or magnetic charge. However, astrophysical black holes do spin around a particular axis.\footnote{See e.g.~Broderick et al.~(2015).} Hence, generalisation of the Strominger-Vafa argument to spinning black holes is desirable (see Section \ref{spinBH}).

\subsubsection{Higher dimensions}
\label{higherdim}
A second idealisation concerns the use of {\it higher dimensions}: the Strominger-Vafa black hole is five-dimensional. Strominger and Vafa (1996:~p.~100)   anticipated, though, that their calculation could be extended to other dimensions:  
\begin{quote}\small
    The five-dimensional problem is considered here because it seems to be the simplest non-trivial case. We expect that similar calculations will reproduce $S_{\tn{BH}}$ for other types of black holes in string theory. 
\end{quote}    
If the calculation only worked in five dimensions but not in other dimensions (especially, in four dimensions), this would be a good reason to question the generality and salience of the argument---it could be a case of `luck', and irrelevant to four dimensions. 
However, there are four-dimensional versions of the argument, as we will see in Section \ref{4dBH}.

\subsubsection{Approximations}
\label{approx}
A third idealising assumption concerns the various {\it approximations} used, as shown in Figure~\ref{SVdiagr}. In particular, the black hole description is only valid at low energies (large distances) and strong 't Hooft coupling. These approximations limit the validity of the supergravity part of the calculation by Strominger and Vafa to the regime of parameters discussed in Section \ref{SVanalyse}. That is, the regime of: (a) large charges, Eq.~\eq{sugraC}; (b) large distances, $r$, compared to the string length.

Regime  (a)  of large charges (compared to the string's elementary charge) amounts to the presence of a large number of D-branes (see equation~\eq{chargesM}) and high open-string momenta. These are reasonable assumptions for a macroscopic black hole.

Regime  (b)  of low energies coincides precisely with the regime for which the semi-classical calculations of the Bekenstein-Hawking entropy, in any of their guises, are also valid. So, it is the regime for which one would hope to find agreement. However, in Section \ref{subleadingS} we will see that subsequent work calculated sub-leading (quantum) corrections to the leading Bekenstein-Hawking result, and thus went beyond a classical limit.\\

\noindent Our preliminary conclusion is that, of the three key idealisations made by the Strominger-Vafa analysis that affect the salience of the argument, the third one is innocuous, but the first two are substantive. If the counting argument could not be extended beyond the extremal case, or for spinning black holes, or if it did not work in four dimensions, the result might well turn out to be a mere curiosity. Subsequent work, however, improved this situation.

\subsection{Four generalisations of the Strominger-Vafa calculation}\label{generalSV}

We now discuss four generalisations of the Strominger-Vafa calculation that shed further light on Figure \ref{SVdiagr} and address the concerns raised in Section \ref{idealised}. They are: (i) Going beyond the extremal limit to the near-extremal black hole; (ii) Adding spin to the Strominger-Vafa black hole; (iii) Doing entropy calculations for four-dimensional black holes;  (iv) Calculating sub-leading corrections to the Bekenstein-Hawking entropy.\footnote{In a lecture on `String theory and the Bekenstein-Hawking black hole entropy' held at Harvard in October 1997, A.~Strominger listed six points of further work: in addition to our topics (i)-(iii) and the greybody factors which we will discuss in Section \ref{svads}, his list also included ``different compactifications'' (Kaplan et al.~(1997), Ferrara et al.~(1996), Behrndt et al.~(1997)) and ``different charges'' (Dijkgraaf et al.~(1997), Balasubramanian et al.~(1996), Breckenridge et al.~(1996)). Our choice above is motivated by the interest in moving away from idealisations and towards more realistic black holes (cf.~Section \ref{idealised}), rather than by the theoretical interest in extending the Strominger-Vafa calculation to other black holes. (Strominger did not list our (iv), the most important contributions to which appeared one month after his presentation.)}

However, these generalisations do more than just address the concerns expressed in Section \ref{idealised}. They should be seen primarily as positive proposals, in particular concerning (with numbering as above): (i) the string-theoretic mechanism for black hole evaporation, and (iv) quantum corrections to the Bekenstein-Hawking entropy formula.

\subsubsection{Non-extremal black holes}\label{NEBHs}

A first microscopic calculation for a {\it near-extremal} five-dimensional black hole was published by Curtis Callan and his then-graduate student Juan Maldacena in February 1996. A near-extremal black hole has a mass $M$ that is larger than its charge by a small amount:
\bea
M=M_0+\d M=|Q|+\d M,
\eea
where $\d M>0$ is the small amount above extremality, and $M_0$ is the mass of an extremal black hole, $M_0=|Q|$. Thus, the non-extremal black hole can be treated as a linear perturbation of an extremal black hole.

An extremal black hole has zero temperature, as we saw in Section \ref{idealised}. For a near-extremal Reissner-Nordstr\"om black hole, the outer and inner horizons are separated by a small radial distance. Since the Hawking temperature is related to the difference between the two horizons (more precisely, in five dimensions it is proportional to $r_+^2 - r_-^2$; see footnote \ref{ftemperature}), there is a small, non-zero temperature: so the black hole evaporates. The modified Bekenstein-Hawking entropy and the Hawking temperature of the five-dimensional black hole are both proportional to the square root of the `excess' mass  $\d M$,\footnote{Cf.~Eqs.~(2.3) and (2.4) in Callan and Maldacena (1996).}
\bea\label{NEE}
{\frac{\d S}{S_0}}&=&\frac{3}{\sqrt{2}}\sqrt{ \frac{\d M}{ M_0}} \, , \\
T_{\tn H}&=&\frac{2}{\pi r_{0}}\,\sqrt{\frac{\d M}{2M_0}}~,\label{NET}
\eea
where $S_0$ is the entropy of the extremal black hole and $r_0$ is its horizon radius. 

Callan and Maldacena's  black hole is a solution to type IIB supergravity. It corresponds to a configuration of D1-D5-branes in type IIB string theory that  is a special case of the Strominger-Vafa system, with additional momentum in the direction opposite to the momentum in the original calculation (see Section \ref{heartofdarkness}). Callan and Maldacena calculated the excess entropy microscopically (i.e.~in the D-brane set-up) by noticing that the excess entropy is given by the square root of the increase in momentum.
When rewriting those numbers in terms of the mass, they reproduced exactly  the result that was familiar from semi-classical gravity, Eq.~\eq{NEE}.\\

\noindent The D-branes in the Callan-Maldacena set-up are non-BPS states: supersymmetry is broken by the added momentum, so the system will decay. The process of decay is imagined as the scattering between open strings moving in opposite directions along the D-brane, which then join to  form a massless closed string (a graviton) that leaves the brane as Hawking radiation. 
Callan and Maldacena  described this process in detail and reproduced the near-extremal value of the Hawking temperature, Eq.~\eq{NET}, from the open string calculation. They also calculated the {\it radiation rate}. This came out proportional to the black hole area and confirmed the black body spectrum predicted by Hawking (up to an overall numerical constant).

So, the calculation represented the microscopic mechanism for black hole evaporation as the emission of closed strings. 
As was the case in the Strominger-Vafa calculation, it was done in a regime in which there is no classical black hole at all, but only a number of intersecting D-branes with strings between them. Still, remarkably, the temperature and the thermal spectrum matched the semi-classical results of Hawking. 

Since supersymmetry is broken in the calculation, however, it was less clear how the necessary extrapolation from weak to strong values of the 't Hooft coupling (rightwards in Figure  \ref{SVdiagr}) could be justified: in other words, why one is justified in expecting that the entropy is `protected'---even though the numerical agreement between the entropies suggests that it is.\footnote{For a  discussion of this point, see also Maldacena and Susskind (1996).}
In any case, these results strengthened confidence about the entropy calculation of Strominger and Vafa, and about inferences to unitary scenarios for black hole evaporation.

Other relevant work was soon done on the entropy of near-extremal black holes. For example, Horowitz, Maldacena and Strominger (1996) constructed a six-parameter family of five-dimensional black hole solutions for type IIB string theory; in the supergravity description, the six parameters gave the mass, charges and asymptotic values of two scalar fields, while in the D-brane description they corresponded to the number of branes, `anti-branes' and momentum modes. 
The Bekenstein-Hawking entropy was derived in several limiting cases from the weak coupling description of the D-brane system, using a still poorly understood duality called `U-duality'. 
In these entropy calculations, the black holes were arbitrarily far from extremality, and even the five-dimensional Schwarzschild solution was included, illustrating that one wished to extend the studies to increasingly physical models.\footnote{See e.g.~also the  extension of these methods to non-extremal rotating four-dimensional black holes by Horowitz, Lowe and Maldacena (1996).}  
 
\subsubsection{Spinning black holes}\label{spinBH}

In February 1996, a month after the Strominger-Vafa preprint, Jason Breckenridge, Robert Myers, Amanda Peet and Cumrun Vafa  extended its argument to extremal spinning and charged black holes. They first obtained a new class of black hole solutions to type IIA supergravity compactified on the internal, five-dimensional manifold $\mbox{K}3 \times S^1$. These are rotating three-charge extremal black holes in five dimensions and a generalisation of the Strominger-Vafa black hole to a black hole with spin. They rotate in two orthogonal planes, while the extremality (or supersymmetry) forces the two angular momenta to be equal and opposite, i.e.~$J_1 = - J_2 = J.$ The Bekenstein-Hawking entropy for these `BMPV' black holes  (named after the publication's authors) is given by:\footnote{See Breckenridge et al.~(1997:~p.~93).}
\begin{equation} \label{BHentropyspin}
S_{\tn{BH}} = 2 \pi \sqrt{\frac{Q_{\tn H} Q_{\tn F}^2}{2} - J^2} \, . 
\end{equation}
For $J=0$, this reduces to the entropy, Eq.~\eqref{SVarelikeLudwig2}, of the Strominger-Vafa black hole. Note that the angular momentum lowers the value of the entropy of the black hole,  because angular momenta cause the horizon to Lorentz contract and hence decrease.\footnote{See Bena, El-Showk and Vercnocke (2013).} 

For large charges and large angular momenta, Breckenridge et al.~were able to reproduce the Bekenstein-Hawking entropy exactly by counting the number of D-branes, with the additional term in the entropy arising from the angular momenta of the D-branes. The successful comparison between supergravity and D-brane computations was again based on ``adiabatic arguments for the invariance of the expression for the entropy under changes in the string coupling.''\footnote{Breckenridge et al.~(1997:~on p.~1 of the preprint version).} Soon after the BPMV paper, Breckenridge, Lowe, Myers, Peet, Strominger and Vafa (1996) generalised the extremal rotating black holes to near-extremal spinning solutions, and they derived the entropy for these black holes again from a microstate counting of D-branes.\footnote{For further reading on microstate counting for rotating black holes, see the review by Peet (2000).}

\subsubsection{Four-dimensional black holes}\label{4dBH}

Arguably, the most important generalisation of the Strominger-Vafa argument is its extension to four dimensional black holes. Both Juan Maldacena and Andrew Strominger (1996) and Clifford Johnson, Ramzi Khuri and Robert Myers (1996) derived the entropy of four-dimensional Reissner-Nordstr\"{o}m extremal black holes in string theory---their preprints actually appeared on the same day, 11 March 1996, on the arXiv server: illustrating once again the focussed group dynamic and the fast pace of developments.

A distinctive feature of these four dimensional black holes is that they carry four electric (or magnetic)  charges---compared to the three charges of the five-dimensional case---in order for the entropy to be non-zero. Their entropy is proportional to the square root of the product of the four charges:\footnote{Cf.~Eq.~(3) of Maldacena and Strominger (1996).}
\begin{equation} \label{4dentropy}
S_{\tn{BH}} =2\pi \sqrt{Q_2 \, Q_6 \, n \, m} \, . 
\end{equation}
In the D-brane picture, Maldacena and Strominger derived this entropy from a microstate counting in type IIA string theory, with the $Q$'s and $n$ and $m$ given by wrapping numbers of D-branes on internal compact manifolds, or momenta in compact directions. 
The derivation of the entropy formula, Eq.~\eqref{4dentropy}, went through the same moves as the Strominger-Vafa argument, and again depended on the Cardy formula, with the extra charge complicating the calculation.
The success of this four-dimensional extension of course added  to the confidence among string theorists that they were on the right track in describing black holes (and gravity more generally) at the quantum level.

\subsubsection{Black hole entropy from M-theory}\label{subleadingS}

As we discussed in Section \ref{ocsd}, Witten's {\it Strings '95} lecture was programmatic: it proposed that 11-dimensional M-theory should reproduce the five known 10-dimensional string theories as well as 11-dimensional supergravity in appropriate limits, and that these limiting theories were connected by various dualities. 
One of the attractive aspects of his conjecture
was that certain quantum aspects of string theory would be retrievable by conducting semi-classical calculations in eleven dimensions, as implied by the proposed duality between 11-dimensional supergravity and the 10-dimensional string theories.
This was because the string theory coupling constant, $g_{\tn s}$, had been `geometrised': it was related to the radius of compactification of the 11th dimension. Weak coupling then  corresponds to a small radius of compactification (effectively reducing the world to 10 dimensions, like in  ordinary string theory), and strong coupling corresponds to a large compactification radius (so that the world is 11-dimensional). 
In the latter situation, the low-energy 11-dimensional supergravity theory may already 
cover a lot of the non-perturbative physics of the string.

After 1995, a number of proposals arose for concrete formulations of M-theory, or for particular sectors thereof, of which the `Matrix Model' by Thomas Banks, Willy Fischler, Stephen Shenker, and Leonard Susskind (1997) and Maldacena's (1998c) AdS/CFT correspondence are best known. 
D-branes played a central role in all of these.  
We will first briefly discuss work by Maldacena, Strominger, and Witten (1997), which consisted of a black hole entropy count conducted in the context of M-theory; the next section will offer a short history of the AdS/CFT correspondence, in the context of black hole entropy countings.

Maldacena, Strominger, and Witten used the world-volume theory of an M5-brane in eleven dimensions to determine the exact entropy of this object from M-theory.
This was then compared to a supergravity calculation for the corresponding black hole, corrected by a one-loop quantum correction term.

An M5-brane is a five-dimensional magnetically charged soliton that has a six-dimensio\-nal world-volume and lives in a space of eleven dimensions.\footnote{The M5-brane is the magnetic dual of the M2-brane, which is often regarded as a fundamental object in M-theory, and which upon compactification turns into a string.} Its full world-volume theory was unknown to Maldacena et al., but they argued that under compactification, it simplified to a particular and familiar quantum field theory. Within this quantum field theory, an exact calculation of the entropy was done---that is, Maldacena and his co-authors computed it to all orders in the coupling constant.
The lowest order reproduced the familiar Bekenstein-Hawking entropy, but they also produced a match for the first order loop correction of the classical value. So, they had computed a correction to the Bekenstein-Hawking entropy due to the quantum nature of the gravity description in this version of M-theory, and this matched with the first order term of the entropy derived from the M5-brane's field theory description. 
This was an impressive result, since it evaluated the first quantum correction to the Bekenstein-Hawking entropy using two very different methods, finding that the answers agreed.

Due to its engagement with M-theory, this calculation by Maldacena et al.~can be seen as an attempt to fill in the bottom-right corner of Figure \ref{SVdiagr}, at least for the specific black hole configuration considered.
Vafa (1998) published a similar calculation for more general four- and five-dimensional black holes three days after the article by Maldacena, Strominger and Witten had appeared.\footnote{Other examples of calculations that go beyond the leading supergravity result, and thus also can be seen as in some sense filling in the lower-right corner of Figure \ref{SVdiagr}, are Li and Martinec (1997); Strominger (1998); Sfetsos and Skenderis (1998); Mohaupt (2001); Sen (2005); and Ryu and Takayanagi (2006).} 
He showed how an exact CFT partition function could be evaluated close to the regime in which the supergravity approximation is valid and he thus obtained the leading quantum correction to the degeneracy of BPS states, i.e.~to black hole entropy, from the cohomology of the moduli space.\footnote{
Further work developed the connection between black holes and the elliptic genus
(see note \ref{topinv}) 
to obtain results valid in the regime of small charges; see e.g Katz et al.~(1999). Also, Ooguri, Strominger, and Vafa (2004) conjectured an expression for the elliptic genus of four-dimensional BPS black holes: this is known as the `OSV conjecture' and allows one to systematically calculate the corrections to the semi-classical black hole entropy. The conjecture relates the elliptic genus of the black hole to the partition function of the topological string. More recently, Haghighat et al.~(2016) derived, from the elliptic genus, a phase diagram for the microscopic entropy, exhibiting two phases: a `black hole' phase that reproduces the entropy of a five-dimensional spinning black hole (including quantum corrections) and a new phase that they interpreted as corresponding to ``small black holes with a stringy scale horizon'' (p.~36). }

\section{Relation to the AdS/CFT Correspondence}\label{rads}

This section will address two issues. In Section \ref{svads},  we will give a brief overview of how elaborations of the Strominger-Vafa entropy calculation and the formulation of the AdS/CFT proposal were related. Then in Section \ref{bhentadscft}, we will discuss how black hole entropy calculations of Strominger-Vafa-like systems are usually conducted in AdS/CFT---these, in fact, provide reinterpretations of the Strominger and Vafa calculation.

\subsection{From Strominger-Vafa to AdS/CFT}\label{svads}

On 27 November 1997, Juan Maldacena posted online the article that contains the celebrated AdS/CFT correspondence. The article postulated a series of conjectured dualities, relating theories of quantum gravity on anti-de Sitter spaces (specifically, string theories and M-theory) and conformal field theories defined on the boundary of those spaces.\footnote{See Maldacena (1998c). For reviews in the physics literature, see Aharony et al.~(2000) and Ammon and Erdmenger (2015). For a general review aimed at philosophers, see De Haro, Mayerson and Butterfield (2016), and for discussions of particular issues such as emergence or the distinction between duality and symmetry, see: Dieks, Dongen, De Haro~(2015), De Haro (2017b), De Haro, Teh, and Butterfield (2017).} Subsequent papers in February of the next year by Steven Gubser, Igor Klebanov, and Alexander Polyakov (1998), and by Edward Witten (1998a), led to a flurry of activity.\footnote{Citation data from INSPIRE (search performed 7 March 2019) reveal that, in the near 3-month period before articles by Gubser et al.~(1998) and Witten (1998a), Maldacena (1998c) had a mere 15 citations. In the three-month period following those two papers (which are themselves highly cited), Maldacena (1998c) was cited 126 times.} 

AdS/CFT promised that the calculation and comparison of physical quantities (specifically: correlation functions of operators) could be performed on either side of the dualities, at reverse regimes for the coupling: either in supergravity or in the conformal field theory, thus promising an inroad to quantum gravity via gauge theory. This promise was strengthened by Witten's (1998a) point that AdS/CFT was a realisation of the holographic principle, postulated earlier by Gerard 't Hooft (1993) and Leonard Susskind (1995) in the context of the information paradox.\footnote{The holographic nature of Maldacena's conjecture was further emphasized by Witten and Susskind (1998).} 
The excitement has continued to this day---Maldacena's (1998c) article has been cited some 17,500 times up to March 2019\footnote{Google scholar count, performed on 6 March 2019.}---and has predominantly shaped the subject of string theory in the last twenty years. 

Maldacena, in an influential review of the subject co-authored with four others (i.e., string theorists Ofer Aharony, Steven Gubser, Hirosi Ooguri and Yaron Oz, 2000), pointed to ``studies of D-branes and black holes in string theory''\footnote{Aharony et al.~(2000:~p.~188).}  as key in the development to his conjecture, along with general issues in field theory, in particular its large $N$-limit. One aspect of studies of string theory black holes that had been important was the calculation of absorption cross-sections of massless fields (scalars and gravitons) for various $p$-branes in supergravity. These were compared to calculations in the world-volume theories of the corresponding D-branes, and found to agree.\footnote{Klebanov (1997); Gubser, Klebanov and Tseytlin (1997); Gubser and Klebanov (1997). 
Earlier, Klebanov and Tseytlin (1996) had calculated the Bekenstein-Hawking entropy for various $p$-branes in supergravity, and successfully compared it qualitatively with the statistical entropy of free massless fields on the world-volume of the corresponding D-branes, treated as an ideal gas of massless particles in $p$ spatial dimensions.}
In particular, Gubser and Klebanov (1997) had used a correspondence between absorption cross-sections and discontinuities in the two-point function of certain operators in the D-brane world-volume theory (or in thermal Green's functions for near-extremal D-branes, i.e.~for D-branes whose mass is close but not equal to their charge) to calculate the contributions of different terms in the D-brane action to the absorption amplitude. They showed that the two-point functions of the stress-energy tensor, relevant for the calculation of absorption of gravitons by the D-brane, were not renormalized beyond one loop, since the relevant contribution came from the trace anomaly, which does not get corrections beyond one loop. This entailed that an earlier-derived agreement of the field theory cross-section with semiclassical supergravity ``would survive all loop corrections."\footnote{Gubser and Klebanov (1997), from the abstract.} 
It pointed to a more profound and direct correspondence: 
this ``successful comparison [...] was the first hint that Green's functions of ${\cal N}=4$ super Yang-Mills theory could be computed from supergravity'', in the words of Aharony et al.~(2000:~p.~202).

These studies were in turn inspired by the work of Strominger and Vafa. For they extended the matching between supergravity and D-brane calculations to quantities different from the entropy: namely, to the absorption cross-sections of black holes.\footnote{Other relevant earlier work is e.g.~Boonstra et al.~(1997) and Sfetsos et al.~(1998), in which the AdS near-horizon geometries of branes play an important role; and  Alexander Polyakov's (1998) idea of the relation between five-dimensional string theory and four-dimensional gauge theory; see Aharony et al.~(2000:~pp.~188, 189, 195).}  Maldacena himself had of course done black hole entropy state counting together with Callan early in 1996 when he was still a graduate student at Princeton (see Section \ref{NEBHs}) and had  already then calculated absorption cross-sections of the near-extremal five-dimensional black hole. Also, Maldacena in 1996 (after submitting his PhD thesis on \emph{Black Holes in String Theory}), together with Strominger, determined this black hole’s ‘greybody factors’, which are a measure for how much the black hole spectrum deviates from Hawking's perfect blackbody radiation: they did not fail to point out the relevance of these findings for the black hole information paradox.\footnote{Maldacena and Strominger (1997a); related work is Das and Mathur (1996), Horowitz and Strominger (1996).} 
The values of the greybody factor were due to frequency-dependent potential barriers outside the horizon, and agreement with a corresponding effect for the D-brane bound state was claimed. 

In June of 1997, a few months before circulating his AdS/CFT preprint, Maldacena lectured at the Amsterdam {\it Strings '97} conference, and his subject was a clear lead-up to the celebrated duality. Indeed, he was interested in studies that used D-branes to `probe' spacetime geometries, in particular those of black holes. The idea was that a D-brane probe would be connected to other D-branes by massive open strings, and that when integrating out these degrees of freedom, an effective action for the D-brane's motion would be obtained. The latter would then be reinterpreted as prescribing a supergravity background, for which the metric was reconstructed from the effective D-brane action. Thus, geometry was to follow from D-brane quantum field theory.\footnote{See Maldacena (1998b) for his Amsterdam conference contribution.}  

This approach had been developed by a number of string theorists: among them Maldacena himself, who had elaborated particularly the case of near-extremal black holes, in line with his earlier work.\footnote{Maldacena (1998a);
other references that developed the approach are Douglas (1998); Banks, Douglas and Seiberg  (1996); Douglas, Kabat, Pouliot and Shenker  (1997).}
Douglas, Polchinski and Strominger (1997) had studied the case of the extremal black hole, that is, the black hole of the Strominger-Vafa analysis, and had shown that there was a perfect match between the D-brane effective action and the black hole geometry up to first order. At second order---at the two-loop term in the D-brane interaction---Douglas et al.~actually reported a failure to match the two descriptions. In Amsterdam, Maldacena lectured that the near-extremal case gave perfect agreement between the gauge theory and supergravity at one loop, and furthermore, that his results showed that the second loop term for the BPS black hole would also likely work out.\footnote{A similar claim was made in Chepelev and Tseytlin (1998).} These results told him that ``supergravity solutions demand a certain behavior for large $N$ diagrams in gauge theories.'' He further observed that ``[t]his correspondence [...] arises just from the physics of black holes, but can, of course, be of use in matrix theory''---in other words, it may offer ``new insights" for a full version of quantum gravity which had ``only [begun] to be explored."\footnote{Maldacena (1998b:~pp.~25-26).} The full exploration of that correspondence would ensue only a few months later, when the AdS/CFT conjecture was firstly published on the physics `arXiv'.

Clearly, black hole studies were essential in the lead-up to AdS/CFT; and in some of these Maldacena was directly involved, often in fact in collaboration with Andrew Strominger, whom he had joined at Harvard. All such studies, as the cases in the previous Section \ref{generalSV} illustrate, shared the comparison of a D-brane-\emph{cum}-string system with an appropriate black hole in supergravity, in the limit of large $g_{\tn s}N$. In different particular cases, different quantities were compared (e.g.~entropy, or cross sections); but the strategy of matching systems to find matching quantities, and, vice versa, to see such matches as an indication of the sameness of the systems, was common among them.\footnote{The issue of the presumed `sameness' of the systems, and its relation to emergence is the subject of our companion paper, `Emergence and Correspondence for String Theory Black Holes'.}

\begin{figure}
\begin{center}
\includegraphics[height=7.9cm]{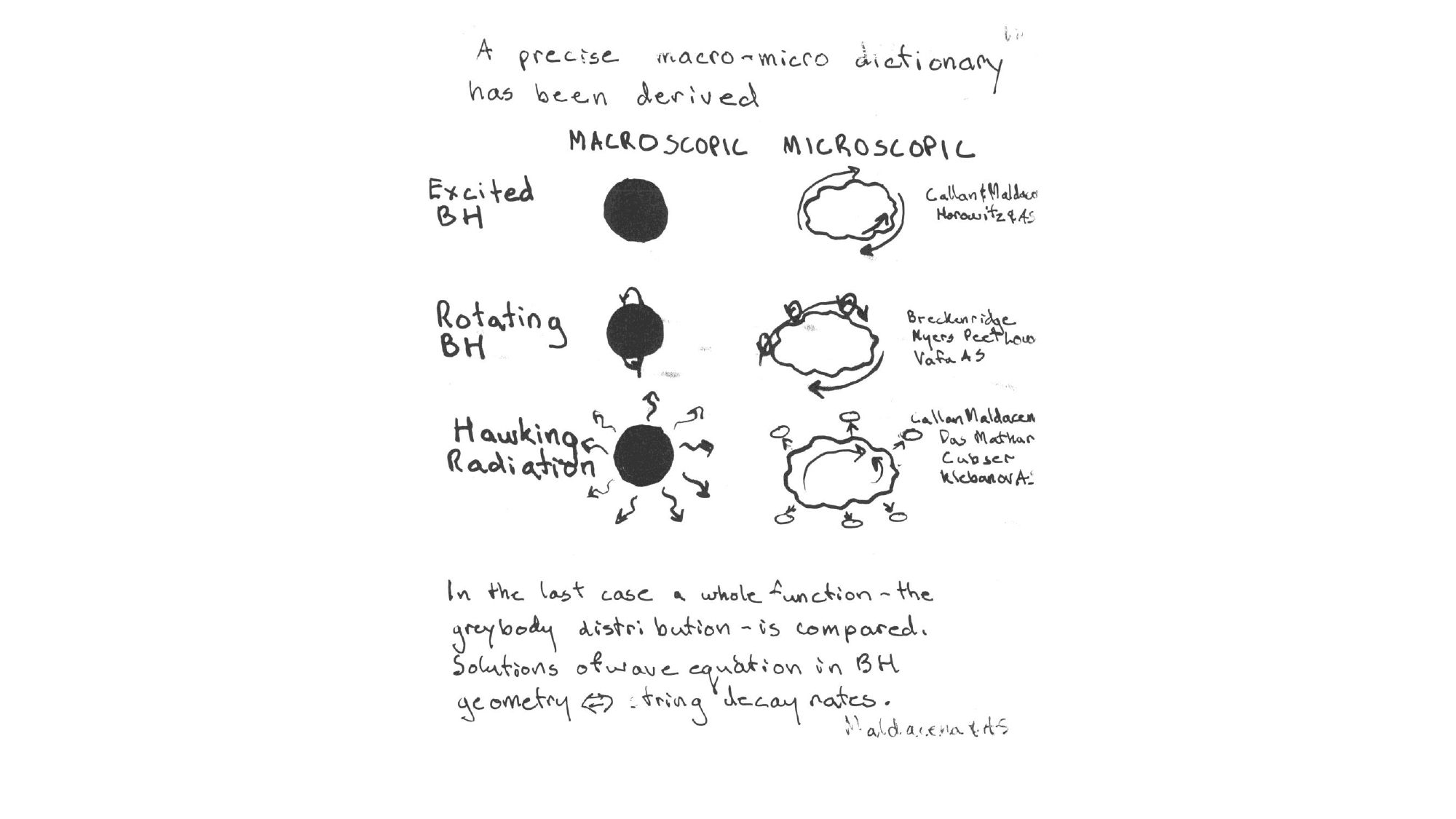}
\caption{\small Transparency from Strominger's lecture on `String theory and the Bekenstein-Hawking black hole entropy' at Harvard in October 1997 which expresses that string theorists have constructed a ``dictionary" between macroscopic black holes and microscopic string systems.  \emph{Source}: A.~Strominger.}
\label{Strominger}
\end{center}
\end{figure}

Of course, this was also the argumentative structure of Strominger and Vafa's analysis. Black holes, thus, brought one to the brink of duality. The point is illustrated by the transparencies of a lecture by Andrew Strominger at Harvard in October 1997 (Figure \ref{Strominger}): above two columns, one depicting black holes, the other string theory-systems, he wrote that a ``macro-micro dictionary has been derived" that compares black holes (``macroscopic'') and stringy systems (``microscopic''). ``In fact," his transparencies continued, ``the effective string picture works better than it should. Agreement persists even when fundamental string theory is strongly coupled! Apparently we stumbled on a low-energy effective quantum description of extremal black holes whose validity transcends its origin."\footnote{A.~Strominger, `String theory and the Bekenstein-Hawking black hole entropy', transparencies of lecture at Harvard, October 1997.} `Effective string theory' was the name given to the field theory that captured the low energy dynamics of D-brane intersections that matched supergravity black holes. Gubser and Klebanov, in August of the same year, believed that entropy counts suggested ``connections" between black holes and 1+1 conformal field theories, realised by such ```effective string' models"; many non-trivial tests confirmed these connections, even if there was still ``little understanding of the `effective string' from first principles''.\footnote{Gubser and Klebanov (1997:~p.~48).}

Maldacena, in his article of November 1997, argued that in an additional ``decoupling limit" of small string length at fixed energies, the black hole spacetimes turn into AdS-spaces. He showed this concretely in the cases of supergravity solutions that matched D3, D1-D5 and M5-brane charges.   
In the quantum system of D-branes and strings, the same limit makes the bulk string theory `decouple' from the field on the D-brane world volume; the remaining two systems then exhibited the exact same excitations, which led Maldacena to conjecture that they were each other's dual, also away from the limit of large~$N$. Together with Strominger, Maldacena had just studied the same decoupling limit for a system of `NS' five-branes, which matched a near-extremal five-dimensional black hole; here, however, decoupling was not complete.\footnote{Maldacena and Strominger (1997b); for the decoupling limit, see also Seiberg (1997). Strominger (1996:~p.~47) studied the idea of decoupling of the D-brane dynamics from supergravity.}

In his publication that contained the AdS/CFT conjecture, Maldacena (1998c:~p.~242) used this decoupling limit to discuss the \emph{near-horizon} physics of the Strominger-Vafa black hole. The geometry on which the low energy string theory lives in this limit is given by AdS$_3$ times an internal manifold, which Maldacena conjectured to be dual to a certain branch of the conformal field theory of the D1-D5-brane used by Strominger and Vafa. He now formulated a \emph{precise conjecture}: type IIB string theory on AdS$_3$ times an internal manifold is \emph{dual} to a (1+1)-dimensional CFT describing the D1-D5 system.\footnote{This also resolved a puzzle regarding the status of the `effective string' living in the D1-D5 intersection. Namely, Vafa and Witten (1994) had pointed out that the partition function of ${\cal N}=4$ super Yang-Mills theory on K3 gives the partition function of a bosonic string, which was initially difficult to understand. After Polchinski's introduction of D-branes, Vafa (1996a, 1996b) identified the relevant string with a bound state of D-branes on the intersection. Furthermore, Callan and Maldacena (1996, cf.~also Horowitz and Strominger, 1996) noticed that the dynamics of the D-brane intersection, described by the 1+1-dimensional CFT, was related to a string theory for the strings attached to the intersection. The AdS$_3$/CFT$_2$ duality suggested a precise formulation for a type IIB string theory of this form. The conjecture focusses on the Higgs branch of the D1-D5 system.\label{bosonicS}} Focussing on the near-horizon region made the asymptotic region of the black hole irrelevant.

Maldacena's previous work on D-brane probes had taught him that the interesting matching of supergravity and D-brane world-volume theory happens near the horizon, and is independent of the physics at infinity.\footnote{Strominger (1998:~p.~8) stated that ``D-brane and supergravity descriptions have an overlapping region of validity for large $gQ$ in the near-horizon small $r$ region.''} Without taking the limit, the open string side (i.e.~the left side) of Figure \ref{SVdiagr} contains both open and closed strings, as does the closed string (i.e.~right) side, which in general complicates the situation when varying the coupling. Yet, the idea behind Polchinski's D-brane construction was that open strings were dual to closed strings. So, to get a duality Maldacena needed to decouple the open and the closed strings. By going close to the D-branes, but keeping the masses of the open strings between the D-branes fixed, the closed strings would become irrelevant, because the open string dynamics between the D-branes would outweigh the closed string dynamics. On the other hand, on the closed string side near the horizon of a (large) black hole, the gravitational interaction would outweigh the open string interactions. So near the horizon, Maldacena realised, closed and open strings decouple (hence `decoupling limit').\footnote{See Maldacena (1998c:~p.~233), Aharony et al.~(2000:~p.~226).}  
This suggested that there could be a duality, without any approximations, in that limit. In the words of the later review by Aharony et al.~(2000,~p.~206),  ``cutting out the near-horizon region of the supergravity geometry and replacing it with the D-branes leads to an identical response to low-energy external probes", i.e.~probes of the D-brane system and probes of the near-horizon black hole geometry (that is, in the AdS space) would exhibit the exact same force fields in the respective appropriate regimes of coupling and scale. 

\subsection{Black hole entropy in AdS/CFT}\label{bhentadscft}

How was black hole entropy counting \`{a} la Strominger-Vafa reconstituted in the new perspective of AdS/CFT? The main change in perspective entailed going from a situation in which the duality relation of the black hole to the intersection of D-branes\footnote{For a discussion of intersections between D-branes, see Sections \ref{ocsd} and \ref{heartofdarkness}.} is uncertain and possibly only approximate (via open-closed string duality), to a situation in which an exact duality is believed to exist, so that the full general relativity metric is to be reconstructed from the dual field theory.\footnote{Maldacena (1998c,~p.~247) wrote: ``When we study non-extremal black holes in AdS spacetimes we are no longer restricted to low energies, as we were in the discussion in higher dimensions.''} 
Before, the reconstruction of quantities such as the entropy relied on principles such as adiabatic invariance across different scales of the coupling, for example as due to the specific BPS nature of the solutions.  

In AdS/CFT, Maldacena (1998c,~p.~242) focussed on the throat-like region close to the horizon of various black holes. In the   Strominger-Vafa scenario, Maldacena took the D1-D5 metric in type IIB string theory compactified on the four-dimensional internal manifolds $T^4$ or $K3$, also called a `black string' in six dimensions:\footnote{Cf.~Eq.~(4.2) in Maldacena (1998c).}
\begin{equation}
    \begin{aligned}
\dd s^2&=f^{-1/2}\,\left(-\dd t^2+\dd x^2\right)+f^{1/2}\left(\dd r^2+r^2\,\dd\Omega^2_3\right) \\
f&=\left(1+\frac{g\,\alpha'\,Q_1}{v\, r^2}\right)\left(1+\frac{g\,\alpha'\,Q_ 5}{r^2}\right) \\
v&:=\frac{V_4}{(2\pi)^4\,\a'^2}~,
\end{aligned}
\end{equation}
with $V_4$ the volume of the internal manifold. Here, $x$ is the coordinate along the D-string (i.e.~the D1-D5 intersection line, cf.~Section \ref{heartofdarkness}). Upon dimensionally reducing the $x$-direction on a compact $S^1$, this becomes the D1-D5 metric of Callan and Maldacena, discussed in Section  \ref{NEBHs}.

As we explained in the previous Section, the `decoupling' limit
of this metric is obtained by going very close to the horizon (i.e.~taking $r\rightarrow0$) while at the same time the typical mass of open strings stretched between the D-branes (viz.~$r/\alpha'$) and other quantities are kept fixed:\footnote{Cf.~Eq.~(4.1) in Maldacena (1998c).}
\bea
\alpha'\rightarrow0,~~~~U:=\frac{r}{\a'}=\mbox{fixed},~~v=\mbox{fixed},~~~~g_6:=\frac{g}{\sqrt{v}}=\mbox{fixed}.
\eea
In this limit, the above metric reduces to:\footnote{Cf.~Eq.~(4.3) in Maldacena (1998c).}
\bea \label{decouplingmetric}
\dd s^2=\a'\left(\frac{U^2}{ g_6\,\sqrt{Q_1Q_5}}\left(-\dd t^2+\dd x^2\right)+g_6\,\sqrt{Q_1Q_5}\,\frac{\dd U^2}{U^2}+g_6\,\sqrt{Q_1Q_5}\,\dd\Omega_3^2\right).
\eea
Though the metric vanishes in the limit $\a'\rightarrow 0$ because of the overall $\a'$ factor, Maldacena justified measuring the metric in units of the string length $l_{\tn s}^2$, so that ``the metric remains constant in $\a'$ units'', by comparing finite energy excitations in the dual gauge theory to finite proper energies on the gravity side.\footnote{Maldacena (1998c:~p.~234).} The coordinate $U$ can be thought of as an energy scale in the dual field theory, where small $U$ corresponds to low energies  in the theory and large $U$ to high energies.\footnote{Maldacena (1998c:~p.~238).}

The above metric  describes AdS$_3\times S^3$.\footnote{This can be seen by applying the  coordinate transformation $U = \ell^2 /(\alpha'\,z)$ with $\ell^2 = \alpha' \,g_6\, \sqrt{Q_1 Q_5}$, under which the   metric \eqref{decouplingmetric}   turns into that of empty   AdS$_3$ in Poincar\'{e} coordinates, with AdS length $\ell$, times a constant three-dimensional sphere whose radius is equal to  $\ell$.} In three-dimensional gravity, different solutions can be obtained from a given metric by doing global identifications. After certain periodic identifications of the coordinates, the metric thus obtained describes  a `BTZ' black hole times a three-sphere.\footnote{This aspect was  made clear by Hyun (1998:~p.~S533) and Strominger (1998:~p.~6). The acronym `BTZ' refers to the authors who introduced the black hole solution in 2+1 gravity a few years before:  Ba\~nados, Teitelboim and Zanelli~(1992).} 
Indeed, using a coordinate transformation,\footnote{See e.g.~Eq.~(2.19) in Carlip and Teitelboim (1995).} the metric \eqref{decouplingmetric}   can be brought to the   form of a rotating BTZ black hole metric:
\begin{equation}
\dd s^2_{\tn{BTZ}}=-\frac{(\rho^2-\rho_+^2)(\rho^2 - \rho_-^2)}{\ell^2\, \rho^2}\,\dd t^2+\rho^2\left(\dd\phi-\frac{\rho_+ \rho_-}{\ell\, \rho^2}\,\dd t\right)^2+\frac{\ell^2\,\rho^2}{(\rho^2-\rho_+^2)(\rho^2 - \rho_-^2)}\,\dd \rho^2+\ell^2\,\dd\Omega_3^2, \label{btz3}
\end{equation}
where $\ell$ is the radius of curvature of AdS, $\phi$ is periodic with period $2\pi$, and $\rho_+$ and $\rho_-$ are the radii of the outer and inner horizon of the   black hole, respectively.

The relation to the BTZ black hole in 2+1-dimensional gravity was relevant for attempts at entropy counting. For, ever since its introduction in 1992, this black hole had been a favourite for attempts to count microstates in quantum gravity. 
For example, Steven Carlip (1995) had proposed a microscopic derivation of the entropy formula of the BTZ black hole, based on a field theory formulation of 2+1 gravity. In 2+1 dimensions, gravity is a fully topological theory, and only the asymptotic behaviour of the metric matters, which greatly simplified the situation. 

The perspective of the AdS$_3$-limit  gave new insight into   the Strominger-Vafa calculation. David Brown and Marc Henneaux (1986) had already analysed three-dimensional general relativity with a negative cosmological constant, of which AdS$_3$ (i.e.,~Eq.~(\ref{decouplingmetric}) when ignoring the internal $S^3$-sphere) was a solution, and had found that the algebra of diffeomorphisms at infinity was given by two copies of the Virasoro algebra.
Strominger (1998:~p.~4) suggested making this into a quantum gravity duality by conjecturing that ``quantum gravity on AdS$_3$ is a conformal field theory with central charge $c={3\ell / 2G}$'', where the central charge is obtained from the Virasoro algebras at infinity. 
In the (1+1)-dimensional CFT with a central charge on the boundary of AdS$_3$, the entropy can be calculated asymptotically using Cardy's formula (cf.~Eq.~\ref{cardyF}):\footnote{Cf.~Eq.~(5.2) in Strominger (1998).}
\bea
S_{\rm{stat}} = 2 \pi \sqrt{\frac{c~n_R}{6}} +  2 \pi \sqrt{\frac{c~n_L}{6}}\, , 
\eea
where $n_R$ and $n_L$ are the eigenvalues of the right- and left-moving lowest Virasoro operators $L_0$ and $\bar L_0$, respectively. Using the AdS$_3$/CFT$_2$ dictionary for the central charge $c = 3 \ell / 2G$,   mass $M \ell = L_0 + \bar L_0 $ and angular momentum $J = L_0 - \bar L_0$, the entropy becomes:\footnote{Cf.~Eq.~(5.3) in Strominger (1998).}
\bea
S_{\tn{BH}}=\pi\,\sqrt{\frac{\ell(\ell\,M+J)}{2G}}+\pi\,\sqrt{\frac{\ell(\ell\,M-J)}{2G}} = \frac{2 \pi \rho_+}{4 G}\,.
\eea
In the last equality the relation between the mass, angular momentum and outer horizon radius was used. This result agrees with the Bekenstein-Hawking area-entropy formula for the BTZ black hole, since   $2 \pi \rho_+$ is the area of the outer event horizon. 

In fact, upon filling in the value of the central charge $c=6Q_1Q_5$ for the CFT in the Strominger-Vafa case, Strominger (1998:~p.~7) showed that the result also agrees with the Strominger-Vafa value of the entropy for the five-dimensional black hole. In other words, the entropy of the Strominger-Vafa black hole is the entropy of the BTZ black hole obtained in the near-horizon limit, which can be calculated by the (1+1)-dimensional CFT dual to BTZ.

\begin{figure}
\begin{center}
\includegraphics[height=6.5cm]{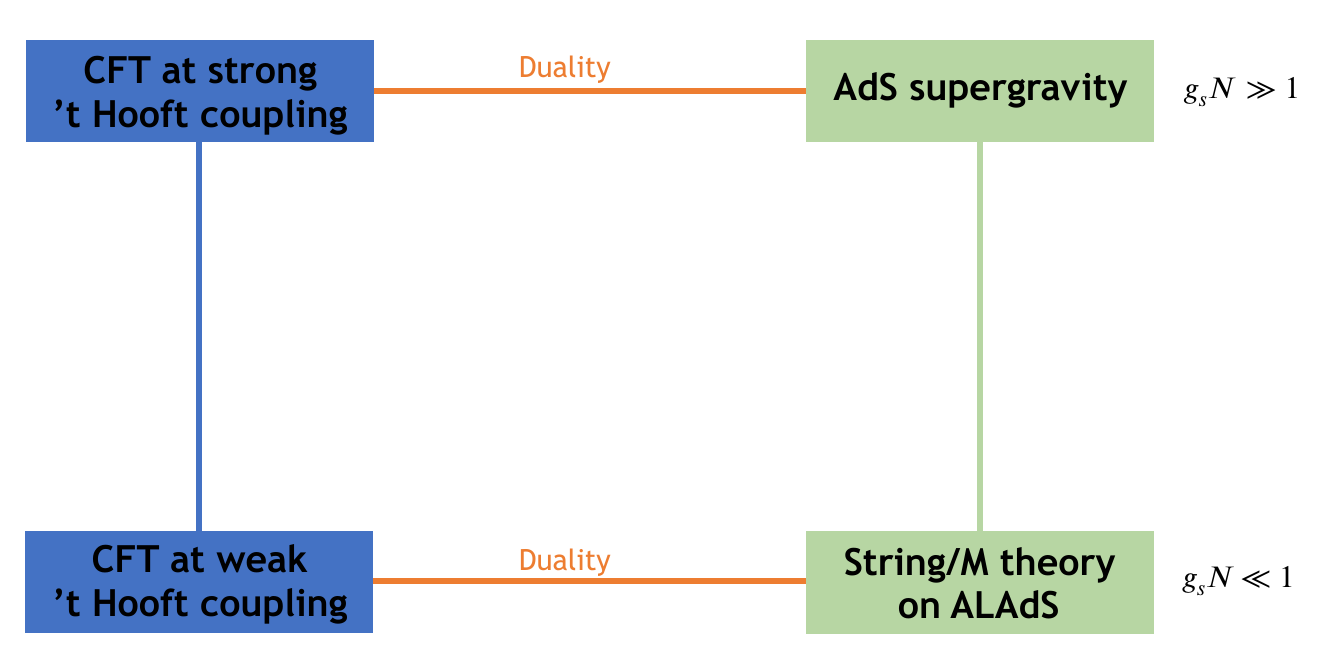}
\caption{\small The standard AdS/CFT picture. The 't Hooft coupling increases going up and there is exact duality at each level between conformal field theory on the left-hand side, and quantum gravity or supergravity on the right-hand side. ALAdS stands for asymptotically locally anti-de Sitter.}
\label{StandardAdSCFT}
\end{center}
\end{figure}

This result gave the Strominger-Vafa calculation a novel and firm footing in the AdS/CFT duality. In our Figure \ref{SVdiagr} of the Strominger-Vafa calculation, open-closed duality could only be assumed to be perturbatively valid, and the entropy result depended on the BPS nature of the solutions on both sides so as to secure invariance of the entropy as the coupling varied.
Yet in the near-horizon limit, AdS/CFT becomes a duality that is exactly valid: it is conjectured to be valid non-perturbatively.

The near-horizon limit focusses on those parts of the system that are certain to be fully equivalent in both descriptions.
Thus we obtain  Figure \ref{StandardAdSCFT}, where one can imagine that `a single theory' applies to the left side; and similarly, to the right-hand side. Namely, on the right we have string theory or M-theory with certain boundary conditions (and supergravity at low energies). On the left, we have a single CFT, with a weakly coupled and a strongly coupled regime; both sides are each other's dual. 

\section{Discussion and Conclusion}\label{dconc}

In this paper, we have analysed the Strominger and Vafa (1996) calculation of black hole entropy in the context of its contemporary literature. Our conceptual analysis of Section \ref{argument} showed how the Strominger-Vafa calculation was rooted in conjectured principles of string theory---such as duality and extrapolation to different values of parameters---which were widely regarded as sound by string theorists. This of course makes clear why the internal consistency of the argument, and its agreement with the overall string theory picture, rendered it a robust and convincing result in the eyes of the string theory community. 
Furthermore, some of the generalisations that we discussed in Section \ref{IG}, especially the success of the calculations for non-extremal, spinning and four-dimensional black holes, made the Strominger-Vafa result physically more significant, which would give it a wider appeal among the larger community of black hole physicists. 

Further theoretical developments spawned by the Strominger-Vafa result involved the construction of new theories: the calculation of quantum corrections to black hole entropy and radiation was used as a way to navigate the  M-theory conjecture, and was instrumental in the formulation of the non-perturbative AdS/CFT conjecture. 
These developments further buttressed the authority awarded to the calculation by Strominger and Vafa. They also illustrate a different point. These attempts can be distinguished from 
`mere generalisations' of the Strominger-Vafa calculation within known theories (e.g.~calculations of entropy for near-extremal or rotating systems): they are attempts at `new theory construction'.

This distinction mirrors the distinction between `theoretical' and  `heuristic' functions of dualities and of theoretical equivalence in theory construction.\footnote{For an elaboration of the distinction between these two functions of dualities, and how these functions contribute to achieving various aims of scientific theories, see De Haro (2019:~Section 3).}
By the {\it theoretical function} of a conjectured inter-theoretic relation, we mean using  that relation to formulate theories which instantiate the relation---theories that are somehow `already given', and `merely need to be worked out'. A conjectured relation's {\it heuristic function}, by contrast, means using it to  guess new theories: which need not necessarily instantiate the conjectured relation.\footnote{A duality may then be a good tool to transfer `cognitive goods' across theories; for a general historical treatment of such transfer, e.g., across disciplines, see Bod et al. (2019).} The works that generalised the Strominger-Vafa scenario in order to make it more physically salient (cf.~Sections \ref{NEBHs}-\ref{4dBH}) reflect a `theoretical function' of the Strominger-Vafa calculation. These works aimed at filling in and strengthening the various aspects of the Strominger-Vafa argument itself: including re-doing the calculation for other already-known theories and similar systems. Here, the Strominger and Vafa result is taken as a specific type of calculation that can be instantiated anew in various generalising examples: namely, a microscopic counting of black hole entropy.

In the case of M-theory (cf.~Section \ref{subleadingS}) and AdS/CFT (cf.~Section \ref{rads}), the (approximate) instantiations of the Strominger-Vafa calculation are not aimed at strengthening or completing that calculation itself, but rather at furthering the construction of a novel theory. Thus they are examples of the heuristic role of the calculation. In the context of M-theory, the calculation aided that theory's development, in the sense that the Maldacena et al.~(1997) paper partially fills in the bottom-right corner of Figure 2; because  their derivation of the Bekenstein-Hawking formula from M-theory also suggests what are the correct degrees of freedom in M-theory. The method was used to calculate quantum corrections to the Bekenstein-Hawking entropy. Thus the Strominger-Vafa calculation is here instantiated only approximately, as is often the case for the heuristic function.  In the case of AdS/CFT, the near-horizon limit led to a new duality, with corresponding new theories: so that the Strominger and Vafa result functioned here too as a heuristic guide. For it suggested details about both the AdS$_3$ configuration and about the (1+1)-dimensional CFT---though the existence of the duality had to be newly inferred by Maldacena, since the heuristic function never leads in an unequivocal manner to new theories. 

These roles of the Strominger-Vafa result illustrate two aspects of its contemporary importance. The theoretical function illustrates that the calculation was considered to be a robust, trustworthy result, whose significance lay in its providing the first microscopic underpinning of black hole entropy, and that one obviously wished to extend. The heuristic function of the Strominger-Vafa result explains why it quickly became absorbed into string theory. For it laid the foundation for other key developments in the field. 

The Strominger-Vafa calculation also left its mark on discussions of the black hole information paradox. The authority awarded to the argument, and its apparently having found a microphysical, quantum-mechanical picture for black holes in a unitary string theory, carried over into the debate on the information paradox: it made the inference to non-unitary evolution of black holes less appealing. For many string theorists and for Strominger and Vafa themselves, as we saw, the result suggested that information could not be lost in black hole evaporation. As Andrew Strominger put it in 2018: 
\begin{quote}\small
[What] ultimately persuaded me---when I would be willing to commit, would be willing to make a bet---is when we did the calculation in string theory and found we got everything out on the nose. Now the person who does the calculation is in a special position because I know I did not cook it---nobody else really knows that, nobody knows that there is not a fudge somewhere and that I did not try a thousand different things. [...] I do not think people were impressed, but [...] Occam's razor [implied that]  it did not make any sense for string theory to supply a way to store exactly all the right amount of information only to destroy it afterwards. We still did not and never have explained how the information gets in and out, we just saw that there is the right amount of information, exactly the right amount. Of course string theory is not the real world but the Hawking paradox exists within string theory, and if it can be solved there, then there is probably a resolution. The rest of the people came around eventually to that.\footnote{Andrew Strominger, interview with Jeroen van Dongen and Sebastian De Haro, 18 November 2018, Harvard University. Cumrun Vafa expressed a similar view: ``the counting of the entropy did not by itself give you a priori a solution of the information paradox, except that [...] we know that D-branes are dual to ordinary string states. A string $S$-matrix ought to be unitary, all the calculations we are doing in string perturbation theory are consistent with unitarity. [...] It would be bizarre if [string theory] has all the properties [in] perturbation theory, but breaks down non-perturbatively'' (interview with S.~De Haro, Harvard University, 30 November 2018). }   
\end{quote}
Thus the Strominger-Vafa account did not give a unitary mechanism for black hole evaporation. But its entropy calculation, grounded in a unitary theory, made advocacy of information loss less attractive to Strominger, and many of his fellow string and field theorists.

Yet we should also note that this opinion was not universally shared. For example, Ted Jacobson, who had been closely following debates on information loss and was not yet convinced of unitary evolution, was not swayed by the calculation done by Strominger and Vafa:
\begin{quote}\small
    [The Strominger and Vafa calculation] did not logically imply to me that black hole evaporation had to be unitary, because it seemed to me, once you turn the coupling up, you can open up a new channel. I thought of the analogy of an electron captured by a nucleus, for example: the evolution of electrons in an atom is not unitary when an electron disappears into the nucleus and becomes another kind of particle, so why couldn't something like that be happening when you turn the coupling up in the context of D-branes?\footnote{T. Jacobson, interview with Sebastian De Haro, 18 May 2017, Seven Pines, Minnesota.}
\end{quote}
Jacobson further expressed the view that in the 1990s, before AdS/CFT, string theory calculations of greybody factors did impress him. But that is not the point we wish to stress here. Jacobson's hesitation illustrates that the \emph{ontological status} of the descriptions deployed in the Strominger-Vafa argument is far from straightforward---are the two sides of the calculation talking about the \emph{same} system?
This concern rightly affects its status as an account of \emph{black hole} entropy and its authority in arbitrating the information paradox. It, of course, may also affect its status as an explanation of black hole entropy.  

However, these issues are beyond the scope of this article; we address it in a separate, complementary article.\footnote{See `Emergence and Correspondence for   String Theory Black Holes'.}  Our intention here has been to lay bare the conceptual lines of inference that ground the black hole entropy counting done by Andrew Strominger and Cumrun Vafa in 1996, and to discuss it in the context of contemporary developments in string theory. As the above questions suggest, however, that result is and should be of immediate interest to historians and philosophers of modern physics: whom we hope to have served with the account presented here.

\section*{Acknowledgements}
\addcontentsline{toc}{section}{Acknowledgements}

We are most grateful for the generous time shared in interviews and in giving feedback to this article and its companion by Andrew Strominger and Cumrun Vafa. We further thank Erik Curiel, Peter Galison, Sean Gryb, Ted Jacobson, Jos Uffink, Erik Verlinde and David Wallace for discussions and essential assistance to our project. We are also grateful for valuable feedback from two referees. SDH's work was supported by the Tarner scholarship in Philosophy of Science and History of Ideas, held at Trinity College, Cambridge, and by a 
fellowship of the Black Hole Initiative at Harvard University; JB and JvD are very grateful for the latter's hospitality in the fall of 2018. MV is supported by the Netherlands Organisation for
Scientific Research (NWO; project number SPI 63-260).

\begin{appendices}

\section{Mathematical Details}\label{mathdet}

In the Appendix we describe  the five-dimensional   extremal black hole solution which Strominger and Vafa studied in more detail. In their paper Strominger and Vafa derived a particular   black hole solution to type IIA supergravity theory, but their notation and conventions are now somewhat out of date and hard to compare with the vast amount of literature on this topic.\footnote{An exception is the paper by Breckenridge, Myers, Peet and Vafa (1997), which \emph{does} use the   same conventions as Strominger and Vafa (1996).} Most reviews and textbooks   present the three-charge extremal black hole solution to type IIB supergravity, when they derive black hole entropy in string theory, but with different notation and normalizations than those used by Strominger and Vafa.  Our aim in this appendix is to compare the Strominger-Vafa black hole with the string theory literature, see e.g.~Maldacena (1996), Skenderis (2000), Peet (2000), Johnson (2003) and Becker et al. (2007). We first work out the type IIA black hole solution by Strominger and Vafa (and check their equations),  then describe the standard type IIB three-charge black hole solution in the literature, and finally compare the two solutions.

 \subsection{The Strominger-Vafa black hole}

As mentioned in the main text in Section \ref{sec:sugraperspective}, Strominger and Vafa start with the five-dimensional type IIA supergravity action, which can be obtained from the ten-dimensional action by compactifying on the internal five-dimensional manifold $\mbox{K3} \times S^1$.\footnote{The type IIA supergravity theory in five dimensions has $\mathcal{N}=4$ supersymmetry, which implies it has   16 real supercharges, of which the BPS states preserve only 4 (see point (i) in Sec. \ref{heartofdarkness}).}
The action \eqref{SVaction} is written in the `string frame', since the term $e^{-2\phi}$ --- which is related to the string coupling $g_{\tn s}$ --- multiplies the Ricci curvature scalar. One can also write the action in the so-called Einstein frame by Weyl rescaling the metric as $g_{\mu\nu} = e^{-4 \phi/3}  \tilde{g}_{\mu\nu}$. In the Einstein frame the action is the standard Einstein-Hilbert action minimally coupled to matter fields:\footnote{Cf.~Eq.~(2.2) in Strominger and Vafa (1996). Strominger and Vafa   mentioned that the action they consider is not the full supergravity action. The full five-dimensional type IIA supergravity action is derived in more detail in Breckenridge et al.~(1997), including all the other  fields arising from the Kaluza-Klein reduction on $\mbox{K3} \times S^1$. The Strominger-Vafa action \eqref{Einsteinframe} can be recovered from   their penultimate equation in section 2 (p.~95)   by turning off three fields, $\sigma = V_{\mu\nu} = X_{\mu \nu} =0$,  and by setting the   Chern-Simons-like term $\frac{1}{8} \epsilon^{\sigma \rho \mu \nu \lambda} \tilde{B}_\sigma F_{\rho \mu} F_{\nu\lambda}$ equal to zero, where $\tilde B$ is the gauge field associated to  $\tilde H$ in Eq.~\eqref{SVmetric}. The $V$ and $X $ field strenghts are also turned off for the  rotating `BMPV' black hole solution, and the dilaton field $\sigma$ can equivalently  be set to zero in that background. The Chern-Simons-like term, however, is nonzero for the   rotating   black hole, but it vanishes for the Strominger-Vafa black hole because of the antisymmetric properties of the epsilon symbol (combined with the fact that  $\tilde B$ only has a $t$-component and $F$ only    a $(tR)$-component, see Eq.~\eqref{SVmetric}).}
  \begin{equation} \label{Einsteinframe}
  S = \frac{1}{16 \pi} \int d^5 x \sqrt{-g} \left ( R - \frac{4}{3} (\nabla \phi)^2 - \frac{1}{4} e^{-4 \phi/3} \tilde{H}^2 - \frac{1}{4} e^{2 \phi/3}F^2 \right) \, . 
  \end{equation}
Here, $\phi$ is the dilaton field, $F$ is a RR  2-form field strength, and $\tilde{H}$ a 2-form field strength  which arises from the  NS-NS three-form after a Kaluza-Klein reduction on the $S^1$. The equations of motion that follow from this action are:  
  \begin{eqnarray}
&  16 \nabla^2 \phi + 2 e^{-4 \phi / 3} \tilde{H}^2- e^{2 \phi/3} F^2  =0 \, , \nonumber \\
& \nabla_\mu \left ( e^{-4\phi/3} \tilde{H}^{\mu\nu}\right) = 0 \, , \quad \nabla_\mu \left ( e^{2\phi/3} F^{\mu\nu}\right) = 0 \, ,  \\
&  R_{\mu \nu} = \frac{4}{3} \partial_\mu \phi \partial_\nu \phi +  \frac{1}{2}e^{-4 \phi/3} \left(\tilde{H}_{\alpha \mu} {\tilde{H}^{\alpha}}_\nu - \frac{1}{6}g_{\mu\nu} \tilde{H}^2 \right) +  \frac{1}{2}e^{2 \phi/3} \left(F_{\alpha \mu} {F^{\alpha}}_\nu - \frac{1}{6} g_{\mu\nu} F^2 \right) \, .  \nonumber
  \end{eqnarray}
Strominger and Vafa   defined the electric charges associated to the field strenghts $\tilde H$ and $F$ as follows:\footnote{Cf.~Eq.~(2.3) in Strominger and Vafa (1996).}
\begin{equation}
 \begin{aligned} \label{SVtwocharges}
 & Q_{\tn H}  \equiv  \frac{1}{4\pi^2} \int_{S^3} e^{-4 \phi/3} \star \tilde{H} \, , \\
&  Q_{\tn F}  \equiv  \frac{1}{16\pi} \int_{S^3} e^{2\phi/3} \star F  \, , 
 \end{aligned}
 \end{equation}
where $\star F$ denotes  the Hodge dual of  $F$. In the current five-dimensional setting, $F$ is a two-form, so $\star F$ is a three-form. Hence it can be integrated over the  three-sphere $S^3$ at infinity. The normalization of the charges are chosen such that $Q_{\tn H}$ and $\frac{1}{2}Q_F^2$ are integers. 

In their paper, Strominger and Vafa considered a special class of solutions to the equations of motion above. They assumed a static, spherically  symmetric metric, a constant dilaton field everywhere in spacetime, and spherically symmetric field strenghts. They worked with the following Ansatz for the solutions:
\begin{equation}
   \begin{aligned} \label{SVmetric}
   & \dd s^2  = - f(R)\, \dd t^2 + \frac{\dd R^2}{ f(R)}+ R^2\, \dd \Omega_3^2~,  \\
  &   \tilde{H} = \tilde{H}_{tR}(R) \dd t \wedge \dd R \, , \quad  F =  F_{tR}(R) \dd t\wedge \dd R \, , \quad  \phi = \phi_h \, . 
  \end{aligned}
    \end{equation}
  Here, $\dd \Omega_3$  is the line element on a three-dimensional unit sphere and $\phi_h$  stands for  the constant value  of the dilaton at the horizon. 
Solving the equations of motion with this Ansatz yields:\footnote{Cf.~Eqs.~(2.6), (2.8) and (2.9) in Strominger and Vafa (1996).}
 \begin{equation}
  \begin{aligned} \label{SVsolutions}
 &   f(R) = \left( 1 - \frac{R_{\tn S}^2}{R^2}\right)^2 \, \quad \text{with} \quad R_{\tn S} =\left (  \frac{8 Q_{\tn H} Q_{\tn F}^2}{\pi^2} \right)^{1/6} \, , \\
& e^{  \phi_h} = \frac{1}{\sqrt{2}}   \frac{4 Q_{\tn F}}{\pi Q_{\tn H}} \, , \quad \tilde{H}_{tR} = \frac{ 4\pi^2 Q_{\tn H}}{ \Omega_3 R^3} e^{4 \phi_h /3} \,, \quad F_{tR} =  \frac{16 \pi Q_F}{  \Omega_3 R^3} e^{-2 \phi_h/3}  \,,
\end{aligned}
\end{equation}
 where $\Omega_3 = 2\pi^2$ is the volume a three-dimensional unit sphere and $R_{\tn S}$ is the horizon radius of the black hole.  The black hole solution is a standard  five-dimensional extremal Reissner-Nordstr\"{o}m solution (as explained in our footnote \ref{RNnote}). The horizon area of the black hole is easily computed to be:
 \begin{equation} \label{area2}
 \text{Area} = \Omega_3 R_{\tn S}^3 = 8 \pi \sqrt{\frac{Q_{\tn H} Q_{\tn F}^2}{2}} \, . 
 \end{equation}
The Bekenstein-Hawking entropy is equal to   the horizon area divided by four, since Strominger and Vafa set   Newton's constant in five dimensions equal to one---see equation \eqref{SVarelikeLudwig2}. Further, the total mass and charge  are given by\footnote{The total mass $M$ and charge $Q$ are related to the mass and charge parameters $m$ and $q$, described in footnote  \ref{RNnote}, through the following equality for the blackening factor,
$
f(R) = 1- \frac{2 m}{R^2} + \frac{q^2}{R^4} = 1- \frac{8 M}{3 \pi R^2} + \frac{  Q^2}{3 \pi^3 R^4 } \, , 
$ for five-dimensional Reissner-Nordstr\"{o}m black holes. Extremality $m=q$ implies the first equation in \eqref{massandcharge}.} 
 \begin{equation} \label{massandcharge}
 M = \frac{1}{4} \sqrt{\frac{3}{\pi}} Q, \qquad Q = \sqrt{3} \pi^{3/2} \left( \frac{8 Q_{\tn H} Q_{\tn F}^2}{\pi^2} \right)^{1/3} \, . 
 \end{equation}
  Note that the overall power of the electric charges matches on both sides of the equation for $Q$. The horizon area can also be   expressed in terms of the total charge: by comparing \eqref{area2} and \eqref{massandcharge}  we see that it is in fact proportional to $Q^{3/2}$.
 Hence,  although it appears that the black hole is parametrized by two charges $Q_{\tn H}$ and $Q_{\tn F}$, it can actually be fully described by a single charge $Q$ (like any   Reissner-Nordstr\"{o}m black hole).

\subsection{The textbook three-charge extremal black hole}
\label{textbookBH}

A three-charge black hole  in five dimensions can be obtained as a solution to ten-dimensional type IIB supergravity theory by a dimensional reduction on a five-dimensional internal manifold. In   textbooks and reviews     the internal manifold is often taken to be the five-torus $T^5$ (although the black hole entropy is the same  for $\mbox{K3} \times S^1$).\footnote{
The  type IIB supergravity theory under consideration has the largest possible number of supersymmetries for a supergravity theory in five dimensions: $\mathcal{N}=8$, hence 32 real supercharges,  of which 4 are preserved by the BPS states.}
The black hole can be constructed with the following configuration of intersecting D$p$-branes: take $N_1$ $D1$-branes wrapped on an $S^1$ of size $R_1$, $N_5$ $D5$-branes wrapped on the five-torus $T^5 = T^4 \times S^1$, and $N_{\tn K}$ momentum modes along the same circle. The resulting five-dimensional metric  in the Einstein frame is   given by:\footnote{Below, we follow Kostas Skenderis' review (2000:~pp.~342-343) of the $5d$ extremal black hole.} 
  \begin{equation}\label{mEf}
  \dd s^2 = - \lambda^{-2/3}\, \dd t^2 + \lambda^{1/3} \left (  \dd r^2 + r^2 \,\dd \Omega_3^2 \right) \, , 
  \end{equation}
  where 
  \begin{equation}\label{qqq}
  \lambda = \left ( 1 + \frac{Q_1}{r^2} \right) \left ( 1 + \frac{Q_5}{r^2}  \right) \left (  1 + \frac{Q_{\tn K}}{r^2}  \right) \, . 
  \end{equation}
  This solution describes an extremal  black hole with three charges $Q_1$, $Q_5$ and $Q_{\tn K}$. The charges are related to the number of branes and the momentum:
  \begin{equation}\label{chargesM}
  Q_1 = \frac{N_1\, (2\pi)^4 g_{\tn s}\, \alpha'^3}{V} \, , \qquad Q_5 = N_5\, g_{\tn s}\, \alpha'\, , \qquad Q_{\tn K} = \frac{N_{\tn K} (2\pi)^4\,g_{\tn s}^2\, \alpha'^4}{R_1^2\, V} \, ,
  \end{equation}
  where $V$ is   the volume of the four-torus $T^4$   and $R_1$ is the radius of the circle $S^1$. 
   The metric is written in so-called isotropic coordinates, which are chosen such that (in the extremal case) the horizon is located at $r=0$.\footnote{For five-dimesional Reissner-Nordstr\"{o}m black holes, described in footnote \ref{RNnote}, the isotropic radial coordinate is defined through  $r^2 = R^2 - R_-^2,$ where $R_-$ is the inner horizon radius. In these isotropic coordinates the horizon is   located at: $r_H^2= R_+^2 - R_-^2 = 2 \sqrt{m^2 - q^2}$, which vanishes in the extremal limit. See Johnson (2003:~pp.~418-420) for further details about $5d$ Reissner-Nordstr\"{o}m black holes.} The horizon area is therefore  given by:
  \begin{equation}
\text{Area} =  \Omega_3 \left (  r^2 \lambda^{1/3} \right)^{3/2}|_{r=0}= 2 \pi^2 \sqrt{Q_1\, Q_5\, Q_{\tn K}} \, . 
  \end{equation}
  Note that this vanishes when any of the three charges vanishes, hence all three charges are needed to support a horizon with a finite area in the supergravity solution. Moreover, the five- and ten-dimensional Newton constant are equal to:
  \begin{equation}\label{gnewton}
  G_5 = \frac{G_{10}}{ 2 \pi R_1 \,V}, \qquad G_{10} = 8 \pi^6\, g_{\tn s}^2\, \alpha'^4 .
  \end{equation} 
  Therefore, the Bekenstein-Hawking entropy can be computed to be:
  \begin{equation} \label{BHentropy}
  S = \frac{\text{Area}}{4\, G_5} =  2 \pi\, \sqrt{N_1\, N_5\, N_{\tn K}} .
  \end{equation}
Note that the entropy only depends on the number of branes, not on the string coupling or string length. Furthermore, the mass  of the three-charge black hole is:
  \begin{equation}
  M = \frac{N_{\tn K}}{R_1} + \frac{N_1 R_1}{g_{\tn s}\alpha'} + \frac{N_5 R_1 V}{g_{\tn s} {\alpha'}^3} \, . 
  \end{equation}
  
 \noindent Finally, how is the   extremal black hole with the three  charges $Q_1, Q_5, Q_{\tn K}$ \eqref{chargesM}  related to the Strominger-Vafa black hole with the two charges $Q_{\tn H}, Q_{\tn F}$ \eqref{SVtwocharges}? 
At the level of the black hole metrics, one can map the two line elements \eqref{SVmetric}-\eqref{SVsolutions} and \eqref{mEf} into each other with the following identifications 
 \begin{equation}
 Q_1 = Q_5 = Q_{\tn K} = \left ( \frac{8 Q_{\tn H} Q_F^2 }{\pi^2}\right)^{1/3} \!\!\!= R_{\tn S}^2 \, \quad \text{and} \quad r^2 = R^2 - R_{\tn S}^2 \, . 
 \end{equation}
 For these tuned values of the charges $Q_1, Q_5, Q_{\tn K}$ the three-charge black hole turns into the standard    five-dimensional extremal Reissner-Nordstro\"{o}m solution.  In particular, the Bekenstein-Hawking entropy  \eqref{BHentropy} becomes equal to the entropy of the Strominger-Vafa black hole, given in  \eqref{SVarelikeLudwig2}. Moreover, the dilaton field becomes a constant, which is what Strominger and Vafa assumed to construct their black hole solution.\footnote{See Maldacena (1996:~p.~36) for an explanation why the dilaton becomes independent of position.}  Therefore, the three-charge extremal black hole can be mapped to the Strominger-Vafa black hole when the three charges are all set equal to a particular combination of the $Q_{\tn H}$ and $Q_{\tn F}$. 
 
\end{appendices}

\section*{References}
\addcontentsline{toc}{section}{References}

\small 

Aharony, O., Gubser, S.S., Maldacena, J.M., Ooguri, H.,~and~Oz, Y.~(2000). `Large $N$ field theories, string theory and gravity', {\it Physics Reports}, 323, pp.~183-386. [hep-th/9905111].\\
\\
Ammon, M.~and Erdmenger, J.~(2015). {\it Gauge/Gravity Duality: Foundations and Applications}. Cambridge: Cambridge University Press.\\
\\
Balasubramanian, V.~and Larsen, F.~(1996). `On D-branes and black holes in four-dimensions', {\it Nuclear Physics B}, 478, pp.~199-208. [hep-th/9604189].\\
\\
Ba\~nados, M., Teitelboim, C.,~and~Zanelli, J.~(1992). `The black hole in three-dimensional space-time', {\it Physical Review Letters}, 69, pp.~1849-1851.  [hep-th/9204099].\\
  \\
Banks, T., Douglas, M.,~and~Seiberg,  N.~(1996). `Probing F-theory with branes',  \emph{Physics Letters B}, 387, pp.~278-281.
[hep-th/9605199].\\
\\
Banks, T., Fischler, W., Shenker, S.H.,~and Susskind, L.~(1997). `M theory as a matrix model: A conjecture', \emph{Physical Review D}, 55, pp.~5112-5128.   
  [hep-th/9610043].\\
\\
Banks, T., O'Loughlin, M.,~and~Strominger, A.~(1993). `Black hole remnants and the information puzzle', {\it Physical Review D}, 47, pp.~4476-4482. 
  [hep-th/9211030].\\
\\
Bardeen, J.M.,  Carter, B.,~and~Hawking, S.W.~(1973). `The four laws of black hole mechanics', \emph{Communications in Mathematical Physics}, {31}, pp.~161-170.\\
 \\ 
Becker, K., Becker, M., and Schwarz, J.H.~(2007). {\it String Theory and M-Theory: A Modern Introduction}. New York: Cambridge University Press.\\
\\
Behrndt, K.~and Mohaupt, T.~(1997). `Entropy of $N=2$ black holes and their M-brane description', {\it Physical Review D}, 56, pp.~2206-2211.
  [hep-th/9611140].\\
 \\
 Bekenstein, J.D.~(1972). `Black holes and the second law', \emph{Lettere al Nuovo Cimento}, {4}, pp.~737-740.
  \\
  \\
Bekenstein, J.D.~(1973). `Black holes and entropy',
  \emph{Physical Review D}, {7}, pp.~2333-2346.
  \\
\\  
Belot, G., Earman, J.,~and Ruetsche, L.~(1999). `The Hawking information loss paradox: The anatomy of controversy', {\it The British Journal for the Philosophy of Science}, 50, pp.~189-229. 
\\
 \\
Bena, I., El-Showk, S.,~and~Vercnocke, B.~(2013). `Black holes in string theory', pp.~59-178 in Bellucci, S. (ed.), {\it Black Objects in Supergravity}, Cham: Springer.\\
\\
Bod, R., Dongen, J. van, Hagen, S. ten, Karstens, B., and Mojet, E. (2019). `The flow of cognitive goods: A historiographical framework for the study of epistemic transfer', \emph{Isis}, 110, in press.\\
\\
Boer, J.~de (1999). `Large $N$ elliptic genus and AdS/CFT correspondence', {\it Journal of High Energy Physics,} 9905, 017. 
  [hep-th/9812240].\\
  \\
Bogomol'nyi, E.B.~(1976). `Stability of classical solutions', \emph{Yadernaya Fizika}, 24, pp.~861-870; translation in: {\it Soviet Journal of Nuclear Physics}, 24, pp.~449-454.\\
\\
Boonstra, H.J., Peeters, B.,~and~Skenderis, K.~(1997). `Duality and asymptotic geometries', \emph{Physics Letters B}, {411}, pp.~59-67.
  [hep-th/9706192]. \\
  \\
Breckenridge, J.C., Lowe, D.A., Myers, R.C., Peet, A.W., Strominger, A.,~and~Vafa, C.~(1996). `Macroscopic and microscopic entropy of near extremal spinning black holes', \emph{Physics Letters B}, {381}, pp.~423-426. 
  [hep-th/9603078].\\
\\ 
Breckenridge, J.C., Myers, R.C., Peet, A.W. and Vafa, C.~(1997). `D-branes and spinning black holes', {\it Physics Letters B},  391, pp.~93-98.
  [hep-th/9602065].\\
\\
Broderick, A.E., Narayan, R., Kormendy, J., Perlman, E.S., Rieke, M.J., and Doeleman, S.S.~(2015). `The event horizon of M87', {\it The Astrophysical Journal}, 805, pp.~1-9.\\
\\
Brown, J.D.~and~Henneaux, M.~(1986). `Central charges in the canonical realization of asymptotic symmetries: An example from three-dimensional gravity', \emph{Communications in Mathematical Physics}, {104}, pp. 207-226.
\\
\\
Callan, C.G., Jr., Martinec, E.J., Perry, M.J. and Friedan, D.~(1985). `Strings in background fields', {\it Nuclear Physics B,} 262, pp.~593-609.\\
\\
Callan, C.G., Jr., Klebanov, I.R.~and Perry, M.J.~(1986). `String theory effective actions', {\it Nuclear Physics B,} 278, pp.~78-90.\\
 \\
Callan, C.G., Giddings, S.B., Harvey, J.A.,~and~Strominger, A.~(1992). `Evanescent black holes', {\it Physical Review D}, 45, pp.~R1005--R1009.
  [hep-th/9111056].\\
\\
Callan, C.G.~and Maldacena, J.M.~(1996). `D-brane approach to black hole quantum mechanics', {\it Nuclear Physics B}, 472, pp.~591-610. 
  [hep-th/9602043].\\
\\
Candelas, P., Horowitz, G.~T., Strominger, A.,~and~Witten, E.~(1985). `Vacuum configurations for superstrings', {\it Nuclear Physics B}, 258, pp.~46-74. 
\\
\\
Cappelli, A., Castellani, E., Colomo, F., and Di Vecchia, P.~(2012). {\it The Birth of String Theory}. Cambridge: Cambridge University Press.\\
\\
Cardy, J.~L.~(1986). `Operator content of two-dimensional conformally invariant theories', {\it Nuclear Physics B}, 270, pp.~186-204. 
\\
\\
Carlip, S.~(1995). `The statistical mechanics of the (2+1)-dimensional black hole', {\it Physical Review D}, 51, pp.~632-637. [gr-qc/9409052].\\
\\
Carlip,   S. and Teitelboim, C.~(1995). 
  `Aspects of black hole quantum mechanics and thermodynamics in (2+1)-dimensions',
  \emph{Physical Review D}, {51}, pp. 622-631.  
  [gr-qc/9405070].\\
  \\
Castellani, E.~(2017). `Duality and ‘particle’ democracy', {\it Studies in History and Philosophy of Science} B: {\it Studies in History and Philosophy of Modern Physics}, 59, pp.~100-108. 
\\
\\
Cecotti, S.~and Vafa, C.~(1993). `On classification of $N=2$ supersymmetric theories', {\it Communications in Mathematical Physics}, 158, pp.~569-644.
  [hep-th/9211097].\\
\\
Chepelev, I.~and Tseytlin, A.A.~(1998). `Long distance interactions of branes: Correspondence between supergravity and super-Yang-Mills descriptions', 
  {\it Nuclear Physics B}, 515, pp.~73-113.
  [hep-th/9709087].\\
\\
Cvetic, M.~and Tseytlin, A.A.~(1996). `Solitonic strings and BPS saturated dyonic black holes', {\it Physical Review D}, 53, pp.~5619-5633.
  [hep-th/9512031].\\
\\
Dai, J., Leigh, R.G., and Polchinski, J.~(1989). `New connections between string theories', {\it Modern Physics Letters A}, 4, pp.~2073-2083. 
\\
\\
Das, S.R.~and Mathur, S.D.~(1996). `Comparing decay rates for black holes and D-branes', {\it Nuclear Physics B}, 478, pp.~561-576.   
  [hep-th/9606185].\\
\\
Dawid, R.~(2013). {\it String Theory and the Scientific Method}.  Cambridge: Cambridge University Press.\\
  \\
De Haro, S.~(2017a). `Dualities and emergent gravity: Gauge/gravity duality', {\it Studies in History and Philosophy of Science} B: {\it Studies in History and Philosophy of Modern Physics}, 59, pp.~109-125. 
[arXiv:1501.06162].\\
 \\
De Haro, S.~(2017b). `Spacetime and physical equivalence'. Forthcoming in Huggett, N.~and W\"uthrich, C.~(eds.), {\it Space and Time after Quantum Gravity}. [arXiv:1707.06581].\\
 \\
De Haro, S.~(2019). `The heuristic function of duality', {\it Synthese},
196 (12), pp.~5169-5203.
[arXiv:1801.09095].\\
\\
De Haro, S.~and Butterfield, J.N.~(2018). `A schema for duality, illustrated by bosonization', pp.~305-376 in: Kouneiher, J.~(ed.), {\it Foundations of Mathematics and Physics One Century after Hilbert}.  Cham: Springer.  [arXiv:1707.06681].\\
\\
De Haro, S. and Dongen, J. van (1998). `Vermoedens', {\it Nederlands Tijdschrift voor Na\-tuur\-kunde}, 64, pp.~284-288.\\
 \\
De Haro, S., Mayerson, D.R.,~and~Butterfield, J.N.~(2016). `Conceptual aspects of gauge/gravity duality', {\it Foundations of Physics}, 46, pp.~1381-1425.
  [arXiv:1509.09231].\\
  \\
De Haro, S.~and~Regt, H.~de~(2018). `A precipice below which lies absurdity? Theories without a spacetime and scientific understanding', {\it Synthese}.  doi:10.1007/s11229-018-1874-9. [arXiv:\-1807.02639]. \\
\\
 De Haro, S., Teh, N.,~and~Butterfield, J.N.~(2017). `Comparing dualities and gauge symmetries', {\it Studies in History and Philosophy of Science} B: {\it Studies in History and Philosophy of Modern Physics}, 59, pp.~68-80.  
  [arXiv:1603.08334].\\
  \\
Dieks, D., Dongen, J. van, De Haro, S.~(2015). `Emergence in holographic scenarios for gravity',
{\it Studies in History and Philosophy of Science} B: {\it Studies in History and Philosophy of Modern Physics}, 52, pp.~203-216. 
[arXiv:1501.04278].\\
\\
Dijkgraaf, R., Verlinde, E.,~and Verlinde, H.~(1997). `5-D black holes and matrix strings', {\it Nuclear Physics B}, 506, pp.~121-142.
  [hep-th/9704018].\\
\\
Dixon, L., Harvey, J.A., Vafa, C.,~and~Witten, E.~(1985). `Strings on orbifolds',  {\it Nuclear Physics B}, 261, pp.~678-686. 
\\
\\
Dixon, L., Harvey, J.A., Vafa, C.,~and~Witten, E.~(1986). `Strings on orbifolds II', {\it Nuclear Physics B}, 274, pp.~285-314. 
\\
\\
Dongen, J. van,~and De Haro, S.~(2004). `On black hole complementarity', {\it Studies in History and Philosophy of Science} B: {\it Studies in History and Philosophy of Modern Physics}, 35, pp.~509-525. 
\\
\\
Douglas, M. (1998). `Gauge fields and D-branes', \emph{Journal of Geometry and Physics}, {28}, pp.~255-262.  
[hep-th/9604198].\\
\\
Douglas, M., Kabat, D., Pouliot, P.,~and Shenker, S.H.~(1997). `D-branes and short distances in string theory', {\it Nuclear Physics B}, 485, pp.~85-127.   
  [hep-th/9608024].\\
\\
Douglas, M., Polchinksi, J.,~and~Strominger, A.~(1997). `Probing five dimensional black holes with D-branes', \emph{Journal of High Energy Physics}, 9712, 003. 
[hep-th/9703031].\\ 
\\
Ellis, G.~and Silk, J.~(2014). `Scientific method: Defend the integrity of physics', {\it Nature}, 516, pp.~321-323. 
\\
\\
Ferrara, S.~and Kallosh, R.~(1996). `Supersymmetry and attractors', {\it Physical Review D}, 54, pp.~1514-1524.
  [hep-th/9602136].\\
\\
Fiola, T.M., Preskill, J., Strominger, A.,~and~Trivedi, S.P.~(1994). `Black hole thermodynamics and information loss in two-dimensions', {\it Physical Review D}, 50, pp.~3987-4014.
  [hep-th/9403137].\\
\\
Fredenhagen, K. and Haag, R. (1990). `On the derivation of Hawking radiation associated with the formation of a black hole', \emph{Communications in Mathematical Physics}, 127, pp. 273-284.\\
\\
Garfinkle, D., Horowitz, G.T.,~and~Strominger, A.~(1991). `Charged black holes in string theory', {\it Physical Review D}, 43, pp.~3140-3143.
\\
\\
Giddings, S.B.,~and~Strominger, A.~(1992). `Dynamics of extremal black holes', {\it Physical Review D}, 46, pp.~627-637.
  [hep-th/9202004].\\
\\  
Green, M.B., Schwarz, J.H.,~and~Witten, E.~(1987). {\it Superstring Theory. Volume 1: Introduction}. Cambridge: Cambridge University Press.\\
\\
Green, M.B.~(1999). `Interconnections between type II superstrings, M theory and N=4 supersymmetric Yang-Mills', {\it Lecture Notes in Physics}, 525, pp.~22-96.
  [hep-th/9903124].\\
\\
Gubser, S.S. and Klebanov, I.R.~(1997). `Absorption by branes and Schwinger terms in the world volume theory', \emph{Physics Letters B}, 413, pp.~41-48. 
[hep-th/9708005]. \\
\\
Gubser, S.S., Klebanov, I.R.,~and~Peet, A.W.~(1996). `Entropy and temperature of black 3-branes', {\it Physical Review D}, 54, pp.~3915-3919. 
  [hep-th/9602135].\\
  \\
Gubser, S.S., Klebanov, I.R.,~and Polyakov, A.M.~(1998). `Gauge theory correlators from noncritical string theory', {\it Physics Letters B}, 428, pp.~105-114. 
  [hep-th/9802109].\\
\\
Gubser, S.S., Klebanov, I.R.,~and~Tseytlin, A.A.~(1997). `String theory and classical absorption by three-branes', {\it Nuclear Physics B}, 499, pp.~217-240.
  [hep-th/9703040].\\
  \\
Guica, M., Hartman, T., Song, W.,~and Strominger, A.~(2009). `The Kerr/CFT Correspondence', {\it Physical Review D,} 80, 124008.\\
  \\
Haghighat, B., Murthy, S., Vafa, C.,~and~Vandoren, S.~(2016). `F-theory, spinning black holes and multi-string branches', {\it Journal of High Energy Physics,} 1601, 009. 
  [arXiv:1509.00455].\\
\\
Hartman, T., Keller, C.A., and Stoica, B. (2014).
`Universal spectrum of 2d conformal field theory in the large $c$ limit',
   \emph{Journal of High Energy Physics}, {1409}, 118. 
  [arXiv:1405.5137].\\
\\
Hawking, S.W.~(1975). 
  `Particle creation by black holes',
  \emph{Communications in Mathematical Physics}, {43}, pp.~199-220.
 \\
 \\
Hawking, S.W.~(1976).
  `Breakdown of predictability in gravitational collapse',
  \emph{Physical Review~D}, {14}, pp.~2460-2473.
 \\
 \\
Hawking, S.W.~and Ellis, G.~(1973). {\it The Large Scale Structure of Space-Time}. Cambridge: Cambridge University Press.\\
\\
Hawking, S.W., Horowitz, G.T.,~and Ross, S.F. (1995).
  `Entropy, area, and black hole pairs',
\emph{Physical Review D}, {51}, pp.~4302-4314.  
  [gr-qc/9409013].\\
\\
Hooft, G.~'t~(1974). `A planar diagram theory for strong interactions', {\it Nuclear Physics B}, 72, pp.~461-473.
\\
\\
Hooft, G.~'t~(1985). `On the quantum structure of a black hole', {\it Nuclear Physics B}, 256, pp.~727-745. 
\\
\\
Hooft, G.~'t~(1993). `Dimensional reduction in quantum gravity', pp. 284-293 in Ali, A., Ellis, J., and Randjbar-Daemi, S. (eds.), {\it Salamfestschrift}, Singapore: World Scientific. [gr-qc/9310026].\\
\\
Horowitz, G.T., Lowe, D.A.,~and Maldacena, J.M.~(1996). `Statistical entropy of nonextremal four-dimensional black holes and U duality', {\it Physical Review Letters}, 77, pp.~430-433.
[hep-th/9603195].\\
  \\
Horowitz, G.T., Maldacena, J.M.,~and Strominger, A.~(1996). `Nonextremal black hole microstates and U-duality', {\it Physics Letters B}, 383, pp.~151-159. 
  [hep-th/9603109].\\
\\
Horowitz, G.T.~and Strominger, A.~(1991). `Black strings and $p$-branes', 
\emph{Nuclear Physics B}, {360}, pp.~197-209.
\\
\\
Horowitz, G.T.~and Strominger, A.~(1996). `Counting states of near-extremal black holes', {\it Physical Review Letters}, 77, pp.~2368-2371. 
[hep-th/9602051].\\
\\
Huggett, N.~(2017). `Target space $\not=$ space', {\it Studies in History and Philosophy of Science} B: {\it Studies in History and Philosophy of Modern Physics}, 59, pp.~81-88. 
[arXiv:1509.06229].\\
\\
Hull, C.M.~and Townsend, P.K.~(1995). `Unity of superstring dualities', {\it Nuclear Physics B}, 438, pp.~109-137.
  [hep-th/9410167].\\
\\
Hyun, S.~(1998). `U-duality between three-dimensional and higher dimensional black holes', {\it Journal of the Korean Physical Society}, 33, pp.~S532-S536.
  [hep-th/9704005].\\
  \\
Jacobson, T.~(1996). {\it Introductory Lectures on Black Hole Thermodynamics}.  Lecture notes, Utrecht University. http://www.physics.umd.edu/grt/taj/776b/lectures.pdf.\\
\\
Jacobson, T.~(2005). 
  `Introduction to quantum fields in curved space-time and the Hawking effect', pp.~38-89 in: Gomberoff, A. and Marolf, D. (eds.), \emph{Lectures on Quantum Gravity}. Boston: Springer.  
  [gr-qc/0308048]. \\
  \\
Johnson, C.V.~(2003). {\it D-Branes}. Cambridge: Cambridge University Press.\\
\\
Johnson, C.V., Khuri, R.R., and Myers, R.C.~(1996). `Entropy of 4-D extremal black holes', {\it Physics Letters B}, 378, pp.~78-86.
  [hep-th/9603061].\\
\\ 
Kaplan, D.M., Lowe, D.A., Maldacena, J.M.,~and Strominger, A.~(1997). `Microscopic entropy of $N=2$ extremal black holes', {\it Physical Review D}, 55, pp.~4898-4902.
  [hep-th/9609204].\\
\\
Katz, S.H., Klemm, A., and  Vafa, C.~(1999). `M-theory, topological strings and spinning black holes', {\it Advances in Theoretical and Mathematical Physics}, 3, pp.~1445-1537. 
  [hep-th/9910181].\\
\\
Kawai, H., Lewellen, D.C., and Tye, S.H.H.~(1986). `A relation between tree
amplitudes of closed and open strings', {\it Nuclear Physics B}, 269, pp. 1-23.
\\
\\
Klebanov, I.R.~(1997). `World volume approach to absorption by non-dilatonic branes', \emph{Nuclear Physics B}, {496}, pp. 231-242. 
  [hep-th/9702076]. \\
  \\
Klebanov, I.R.~and Tseytlin, A.A.~(1996). `Entropy of near extremal black p-branes', {\it Nuclear Physics B}, 475, pp.~164-178.  
  [hep-th/9604089].\\
  \\
Landweber, P.S.~(1986). \emph{Elliptic curves and modular forms in algebraic topology}. Berlin: Springer.\\
\\
Larsen, F.~and Wilczek, F. (1996). `Internal structure of black holes', {\it Physics Letters B}, 375, pp.~37-42.
  [hep-th/9511064].\\
\\
Li, M.~and Martinec, E.J.~(1997).  `On the entropy of matrix black holes', {\it Classical and Quantum Gravity}, 14, pp.~3205-3213. 
[hep-th/9704134].\\
\\
Maldacena, J.M.~(1996). \emph{Black Holes in String Theory}. Ph.D. Thesis, Princeton University. [hep-th/9607235].\\
\\
Maldacena, J.M.~(1998a). `Probing near extremal black holes with D-branes', {\it Physical Review D}, 57, pp.~3736-3741. 
[hep-th/9705053].\\
\\
Maldacena, J.M.~(1998b). `Branes probing black holes', {\it Nuclear Physics B, Proceedings Supplements}, 68, pp.~17-27.
  [hep-th/9709099].\\
\\
Maldacena, J.M.~(1998c). `The large $N$ limit of superconformal field theories and supergravity', 
{\it Advances in Theoretical and Mathematical Physics}, 2, pp.~231-252. 
  [hep-th/9711200].\\
\\
Maldacena, J.M. and Strominger, A.~(1996). `Statistical entropy of four-dimensional extremal black holes', {\it Physical Review Letters}, 77, pp.~428-429. 
[hep-th/9603060].\\
\\
Maldacena, J.M. and Strominger, A.~(1997a). `Black hole grey body factors and D-brane spectroscopy', {\it Physical Review D}, 55, pp.~861-870.
  [hep-th/9609026].\\
\\  
Maldacena, J.M. and Strominger, A.~(1997b). `Semiclassical decay of near-extremal fivebranes', \emph{Journal of High Energy Physics}, 9712, 008. 
[hep-th/9710014]. \\
\\
Maldacena, J.M., Strominger, A., and Witten, E.~(1997). `Black hole entropy in M-theory', {\it Journal of High Energy Physics}, 9712, 002.  
  [hep-th/9711053].\\
\\
Maldacena, J.M.~and Susskind, L.~(1996). `D-branes and fat black holes', {\it Nuclear Physics B}, 475, pp.~679-690. 
[hep-th/9604042].\\
  \\
Matsubara, K.~(2013). `Realism, underdetermination and string theory dualities', {\it Synthese,} 190, pp.~471-489.\\
  \\
Maudlin, T. (2017). `(Information) paradox lost'. [arXiv:1705.03541].\\
\\
McClintock, J.~E., Shafee, R., Narayan, R., Remillard, R.~A., Davis, S.~W., and Li, L.~X.~(2006). `The spin of the near-extreme Kerr black hole GRS 1915+ 105'. {\it The Astrophysical Journal,} 652 (1), 518.\\
\\
Mohaupt, T.~(2001). `Black hole entropy, special geometry and strings',
{\it Fortschritte der Physik}, 49, pp.~3-161.
  [hep-th/0007195].\\
\\
Ooguri, H., Strominger, A.,~and~Vafa, C.~(2004). `Black hole attractors and the topological string', {\it Physical Review D}, 70, 106007.
  [hep-th/0405146].\\
\\
Peet, A.W. (2000). `TASI lectures on black holes in string theory', pp.~353-433 in: Harvey, J.A., Kachru, S., and Silverstein, E. (eds.), {\it Strings, branes and gravity. Proceedings TASI '99}. 
[hep-th/0008241].\\
\\
Polchinski, J.~(1995). `Dirichlet branes and Ramond-Ramond charges', {\it Physical Review Letters}, 75, pp.~4724-4727. 
[hep-th/9510017].\\
\\
Polchinski, J.~(2017). `Memories of a theoretical physicist'. [arXiv:1708.09093].\\
\\
Polchinski, J., Chaudhuri, S.,~and~Johnson, C.V.~(1996). `Notes on D-branes'.   [hep-th/9602052].\\
\\
Polchinski, P.~and~Strominger, A.~(1994). `A possible resolution of the black hole information puzzle', {\it Physical Review D}, 50, pp.~7403-7409.
  [hep-th/9407008].\\
\\
Polyakov, A.M.~(1998). `String theory and quark confinement', {\it Nuclear Physics B. Proceedings Supplements}, 68, pp.~1-8.
  [hep-th/9711002].\\
\\
Prasad, M.K.~and Sommerfield, C.M.~(1975). `Exact classical solution for the 't Hooft monopole and the Julia-Zee dyon', {\it Physical Review Letters}, 35, pp.~760-762. 
\\
\\
Read, J.~and M\o ller-Nielsen, T.~(2018). `Motivating Dualities', {\it Synthese}. doi:10.1007/s11229-018-1817-5.  http://philsci-archive.pitt.edu/14663.\\
\\
Rickles, D.~(2011). `A philosopher looks at string dualities', {\it Studies in History and Philosophy of Science} B: {\it Studies in History and Philosophy of Modern Physics}, 42, pp.~54-67.\\
\\
Rickles, D.~(2014). {\it A Brief History of String Theory. From Dual Models to M-Theory}. Berlin: Springer.\\
\\
Ryu, S.~and~Takayanagi, T.~(2006). `Holographic derivation of entanglement entropy from AdS/CFT', {\it Physical Review Letters}, 96, 181602.
  [hep-th/0603001].\\
\\
Seiberg, N. (1997). `New theories in six-dimensions and matrix theory description of M theory
on $T_{5}$ and $T_{5}/Z^{2}$', \emph{Physics Letters B}, 408, pp.~98-104. [hep-th/9705221].\\
\\
Seiberg, N.~and~Witten, E.~(1994). `Electric-magnetic duality, monopole condensation, and confinement in N=2 supersymmetric Yang-Mills theory', {\it Nuclear Physics B}, 426, pp.~19-52. 
  [hep-th/9407087].\\
\\
Sen, A.~(1995). `Extremal black holes and elementary string states', {\it Modern Physics Letters A}, 10, pp.~2081-2094. 
  [hep-th/9504147].\\
\\
Sen, A.~(2005). `Black hole entropy function and the attractor mechanism in higher derivative gravity', {\it Journal of High Energy Physics}, 0509, 038.
  [hep-th/0506177].\\
\\
Sfetsos, K.~and Skenderis, K.~(1998). `Microscopic derivation of the Bekenstein-Hawking entropy formula for nonextremal black holes', {\it Nuclear Physics B}, 517, pp.~179-204. 
  [hep-th/9711138].\\
\\
Skenderis, K.~(2000). `Black holes and branes in string theory', pp.~325-364 in: Kowalski-Glickman, J. (ed.), {\it Towards Quantum Gravity}. Berlin: Springer. 
  [hep-th/9901050].\\
\\
Strominger, A.~(1990).
  `Heterotic solitons',
 \emph{Nuclear Physics B}, {343}, pp.~167-184.
  \\
\\
Strominger, A.~(1995). `Les Houches lectures on black holes'. 
  [hep-th/9501071].\\
 \\
 Strominger, A.~(1996). `Open p-branes'. {\it Physics Letters B}, 383, pp.~44-47.
  [hep-th/9512059].\\
\\
Strominger, A.~(1998). `Black hole entropy from near-horizon microstates', {\it Journal of High Energy Physics}, 9802,  009. 
  [hep-th/9712251].\\
\\
Strominger, A.~and~Vafa, C.~(1996). `Microscopic origin of the Bekenstein-Hawking entropy', {\it Physics Letters B}, 379, pp.~99-104. 
  [hep-th/9601029].\\
\\
Susskind, L.~(1995). `The world as a hologram', {\it Journal of Mathematical Physics}, 36, pp.~6377-6396. 
  [hep-th/9409089].\\
\\  
Susskind, L.~(1998) [1993]. `Some speculations about black hole entropy in string theory', pp.~118-131 in: Teitelboim, C.~and Zanelli, J.~(eds.), {\it The Black Hole, 25 Years After. Proceedings}. Singapore: Wold Scientific. [hep-th/9309145].\\
\\
Susskind, L.~(2008). {\it The Black Hole War: My Battle with Stephen Hawking to Make the World Safe for Quantum Mechanics}. New York: Back Bay Books.\\
\\
Susskind, L.~and~Witten, E.~(1998). `The holographic bound in anti-de Sitter space'. [hep-th/9805114].\\
  \\
Townsend, P.K.~(1995). `The eleven-dimensional supermembrane revisited', {\it Physics Letters B}, 350, pp.~184-187.
  [hep-th/9501068].\\
\\  
Townsend, P.K.~(1997). {\it Black holes}. Lecture notes, Cambridge University. [gr-qc/9707012].\\
\\
Vafa, C.~(1985). {\it Symmetries, Inequalities and Index Theorems}. Ph.D. Thesis, Princeton University.\\
\\
Vafa, C.~(1996a). `Gas of D-branes and Hagedorn density of BPS states', {\it Nuclear Physics B}, {463}, pp.~415-419.
  [hep-th/9511088].\\
\\  
Vafa, C.~(1996b). `Instantons on D-branes', {\it Nuclear Physics B}, 463, pp.~435-442. 
  [hep-th/9512078].\\
\\
Vafa, C.~(1998). `Black holes and Calabi-Yau threefolds', {\it Advances in Theoretical and Mathematical Physics}, 2, pp.~207-218.   
  [hep-th/9711067].\\
  \\
Vafa, C.~and~Witten, E.~(1994). `A strong coupling test of S-duality', {\it Nuclear Physics B}, 431, pp.~3-77. 
  [hep-th/9408074].\\
\\
Wald, R.M.~(2001).
  `The thermodynamics of black holes', 
  \emph{Living Reviews in Relativity}, {4}:6. doi:10.12942/lrr-2001-6.
  [gr-qc/9912119]. \\
\\  
Wallace, D.~(2017). `Why black hole information loss is paradoxical'. [arXiv:1710.03783]. \\
\\
Wallace, D.~(2019a). 
`The case for black hole thermodynamics, I: Phenomenological thermodynamics', 
\emph{Studies in History and Philosophy of Science} B: {\it Studies in History and Philosophy of Modern Physics}, 64, pp.~52-67. 
[arXiv:1710.02724]. \\
\\
Wallace, D.~(2019b). `The case for black hole thermodynamics, II: Statistical mechanics', \emph{Studies in History and Philosophy of Science Part} B: {\it Studies in History and Philosophy of Modern Physics},  https://doi.org/10.1016/j.shpsb.2018.10.006. [arXiv:1710.02725].\\
\\
Witten, E.~(1987). `Elliptic genera and quantum field theory', {\it Communications in Mathematical Physics}, 109, pp.~525-536. 
\\
\\
Witten, E.~(1995). `String theory dynamics in various dimensions', {\it Nuclear Physics B}, 443, pp.~85-126.
  [hep-th/9503124].\\
\\
Witten, E.~(1996a). `Bound states of strings and $p$-branes', {\it Nuclear Physics B},  460, pp.~335-350. 
  [hep-th/9510135].\\
\\
Witten, E.~(1996b). `Five-branes and M-theory on an orbifold', {\it Nuclear Physics B}, 463, pp.~383-397.
  [hep-th/9512219].\\
\\
Witten, E.~(1998a). `Anti-de Sitter space and holography', {\it Advances in Theoretical and Mathematical Physics}, 2, pp.~253-291. 
  [hep-th/9802150].\\
  \\
Witten, E.~(1998b). `Magic, mystery, and matrix', {\it Notices of the American Mathematical Society}, 45, pp.~1124-1129. 
\\
\\
Zwiebach, B.~(2009). {\it A First Course in String Theory}. Cambridge: Cambridge University Press.

\end{document}